\documentclass[11pt,letterpaper]{article}
\usepackage{latexsym}
\usepackage{amsmath}
\usepackage{latexsym,amssymb}
\usepackage{graphicx}
\usepackage{bm}

\newcommand{\psistate}{\left|\Psi\right>}
\newcommand{\pone}{E^{x}_{\;X}}
\newcommand{\pfour}{E^{y}_{\;Y}}
\newcommand{\pfive}{E^{\zeta}_{\;Z}}
\newcommand{\cone}{A^{X}_{\;x}}
\newcommand{\cfour}{A^{Y}_{\;y}}
\newcommand{\cfive}{A^{Z}_{\;\zeta}}

\newcommand{\pprod}{\pone\pfour}
\newcommand{\ccone}{\frac{\sin(A^{X}_{\;x}\,\delta_{x})}{\delta_{x}}}
\newcommand{\ccfour}{\frac{\sin(A^{Y}_{\;y}\,\delta_{x})}{\delta_{x}}}
\newcommand{\ccfive}{\frac{\sin(A^{Z}_{\;\zeta}\,\delta_{\zeta})}{\delta_{\zeta}}}

\newcommand{\pc}{E_{3}}
\newcommand{\pb}{E_{2}}
\newcommand{\bee}{a_{2}}
\newcommand{\cee}{a_{3}}
\newcommand{\conefactor}{\left(\frac{\cone\sqrt{\Delta}}{\sqrt{\pfive}}\right)}
\newcommand{\cfourfactor}{\left(\frac{\cfour\sqrt{\Delta}}{\sqrt{\pfive}}\right)}
\newcommand{\cfivefactor}{\left(\frac{\cfive\sqrt{\Delta}\sqrt{\pfive}}{\sqrt{\pprod}}\right)}
\newcommand{\bfactor}{\left(\frac{\bee\sqrt{\Delta}}{\sqrt{\pc}}\right)}
\newcommand{\cfactor}{\left(\frac{\cee\sqrt{\Delta}\sqrt{\pc}}{\pb}\right)}

\begin{document}
\title{\bf{\Large The Evolution of $\bm{\Lambda}$ Black Holes in the Mini-Superspace Approximation of Loop Quantum Gravity}}
\author {{\small J. Brannlund \footnote{johanb@mathstat.dal.ca}} \\
\it{\small Department of Mathematics and Statistics, Dalhousie University} \\
\it{\small Halifax, Nova Scotia, B3H 
3J5, Canada }
 \and
{\small S. Kloster \footnote{stevek@sfu.ca}} \\
\it{\small Centre for Experimental and Constructive Mathematics,  Simon Fraser University} \\
\it{\small Burnaby, British Columbia, V5A 1S6, Canada}
\and
{\small A. DeBenedictis \footnote{adebened@sfu.ca}} \\
\it{\small Pacific Institute for the Mathematical Sciences,} \\
\it{\small Simon Fraser University Site} \\
\it{\small and}\\
\it{\small Department of Physics, Simon Fraser University}\\
\it{\small Burnaby, British Columbia, V5A 1S6, Canada }
}
\date{{\small March 14, 2009}}
\maketitle

\begin{abstract}
\noindent Using the improved quantization technique to the mini-superspace approximation of loop quantum gravity, we study the evolution of black holes supported by a cosmological constant. The addition of a cosmological constant allows for classical solutions with planar, cylindrical, toroidal and higher genus black holes. Here we study the quantum analog of these space-times. In all scenarios studied, the singularity present in the classical counter-part is avoided in the quantized version and is replaced by a bounce, and in the late evolution, a series of less severe bounces. Interestingly, although there are differences during the evolution
between the various symmetries and topologies, the evolution on the other side
of the bounce asymptotes to space-times of Nariai-type, with the exception of the planar black hole analyzed here, whose $T$-$R$=constant subspaces seem to continue expanding in the long term evolution.
For the other cases, Nariai-type universes are attractors in the quantum evolution,
albeit with different parameters. We study here the quantum evolution of each symmetry in detail.
\end{abstract}

\vspace{3mm}

\section{Introduction}
It has long been thought that any viable theory of quantum gravity should be able to resolve the singularities which afflict the classical theory. Of primary importance are the space-like singularities which are manifest in many cosmological models and black hole interiors. It is hoped that a quantum theory of gravity not only replaces the singularity with something more benign, but also allows us to answer the question of what happens beyond the corresponding singular geometry.

Today, one promising theory of quantum gravity is loop quantum gravity (LQG). (For a review, the interested reader is referred to  \cite{ref:lrrLQG}-\cite{ref:rev2} and references therein.) Loop quantum gravity has already been successful in answering the question of the origin of black hole entropy \cite{ref:entrev1}-\cite{ref:KBD}. In the paradigm of LQG this entropy has been shown to arise from counting the number of quantum micro-states which yield a macroscopic black hole of fixed area. By setting the Immirzi parameter to one specific value, the area/4 law is recovered for all black holes. The sub-leading contribution to the entropy has been found to be genus dependent \cite{ref:KBD}, which is in agreement with studies done utilizing non-LQG techniques \cite{ref:vanzo}, \cite{ref:mans}, \cite{ref:liko}. There has also been progress on this front from the spin-foam approach \cite{ref:manuel1}, \cite{ref:manuel2}.

Regarding the issue of singularity resolution, there has been much work done in the framework of loop quantum cosmology (LQC) \cite{ref:LQClrr} and, more recently, with the Schwarzschild black hole \cite{ref:modestobhsing}-\cite{ref:CGPint}. In these studies, the full quantum theory is technically formidable to deal with and therefore one resorts to first symmetry reducing the system before quantization, in the hopes that the essential features in the full quantum system remain intact in the reduced analysis. Studies so far, both in LQC and black hole interiors, indicate that the singularities are replaced by quantum bounces and therefore gravitation becomes \emph{repulsive} at extremely high energies. 

In the black hole sector, detailed studies to date have been performed only in the case of the Schwarzschild black hole. As mentioned above, the Schwarzschild singularity is replaced by a bounce in the quantized version. Aside from the elimination of the singularity it is interesting to ask what happens beyond the bounce, as it is now possible to evolve the equations on the other side of the classical singularity. For the Schwarzschild black hole the evolution is \emph{not} symmetric and one does not simply obtain a time-reversal of the original black hole interior on the other side of the bounce. It has been shown that instead, the solution asymptotes to a second non-singular black hole \cite{ref:BandV} in the case where the ``length'' of the holonomy path is taken to be constant (the ``fixed-delta'' scheme). In the scenario where the holonomy path is not constant (the ``improved quantization scheme'') the evolution tends to a Nariai space-time \cite{ref:BandV}. We will discuss the two schemes in more detail below. A particularly in-depth analysis of various evolution schemes, comparing their strengths and weaknesses is in \cite{ref:chiou2}.

We wish to study here other black holes and ask the question of whether or not the singularity is avoided and, if so, what is the subsequent evolution of the geometry beyond the classical singularity. We choose to study the $\Lambda$ black holes. That is, black holes supported by a cosmological constant. The reason for this is that the admission of a cosmological constant allows for the existence of black holes of differing symmetries and topologies in general relativity. For example, one may have planar, cylindrical, toroidal and higher genus black holes \cite{ref:vanzo}, \cite{ref:liko}, \cite{ref:rg1}-\cite{ref:mena}. (An interesting analysis of super-symmetric anti-de Sitter horizons may be found in \cite{ref:BL}.) Such a study is of interest for several reasons. The presence of a cosmological constant may be expected to play a non-trivial role on the large-scale evolution and therefore could affect the late-time evolution of the black hole interior (although it is not expected to play a significant role on the scale of the bounce itself). Secondly, if the singularities in these geometries are all avoided, then it will be evidence for the indication that space-like singularities are \emph{generically} resolved within loop quantum gravity, and not just a product of the spherical symmetry or topology. Since the field of quantum gravity has a notorious lack of experimental guidance, it is the internal consistencies such as this which may ultimately decide if a theory is viable. Finally, the study of the final fate of the black holes is also interesting in its own right. In this paper we study the evolution and late-time behavior of $\Lambda$ black holes of various geometries and topologies. We find that there are qualitative and quantitative differences during the evolution. However, in all cases the classical singularity is avoided and, at late time, the evolution tends to Nariai-type universes with the exception of the particular planar metric studied here. The values of the parameters in the asymptotic space-time and, of course, the symmetry of $T=\mbox{constant}$ surfaces, depends on the initial space-time. Although the effective quantum theory is an approximation to the full theory, there is evidence in the case of Loop Quantum Cosmology that the effective theory similar to that presented here provides an accurate description of the quantum dynamics \cite{ref:APS}, \cite{ref:APS2}, \cite{ref:vander2}.

This paper is organized as follows: In section 2 we review the formalism of the mini-superspace approximation for black holes in loop quantum gravity. In the sub-sections we begin by briefly reviewing the previously studied Schwarzschild case followed by detailed studies of the Schwarzschild-(anti)deSitter black hole, the planar black hole, and finally, the higher genus toroidal and hyperbolic black holes. Concluding remarks are provided in section 3.

\section{Quantum evolution}
In the canonical formulation of loop quantum gravity, one begins with a $3+1$ decomposition of space-time and then re-writes the resulting action in terms of the Ashtekar variables; namely, a generalized $SU(2)$ connection $A^{i}_{\;a}$ and a densitized triad $E^{a}_{\;i}$, which is related to the three-space metric, $q_{ab}\,$, via \footnote{Throughout this manuscript, units are chosen such that $G$=$c$=$\hbar= 1$.}  
\begin{equation}
qq^{ab}=E^{a}_{\;i}E^{b}_{\;j}\delta^{ij}\;. \label{eq:mettriadreln}
\end{equation}
These two variables then become the configuration and momentum variables respectively to be quantized, resulting in what are sometimes known as the quantum Einstein equations:
\begin{align}
 \hat{G}_{i}\psistate:=&\widehat{D_{a}E^{a}_{\;i}}\left|\Psi\right> =0, \label{eq:qE1} \\
 \hat{V}_{a}\psistate =& \left[\widehat{E^{b}_{\;i}F^{i}_{\;ab}} -(1-\gamma^{2}) \widehat{K^{i}_{\;a}G_{i}}\right]\psistate=0, \label{eq:qe2} \\
\hat{S}\psistate=&\left\{\frac{\hat{E}^{a}_{\;i}\hat{E}^{b}_{\;j}}{\sqrt{\det(\hat{E})}}  \left[\epsilon^{ij}_{\;\;\;k}\hat{F}^{k}_{\;ab} -2(1+\gamma^{2}) \widehat{K^{i}_{\;[a}K^{j}_{\;b]}}\right] +2{\Lambda}\sqrt{\det(\hat{E})}  \right\}\psistate=0, \label{eq:qe3}
\end{align}
 with $G_{i}:=\partial_{a}E^{a}_{\;i} +\epsilon_{ij}^{\;\;\;k}A^{j}_{\;a}E^{a}_{\;k}$, $K^{i}_{\;a}:= \frac{1}{\sqrt{\det(E)}}\delta^{ij} E^{b}_{\;j}K_{ab}\,$ ($K_{ab}$ being the extrinsic curvature of the 3-surface) and
$F^{i}_{\;ab}:= \partial_{a}A^{i}_{\;b} -\partial_{b} A^{i}_{\;a} +\epsilon^{i}_{\;\;jk}A^{j}_{\;a}A^{k}_{\;b}$, and $\gamma$ the Immirzi parameter.\\

These equations are the quantum analogs of the now well known Gauss, vector and Hamiltonian (or scalar) constraints respectively. As mentioned in the introduction, the technical difficulty in finding and interpreting solutions to these equations are quite daunting. Therefore, in order to make some progress, one often first deals with a classical model, reducing the system to the desired symmetry, and then studies the quantum evolution of the reduced system. In the Schwarzschild interior and in LQC, the resulting Hamiltonian evolution leads to a \emph{difference} equation for the quantum states, which is well behaved at the classical singularity (see, for examples, \cite{ref:AandB}, \cite{ref:LQC_difference}). However, enough progress has been made that there are now reasonably well defined prescriptions for generating \emph{classical} Hamiltonians which are thought to encode the quantum dynamics of symmetry reduced systems \cite{ref:modestolqbh}. We exploit this prescription here in the study of $\Lambda$ black holes. The general scheme is as follows: One begins with a classical pair $\left(A^{i}_{\;a}\,,\,E^{b}_{\;k}\right)$ which a priori respect the desired spatial symmetry. One then partially gauge fixes the system by satisfying the Gauss constraint and writes the Hamiltonian constraint in terms of the remaining variables. In the resulting constraint the classical $SU(2)$ connection components are replaced with effective holonomies of the connection (this is the effective quantum prescription) and the resulting quantum corrected Hamilton equations are solved for the connection and triad components. All the systems studied have the near horizon as their initial data surface. In this way we avoid large quantum effects at the horizon. This is similar to the procedure used in \cite{ref:BandV} as well as in \cite{ref:chiou2} (more on this below). 

There have been several proposals for the effective quantum prescription to be used in such an effective Hamiltonian, mainly inspired by the much more studied systems of loop quantum cosmology. It has been shown that one may mimic the quantum dynamics by making the substitution $a_{I}\rightarrow \sin(a_{I}\delta)/\delta$, where $\delta$ represents the length of the holonomy path of the $a_{I}$ component of the connection in the principal (coordinate) directions. The early studies took $\delta$ to be constant valued. However, it was later shown in LQC that using a fixed $\delta$ prescription does not yield a good semi-classical limit \cite{ref:APS}. If however the holonomy path-length $\delta$ is made to depend on the momenta, $E_{I}$, a good semi-classical limit can be achieved. There are several candidates for this improved prescription and we use here the analog of the ``$\overline{\mu}{\,}^{\prime}$'' prescription of LQC \cite{ref:chiou} (details are provided below). We should mention before proceeding that one needs to use caution when interpreting results within the mini-superspace approximation. There is no guarantee that this level of approximation always mimics the full quantum theory. An interesting recent analysis of this is in \cite{ref:newbojo}. 

\subsection{A brief review of the Schwarzschild interior}
Here we present a very brief overview of the previously studied Schwarzschild black hole. For full details the interested reader is referred to the pioneering papers \cite{ref:modestobhsing} - \cite{ref:BandV}. One may choose a spherically symmetric connection and triad pair as \cite{ref:zhou}:
\begin{subequations}
 \begin{align}
A=&a_{3} \tau_{R}\,dR + \left(a_{1} \tau_{\theta} + a_{2} \tau_{\phi}\right)d\theta +\left(a_{1}\tau_{\phi} -a_{2}\tau_{\theta}\right)\sin\theta\,d\phi +\tau_{R}\cos\theta\,d\phi\,, \label{eq:sphereA} \\
E=& E_{3} \tau_{R} \sin\theta \frac{\partial}{\partial R} + \left(E_{1}\tau_{\theta} +E_{2}\tau_{\phi}\right)\sin\theta \frac{\partial}{\partial \theta} +\left(E_{1}\tau_{\phi}-E_{2}\tau_{\theta}\right) \frac{\partial}{\partial\phi}\,. \label{eq:sphereE} 
 \end{align}
\end{subequations}
Here the $\tau$ represent the standard Pauli $su(2)$ basis and $a_{J}$ and $E_{J}$ are functions of the interior time only. The classical Gauss constraint yields only one non-trivial condition:
\begin{equation}
 2\sin\theta\left(a_{1} E_{2} - E_{1} a_{2}\right) =0, \nonumber
\end{equation}
which is customarily satisfied by setting $a_{1}=E_{1}=0$. Setting $E_1 = 0$ still leaves the triad
non-degenerate and it can be checked that the classical vector contraint is also satisfied. The classical Hamiltonian constraint now reads as:
\begin{equation}
 H=-\int\frac{N}{8\pi\gamma^{2}\sqrt{E_{3}}}\left[\left(a_{2}^{2}+\gamma^{2}\right)E_{2} +2a_{3}a_{2}E_{3}\right]dR\,d\Omega\,. \label{eq:sphereham1}
\end{equation}
The $R$ integral is performed over a finite length, $L_{0}$, and the resulting factor of $L_{0}$ is absorbed into a re-scaling of $a_{3}$ and $E_{2}$ yielding:
\begin{equation}
 H=-\frac{N}{2\gamma^{2}\sqrt{E_{3}}}\left[\left(a_{2}^{2}+\gamma^{2}\right)E_{2} +2a_{3}a_{2}E_{3}\right]. \label{eq:sphereham}
\end{equation}
In \cite{ref:AandB} the lapse function, $N$, was taken as $N=\frac{\gamma^{2} \sqrt{E_{3}}}{a_{2}}$ and the Hamilton's equations ($\dot{\pb}=\left\{\pb,\,H\right\}=-\partial H/\partial \bee$ etc.) yield the following equations of motion:
\begin{align}
 \dot{E}_{2}=& -\frac{1}{2}\left(a_{2} +\frac{\gamma^{2}}{a_{2}}\right),\;\;\; \dot{E}_{2}=\frac{1}{2} \left(E_{2} -\frac{\gamma^{2}}{a_{2}^{2}}E_{2}\right)\,, \nonumber \\
\dot{a}_{3}=&-2a_{3}, \;\;\;\;\;\;\;\;\;\;\;\;\;\;\;\;\;\;\; \dot{E}_{3}= 2 E_{3}\,, \nonumber
\end{align}
with the dot representing the derivative with respect to the Hamiltonian affine parameter, $T$.

The solution to these equations yields, after a re-definition $T\rightarrow T^{\prime}:=e^{T}$:
\begin{align}
a_{2}=&\pm\gamma\sqrt{\frac{2M}{T^{\prime}}-1}, \;\;\; E_{2}=E_{2}^{(0)}\sqrt{2MT^{\prime}-T^{\prime\,2}}\,, \nonumber \\
a_{3}=&\mp\gamma\, E_{2}^{(0)} \frac{M}{T^{\prime\,2}}, \;\;\;\;\;\; E_{3}=\pm T^{\prime\,2}\, ,
\end{align}
with $E_{2}^{(0)}$ and $M$ constants. By comparing this solution with (\ref{eq:mettriadreln}) it can easily be seen, after absorbing $E_{2}^{(0)}$ in a coordinate re-scaling, that the densitized triad is compatible with a metric which admits the following well-known line element:
\begin{equation}
 d\sigma^{2}= \left(\frac{2M}{T^{\prime}}-1\right)\,dR^{2} +T^{\prime\,2}d\theta^{2} +T^{\prime\,2}\sin^{2}\theta\,d\phi^{2}\,.
\end{equation}
This, of course, is the spatial line element corresponding to the Schwarzschild interior solution.

The analogous quantum system has been studied in, for example, \cite{ref:AandB} where it was shown that a discrete eigenvalue flow equation arises. In a different approach \cite{ref:modestoint}, \cite{ref:BandV}, an effective Hamiltonian was utilized which encodes the quantum corrections. In both types of studies it has been demonstrated that the classical Schwarzschild singularity is avoided. In the effective Hamiltonian scheme, the quantization prescription described previously yields \cite{ref:BandV}:
\begin{equation}
 H=-\frac{\delta_{2}}{2\gamma \sin(a_{2}\delta_{2})}\left[\left(\frac{\sin^{2}(a_{2}\delta_{2})}{\delta_{2}^{2}}+\gamma^{2}\right)E_{2} +2\frac{\sin(a_{3}\delta_{3})}{\delta_{3}}\frac{\sin(a_{2}\delta_{2})}{\delta_{2}}E_{3}\right], \label{eq:qsphereham}
\end{equation}
where the $\delta_{I}$ represent the length of the holonomy path in the $I$-th (angular or radial) direction. In the limit that the $\delta_{I}$'s vanish, one recovers the classical Hamiltonian. The resulting equations of motion may be found in \cite{ref:modestoint}, \cite{ref:BandV} or in the $\Lambda \rightarrow 0$ limit of the following section (albeit in a different gauge).

Studies where the fixed $\delta$ prescription ($\delta_{2}=\delta_{3}=$const.) has been used have found that the singularity is replaced by a quantum bounce, and that on the other side of the bounce the system evolves to another (non-singular) black hole \cite{ref:modestoint}, \cite{ref:BandV}. In the non-fixed $\delta$ prescription, where the $\delta$'s are functions of the triad (the scheme we utilize below), it was shown that the evolution on the other side of the bounce asymptotes to the Nariai space-time \cite{ref:BandV}.

\subsection{The Schwarzschild-(anti)deSitter interior}
The quantization of the Schwarzschild-(anti)deSitter interior is very similar to the above Schwarzschild system. We present it here mainly for completion and as a segue to the other $\Lambda$ black holes studied below. 

The interior Schwarzschild-(anti)deSitter line element is given by:
\begin{equation}
 ds^{2}=-\frac{dT^{2}}{\left(\frac{\Lambda}{3}T^{2} + \frac{2M}{T} -1\right)} + \left(\frac{\Lambda}{3}T^{2} + \frac{2M}{T} -1\right)\,dR^{2} + T^{2}\, d\theta^{2} + T^{2}\sin^{2}\theta\, d\phi^{2}\,, \label{eq:desittermet}
\end{equation}
where $\Lambda$ is the cosmological constant and $M$ the black hole mass. Due to the spherical symmetry, the form of the connection and triad are taken as in (\ref{eq:sphereA}) and (\ref{eq:sphereE}) and the Hamiltonian constraint is therefore
\begin{equation}
 H=-\frac{N}{2 \gamma^{2}\sqrt{E_{3}}}\left[\left(a_{2}^{2}+\gamma^{2}\right)E_{2} +2a_{3}a_{2}E_{3}- {\Lambda}E_{2}E_{3}\right]\,.
\end{equation}
 
We choose a different gauge (choice of $N$) here, however, and pick the lapse function as
\begin{equation}
 N=\frac{\gamma^{2}\sqrt{E_{3}}}{E_{2}}\,, \label{eq:sadslapse}
\end{equation}
which amounts to working directly in Schwarzschild ($T$-domain) coordinates. That is, with this choice, and utilizing (\ref{eq:mettriadreln}) and (\ref{eq:desittermet}), the classical Schwarzschild-(anti)deSitter line element is given by
\begin{equation}
 ds^{2}=- \frac{{E}_{3}}{{E}_{2}{}^{2}}dT^{2} + \frac{{E}_{2}{}^{2}}{{E}_{3}}dR^{2} + {E}_{3}\,d\theta^{2} + {E}_{3}\sin^{2}\theta\,d\phi^{2}\,,\label{eq:ashtekardesitter}
\end{equation}
(classically $\gamma$ can be set equal to unity without loss of generality) with:
\begin{equation}
 {E}_{2}=T\sqrt{\frac{\Lambda}{3}T^{2} + \frac{2M}{T} -1}\;,\;\;\; {E}_{3}=T^{2}\;. \nonumber
\end{equation}
From these expressions it can be seen that the classical singularity occurs at $T=0$ where both $\pc=0$ and $\pb=0$. The horizon(s) are present where only $\pb=0$. One advantage of this choice of $N$ is that, as we shall show below, the asymptotic form of the metric
for large negative $T$ is easier to find since $E_2$ approaches a linear function of $T$.

We now go over to the effective quantum system. In the current gauge the corresponding effective quantum Hamiltonian is given by
\begin{equation}
 H=-\frac{1}{2}\left[\left(\frac{\sin^{2}(a_{2}\delta_{2})}{\delta_{2}^{2}}+\gamma^{2}\right) +2\frac{\sin(a_{3}\delta_{3})}{\delta_{3}} \frac{\sin(a_{2}\delta_{2})}{\delta_{2}}\frac{E_{3}}{E_{2}}- {\Lambda}E_{3}\right]\,. \label{eq:sadsqham}
\end{equation}
We now address the issue of choosing the $\delta$'s. As shown in a cosmological setting in \cite{ref:APS} \cite{ref:chiou} and for black hole interiors in \cite{ref:BandV}, a reasonable semi-classical limit might not be obtained utilizing a fixed $\delta$. Instead, one can relate the length of the holonomy paths (the $\delta$'s) to the classical area. For example, in the $R-\theta$ plane, the proper area spanned by the holonomy loop is given by:
\begin{equation}
 {\sf{A}}_{R\theta}=E_{2}\delta_{2}\delta_{3} = \Delta\,, \label{eq:rtarea}
\end{equation}
whereas, in the $\theta-\phi$ plane a similar relation is given by:
\begin{equation}
 {\sf{A}}_{\theta\phi}=\delta_{2}^{2}  E_{3} = \Delta\,. \label{eq:tparea}
\end{equation}
The final equality in (\ref{eq:rtarea}) and (\ref{eq:tparea}) comes from setting the area equal to the smallest area gap predicted from loop quantum gravity, which we denote as $\Delta$. These relationships yield:
\begin{equation}
 \delta_{2}=\sqrt{\frac{\Delta}{E_{3}}}\,, \;\;\;\; \delta_{3}=\frac{\sqrt{E_{3}\, \Delta}}{E_{2}}\,,
\end{equation}
which are to be substituted in (\ref{eq:sadsqham}) to complete the quantum prescription. The resulting equations of motion are:

\begin{align}
&\dot{\cee}-\frac{1}{\sqrt{\pc}\pb\Delta^{3/2}}\left[-2\sqrt{\pb\pc\Delta}\sin\bfactor\sin\cfactor \right. \nonumber \\
&\pb\Delta\cos\bfactor\sin\cfactor -\cee\pc\Delta\sin\bfactor\cos\cfactor \nonumber \\
&+\bee\pb\Delta\sin\bfactor\cos\bfactor -\pb\sqrt{\pc \Delta} +\pb\sqrt{\pc\Delta}\cos^{2}\bfactor \nonumber \\
&+\Lambda\pb\Delta\sqrt{\pc}=0\,, \label{eq:sadseom1}
\end{align}
\begin{align}
&\dot{\pc}-\frac{2\pc{}^{3/2}}{\pb\sqrt{\Delta}}\sin\bfactor\cos\cfactor =0\,, \label{eq:sadseom2}
\end{align}
\begin{align}
 &\dot{\bee}-\frac{\cee\pc{}^{3/2}}{\sqrt{\Delta}\pb{}^{2}}\sin\bfactor\cos\cfactor =0\,, \label{eq:sadseom3}
\end{align}
\begin{align}
 \dot{\pb}-\frac{\sqrt{\pc}\cos\bfactor}{\sqrt{\Delta}}\left[\sin\cfactor+\sin\bfactor\right]=0\,,
\end{align}
where the over-dot denotes differentiation with respect to $T$.
These equations, along with $H=0$  are evolved using a surface within the horizon as the initial data surface (the inner horizon in the case of positive cosmological constant). Near this surface, it is expected that the classical Schwarzschild-(anti)deSitter solution is valid and therefore provides the initial conditions for the subsequent evolution. This should be a valid approximation as long as the black hole is sufficiently large as the curvature will be small here. However, the horizon itself can eventually pick up quantum corrections, as is hinted at by the fact that $\delta_{3}$ becomes infinite there, and the $\delta$'s arise from the effective quantization. The study by Chiou \cite{ref:chiou2} discusses in detail the effects of the quantum correction on the spherical, $\Lambda=0$ horizon where the black hole singularity is resolved and joins smoothly to the diffused horizon of the consecutive black hole.  We study the long-time evolution in scenarios both with positive and negative cosmological constant. 

\begin{figure}[h!t]
\centering
\vspace{-1.6cm} a) \includegraphics[height=1.25in, width=2.0in]{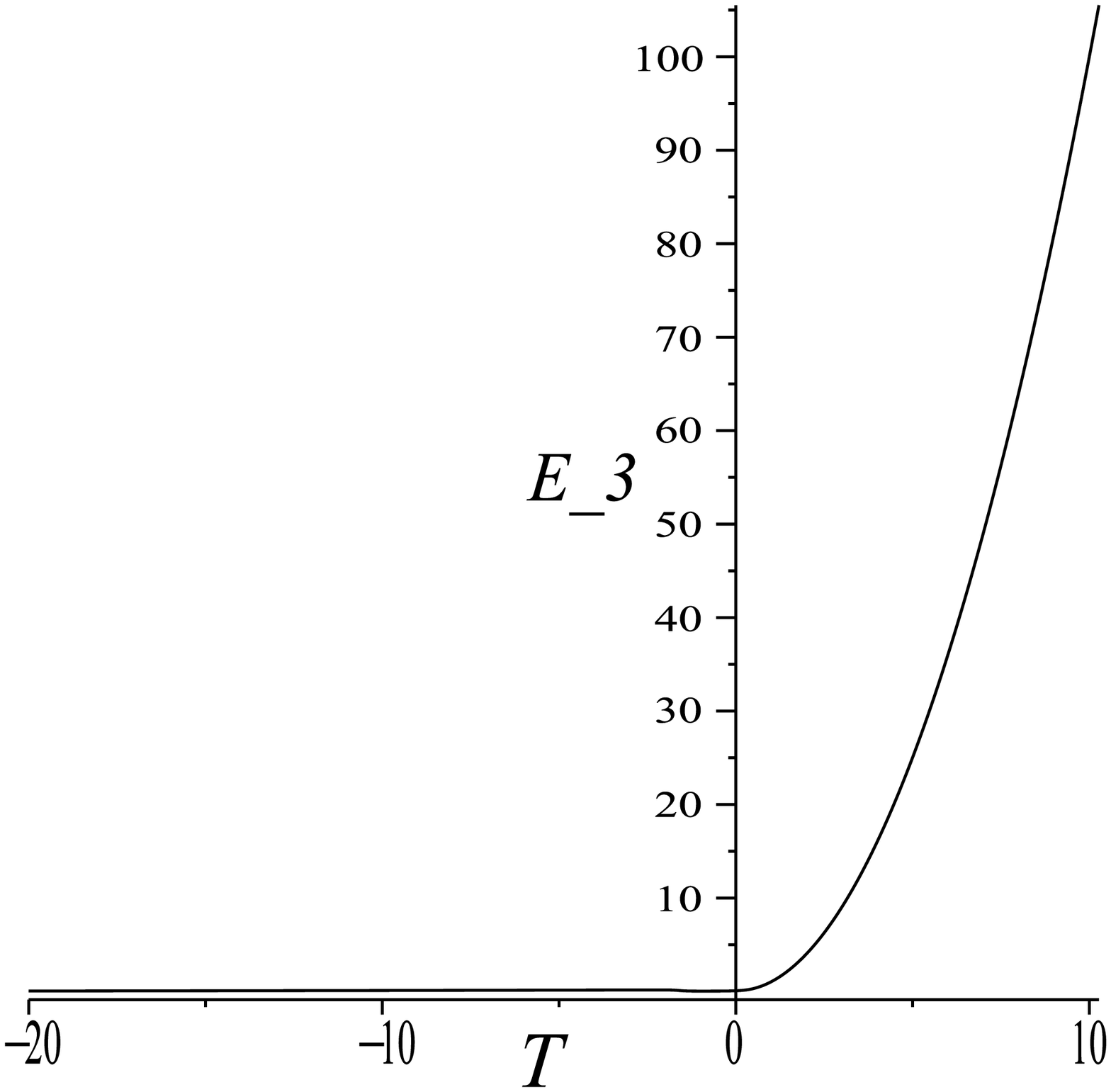}\vspace{0.3cm}
\hspace{0.3cm} b) \includegraphics[height=1.25in, width=2.0in]{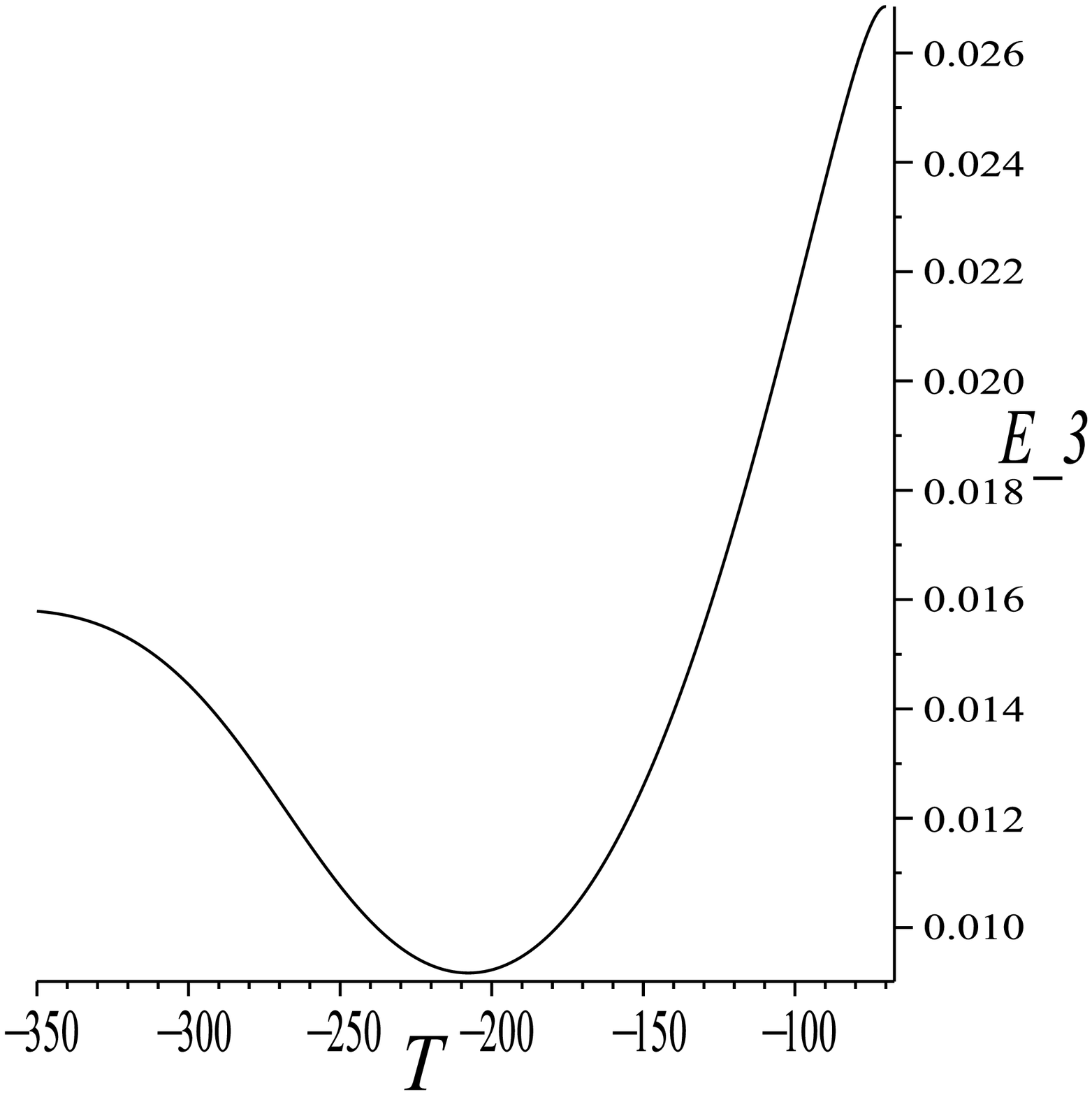} \vspace{0.3cm} \\
c) \includegraphics[height=1.25in, width=2.0in]{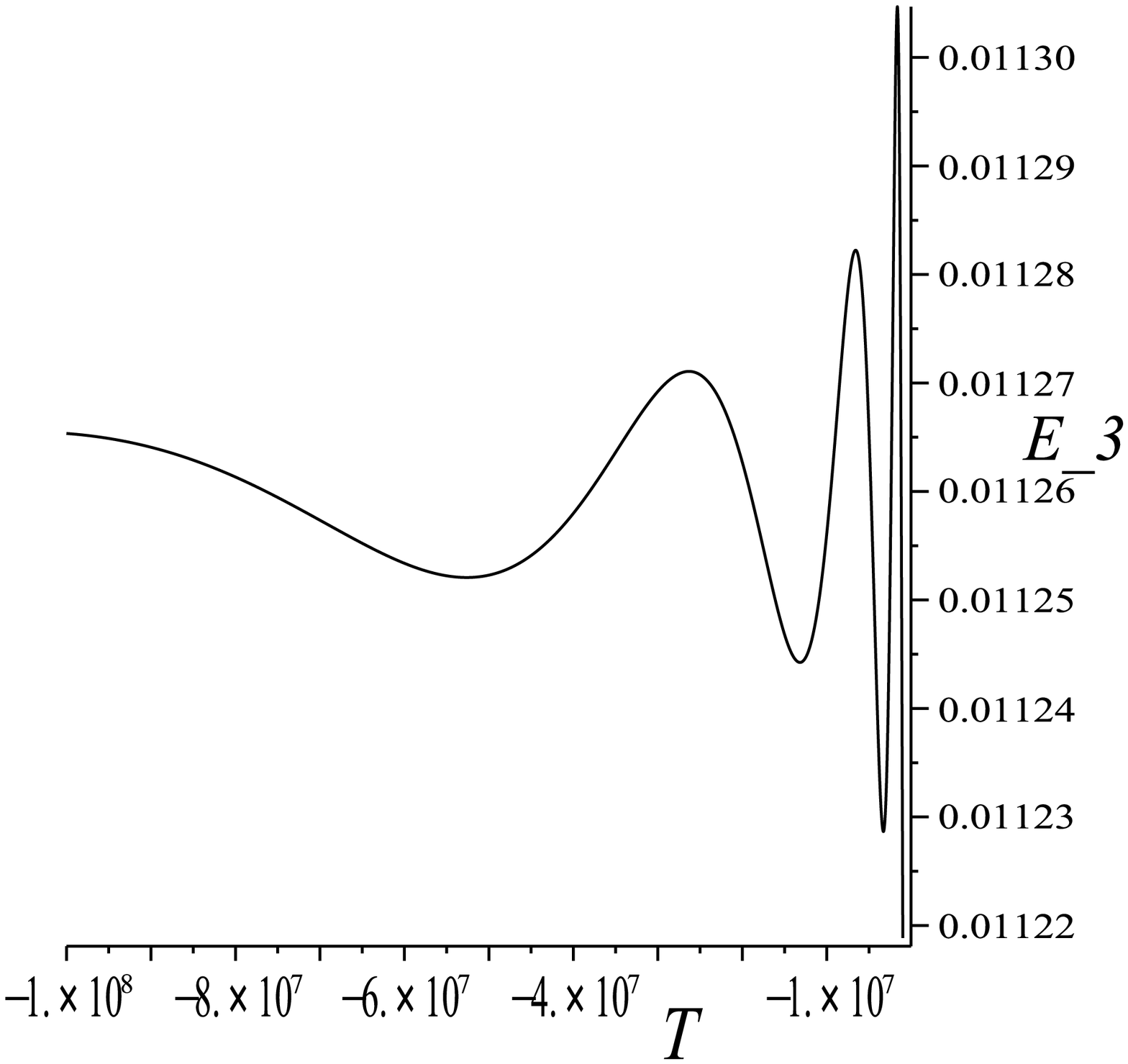} \vspace{0.3cm}
\hspace{0.3cm} d) \includegraphics[height=1.25in, width=2.0in]{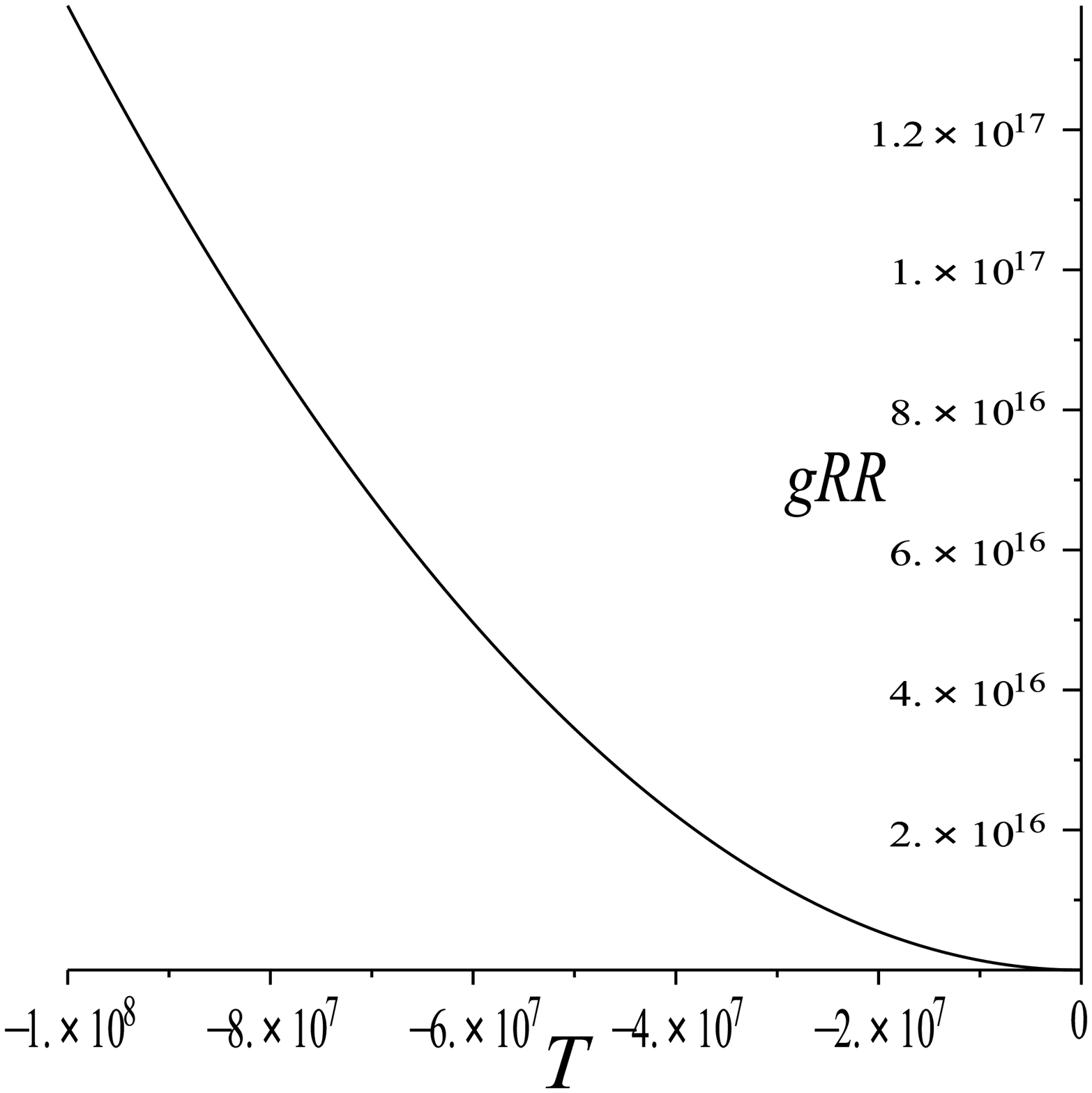}\vspace{0.3cm} \\
\hspace{0.0cm} e) \includegraphics[height=1.25in, width=2.0in]{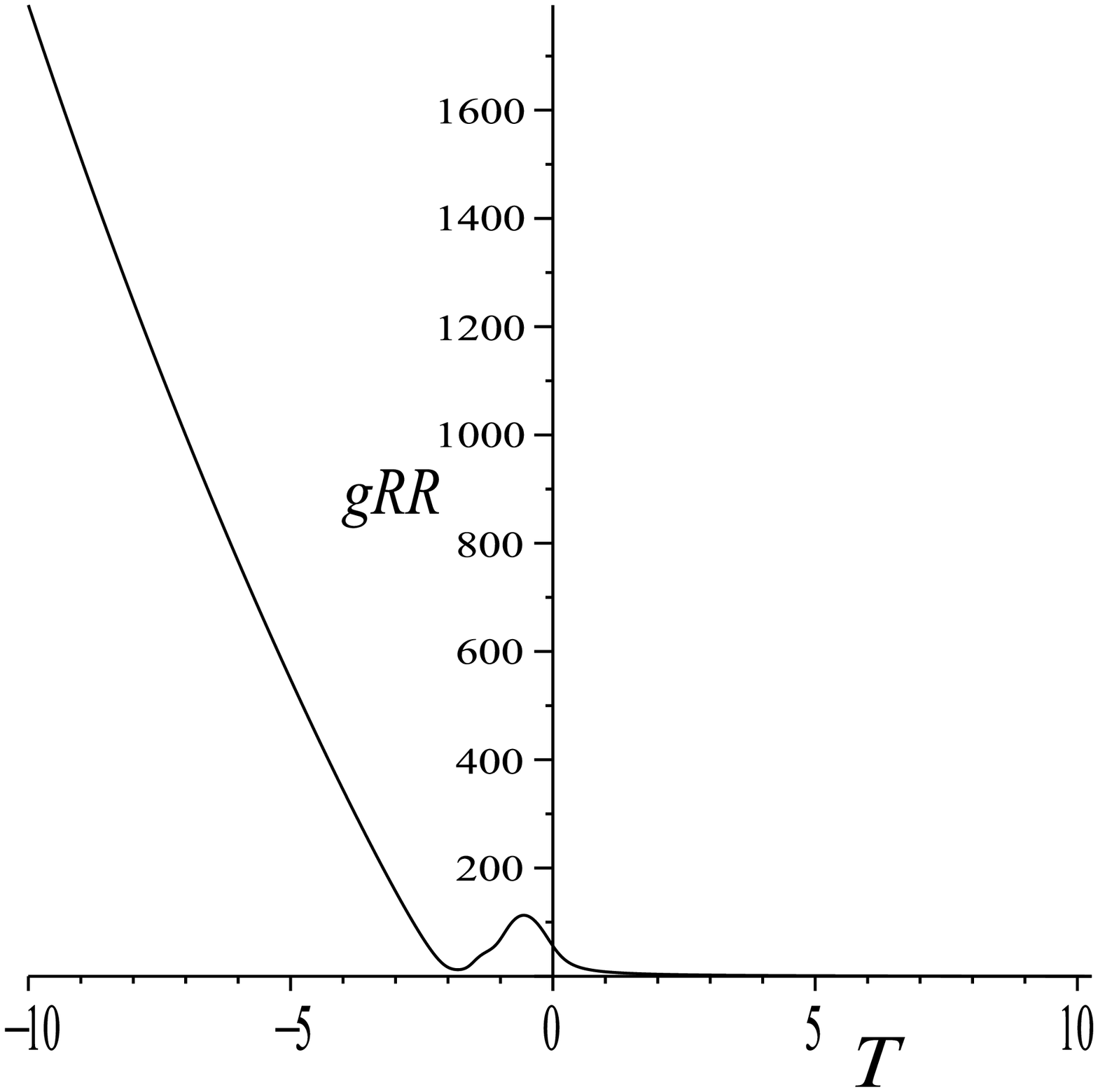}
\hspace{0.3cm} f) \vspace{0.3cm} \includegraphics[height=1.5in, width=2.0in]{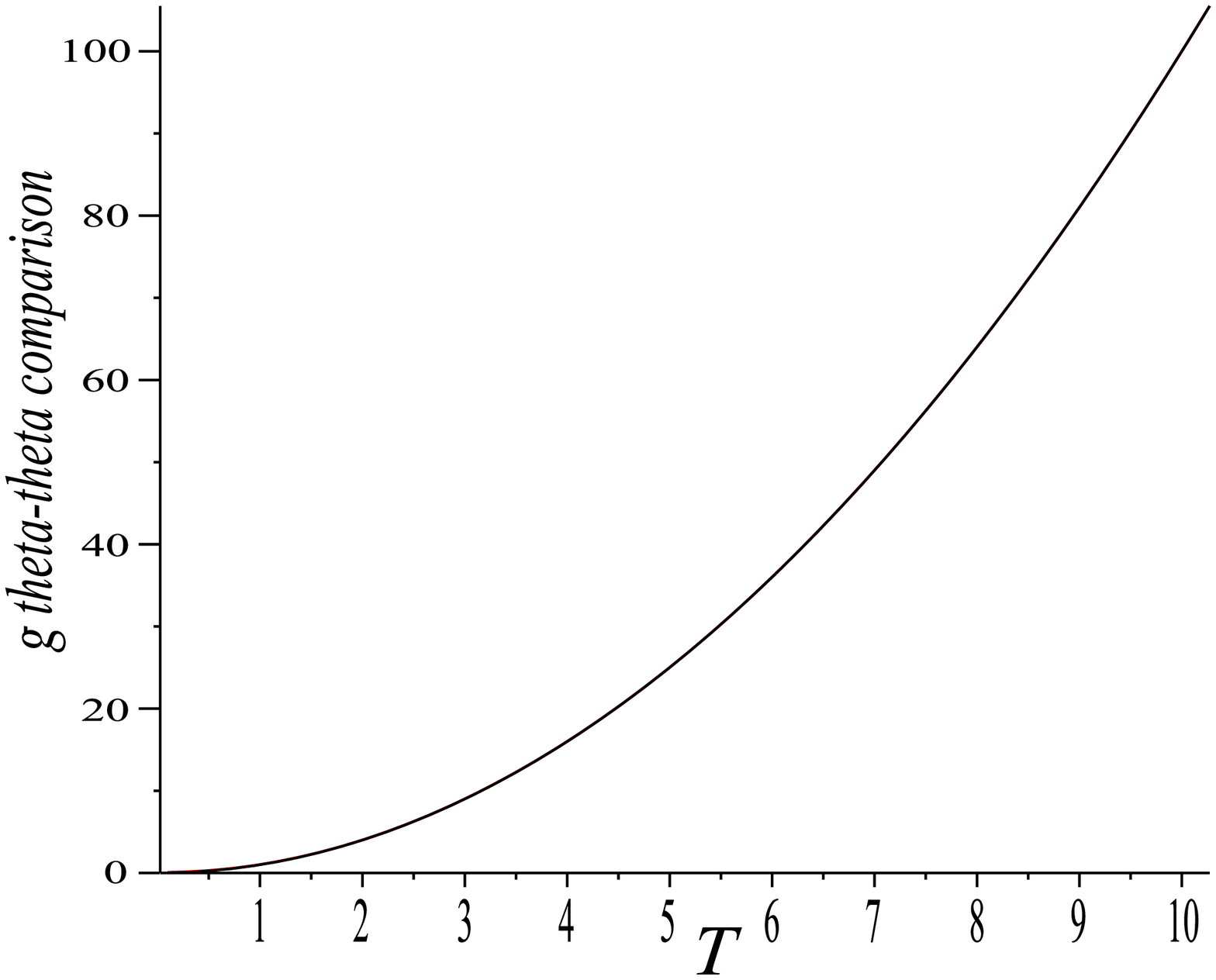}\vspace{0.3cm}\\
\hspace{0.0cm} g) \includegraphics[height=1.25in, width=2.0in]{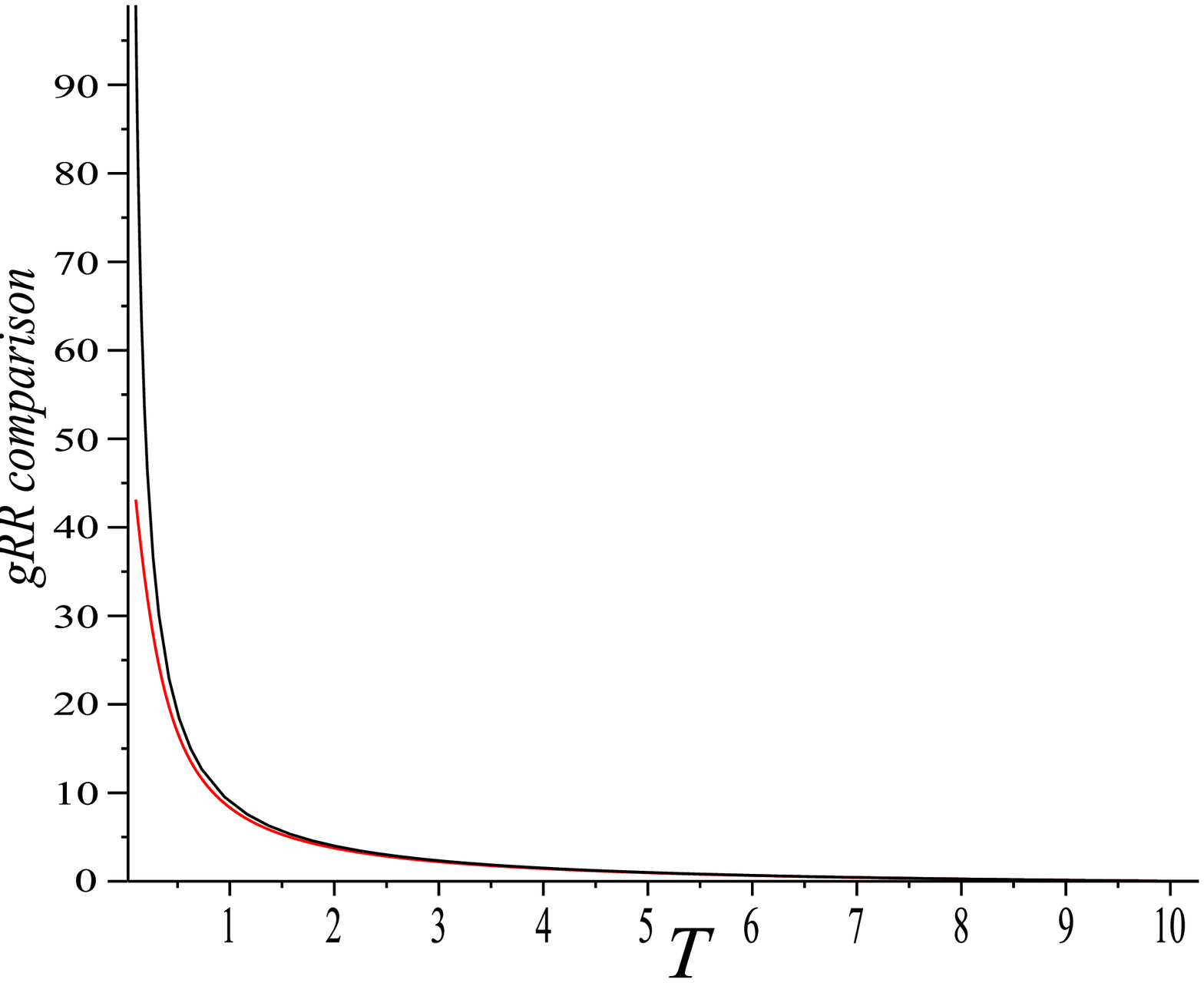} \vspace{0.3cm} 
\hspace{0.3cm} h) \includegraphics[height=1.25in, width=2.0in]{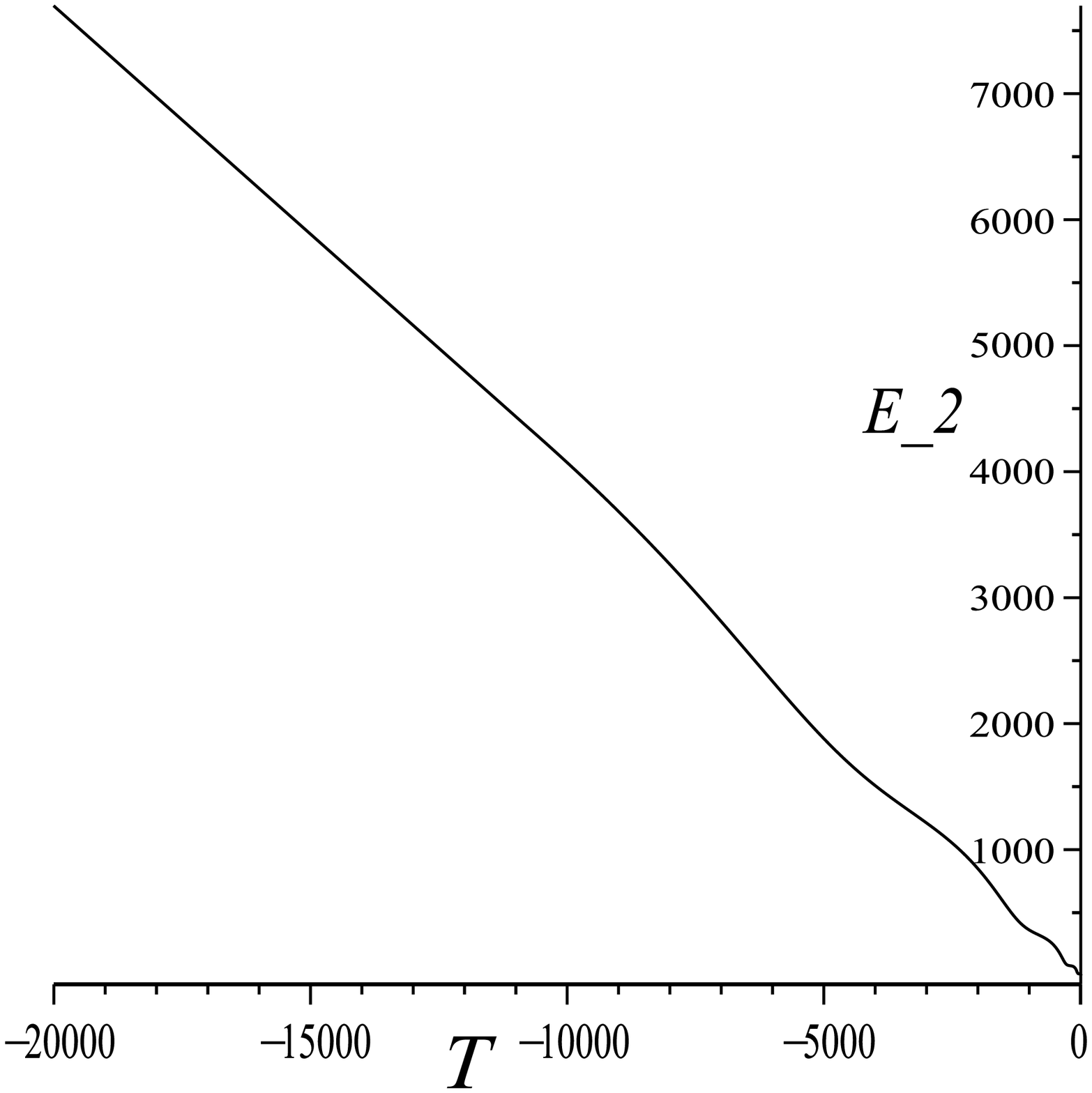}
\caption{{\small The quantum evolution for a Schwarzschild-deSitter black hole interior. a) The triad component $E_{3}$ plotted in the vicinity of the classical singularity. b) The triad component $E_{3}$ plotted from $T=-350$ to $T\approx -70$. c) The triad component $\pfive$ up to $T=-1\times10^{8}$. d) The triad combination $E_{2}^{2}/E_{3}$ which is equal to the classical metric component $g_{RR}$. e) A close-up of the previous plot, near the classical singularity. f) A comparison of the classical (black) vs. quantum (red) evolution of $E_{3}$ for positive $T$. g) A comparison of the classical (black) vs. quantum (red) evolution of $E_{2}^{2}/E_{3}$ for positive $T$. h) The triad component $E_{2}$. The inner horizon is located at $T\approx 10.4$ and the following parameters were used: $\gamma=0.274$, $M=5$, $\Lambda\approx 0.001$. }}
\label{fig:sphds}
\end{figure}

In figure \ref{fig:sphds} we show a sample evolution with positive cosmological constant. From figure \ref{fig:sphds}a) it can be seen that $T=0$, where the classical $\pc=0$ heralds the presence of a singularity, the corresponding quantum $\pc$ is small but \emph{non-vanishing}. In fact, the evolution can be extended beyond $T=0$ where it can be seen that a series of bounces occur (see figures \ref{fig:sphds}b) and c) for different domains.) These are analogous to the series of black holes discussed in \cite{ref:chiou2} where the analysis revealed a fractal-like structure for the spherical black hole without cosmological constant.

\begin{figure}[h!t]
\centering
\vspace{-1.6cm} a) \includegraphics[height=1.25in, width=2.0in]{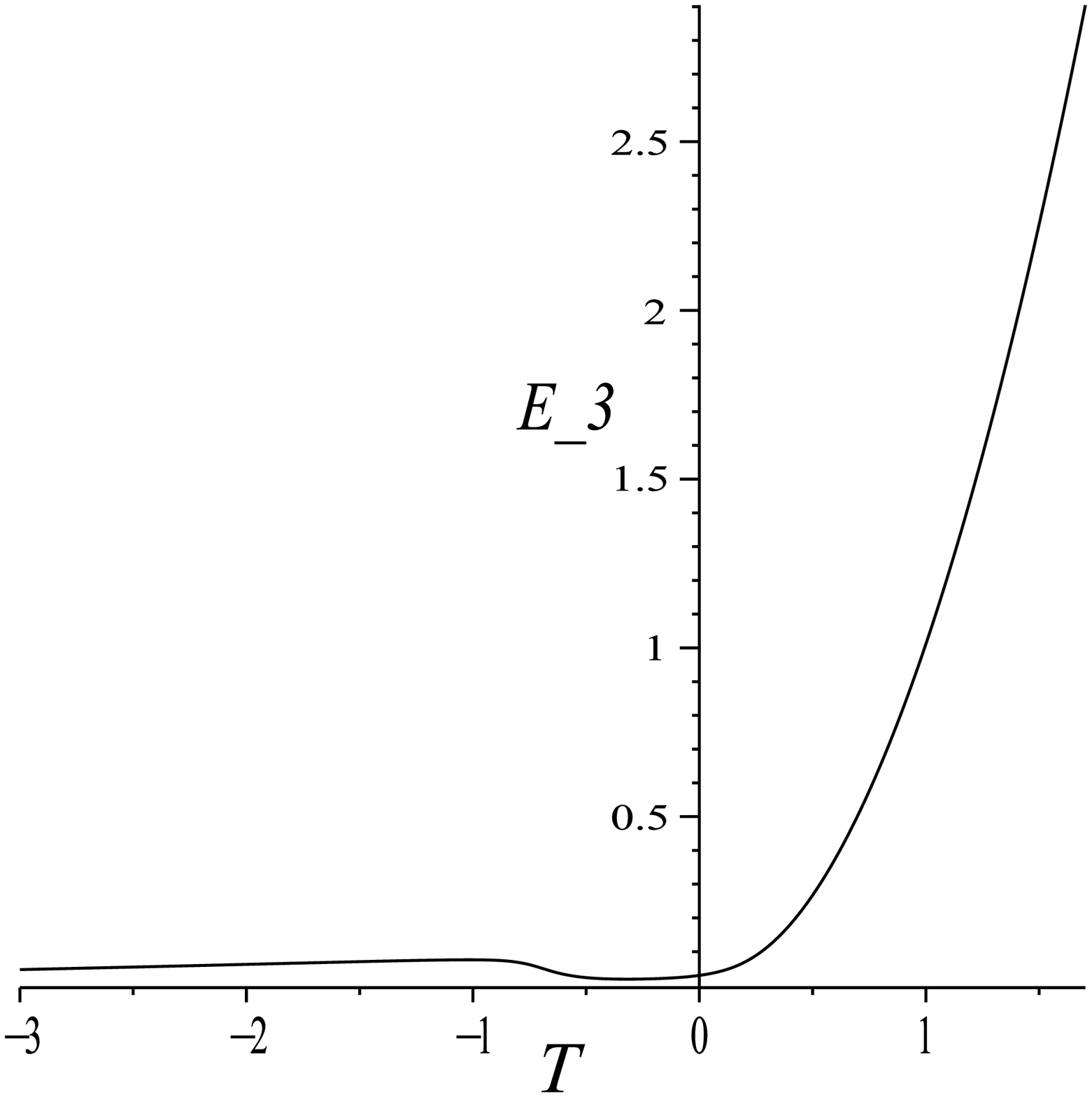}\vspace{0.3cm}
\hspace{0.3cm} b) \includegraphics[height=1.25in, width=2.0in]{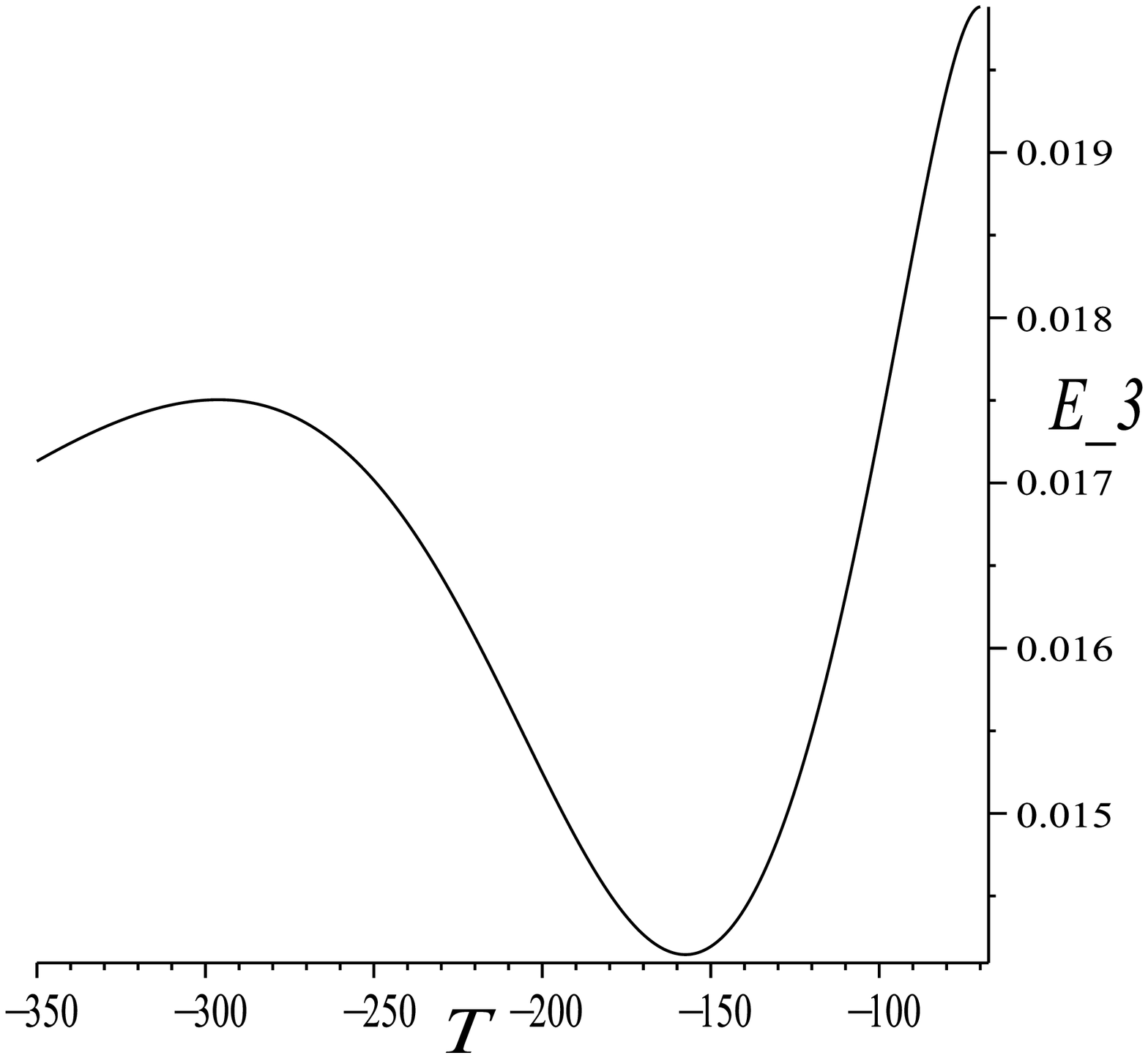} \vspace{0.3cm} \\
c) \includegraphics[height=1.25in, width=2.0in]{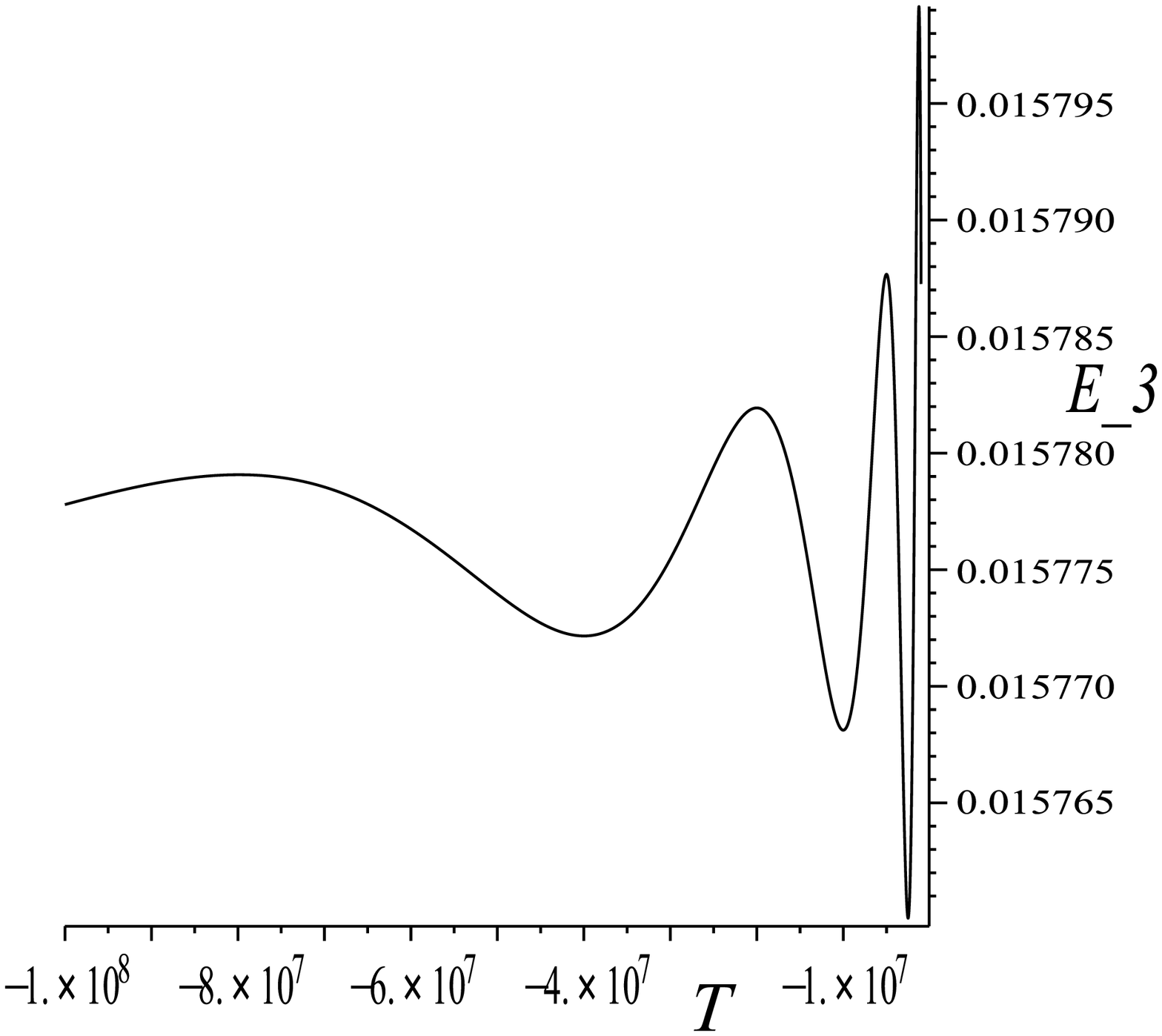} \vspace{0.3cm}
\hspace{0.3cm} d) \includegraphics[height=1.25in, width=2.0in]{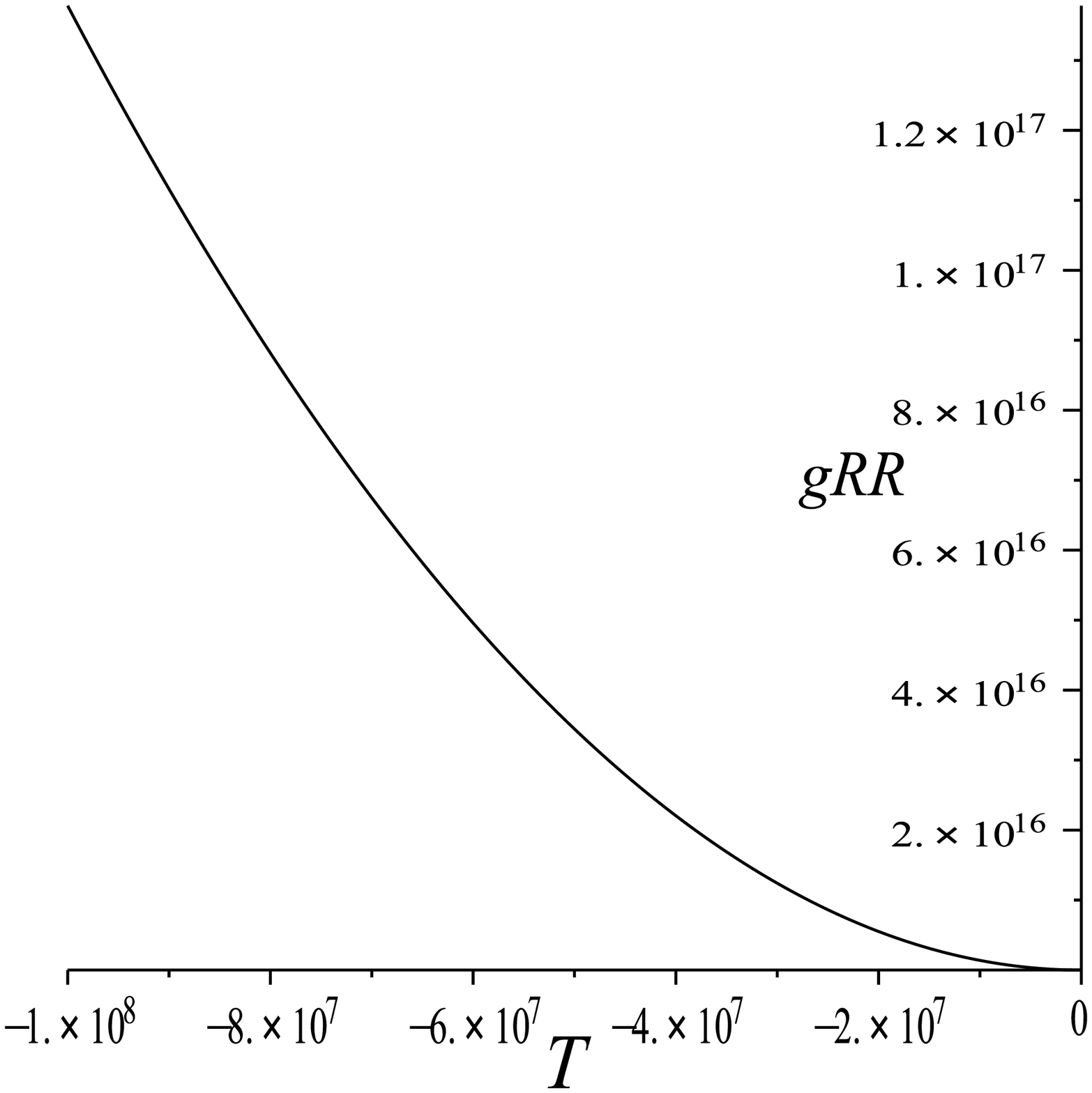}\vspace{0.3cm} \\
\hspace{0.0cm} e) \includegraphics[height=1.25in, width=2.0in]{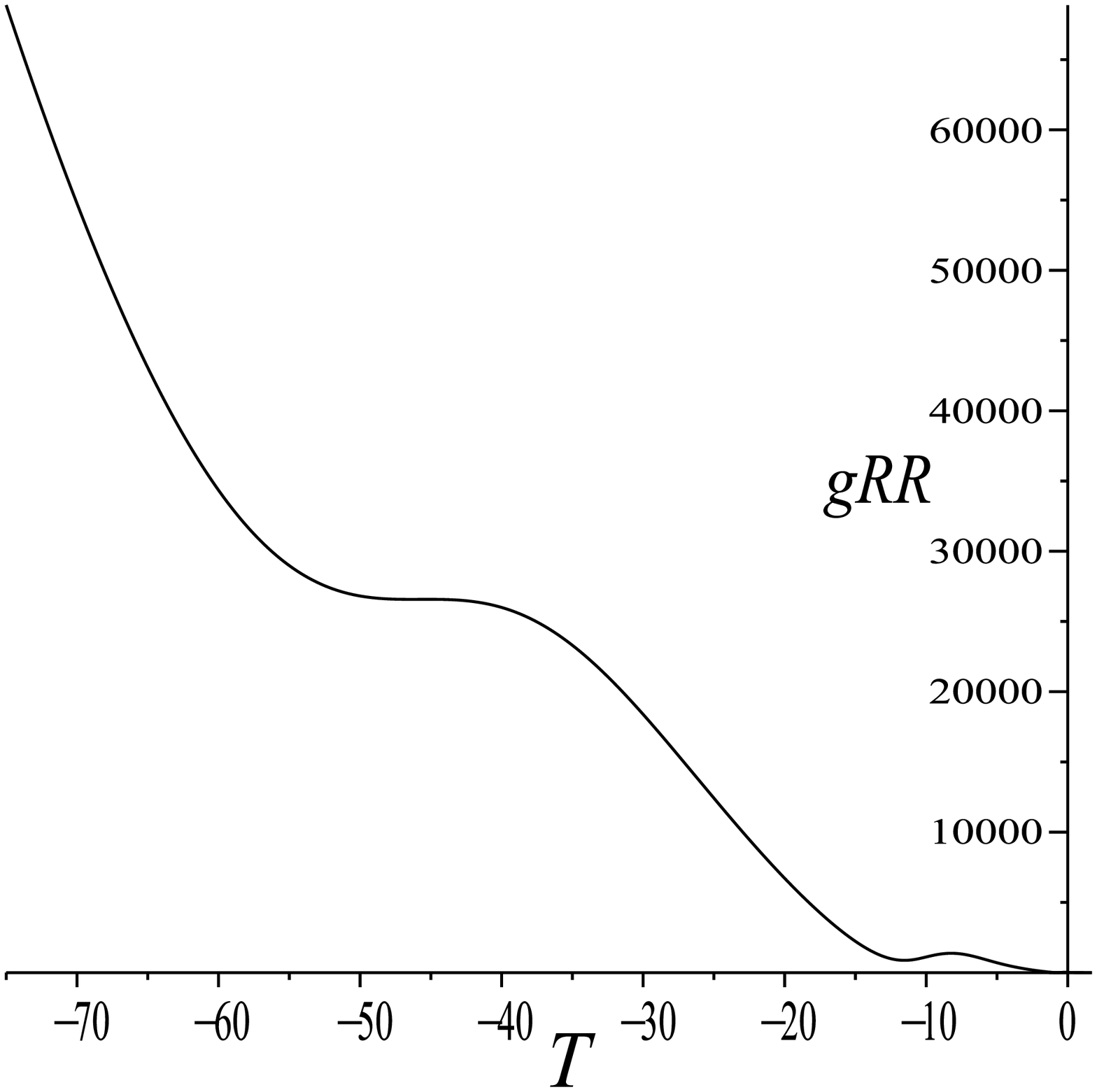}
\hspace{0.3cm} f) \vspace{0.3cm} \includegraphics[height=1.5in, width=2.0in]{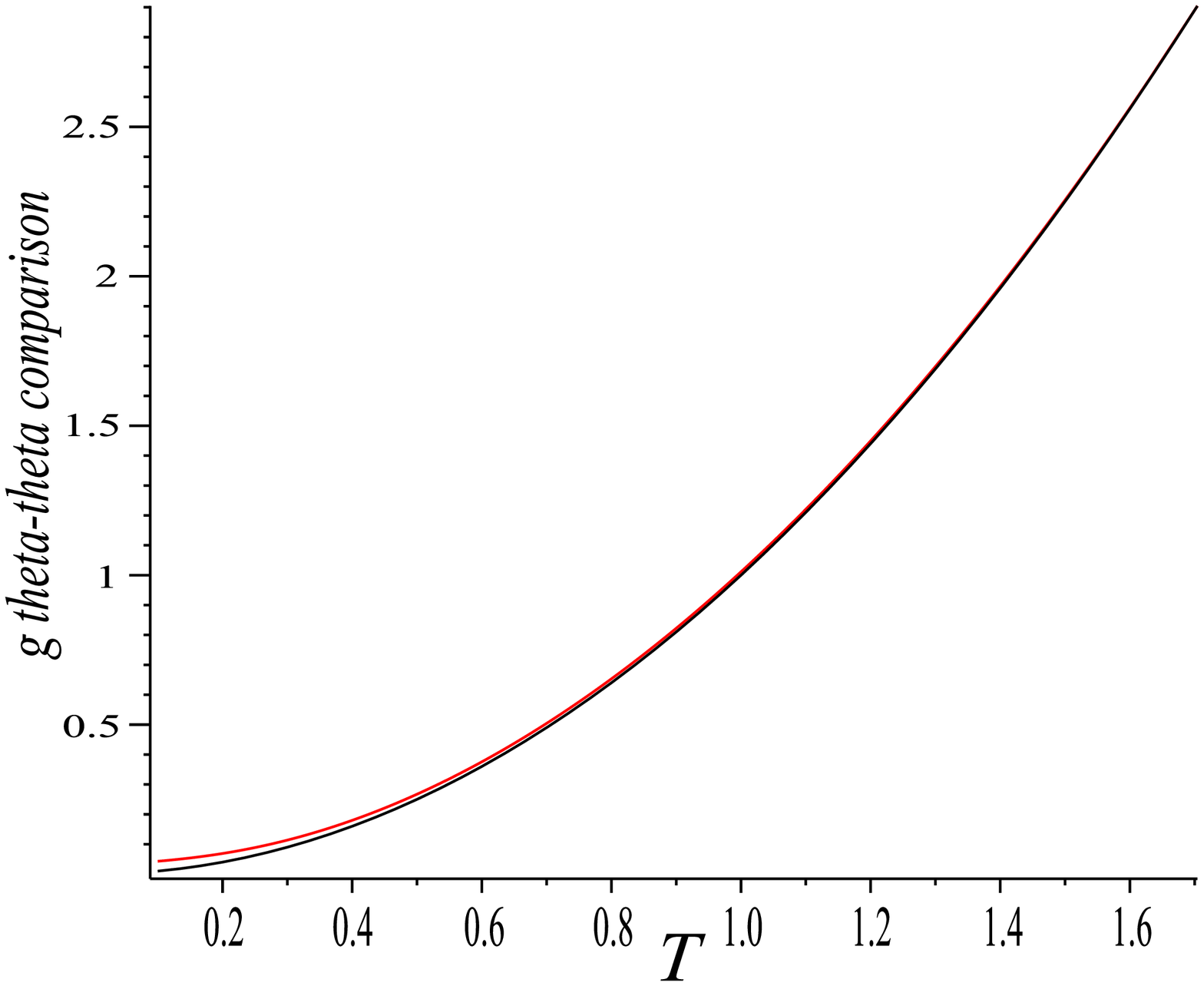}\vspace{0.3cm}\\
\hspace{0.0cm} g) \includegraphics[height=1.25in, width=2.0in]{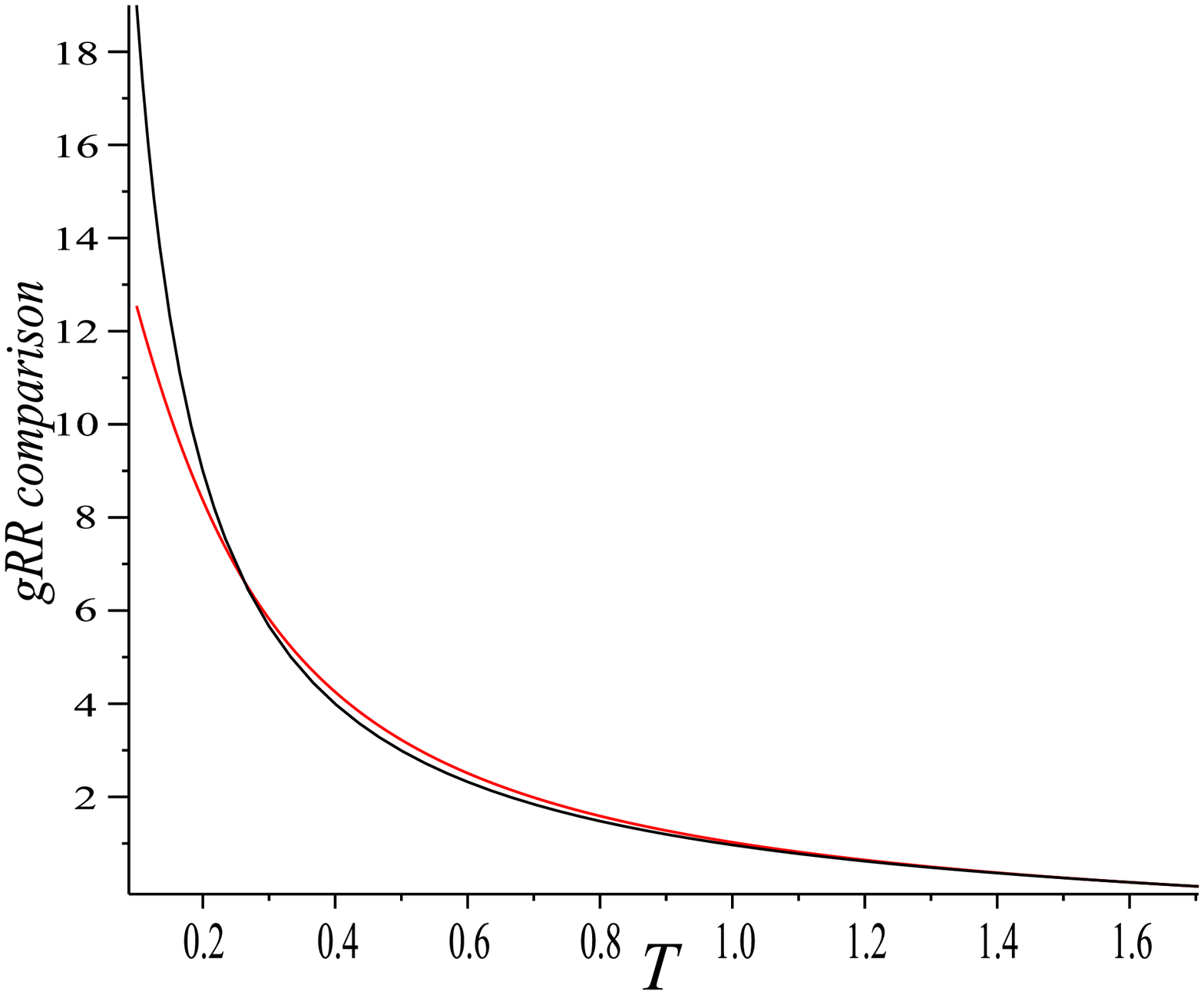} \vspace{0.3cm} 
\hspace{0.3cm} h) \includegraphics[height=1.25in, width=2.0in]{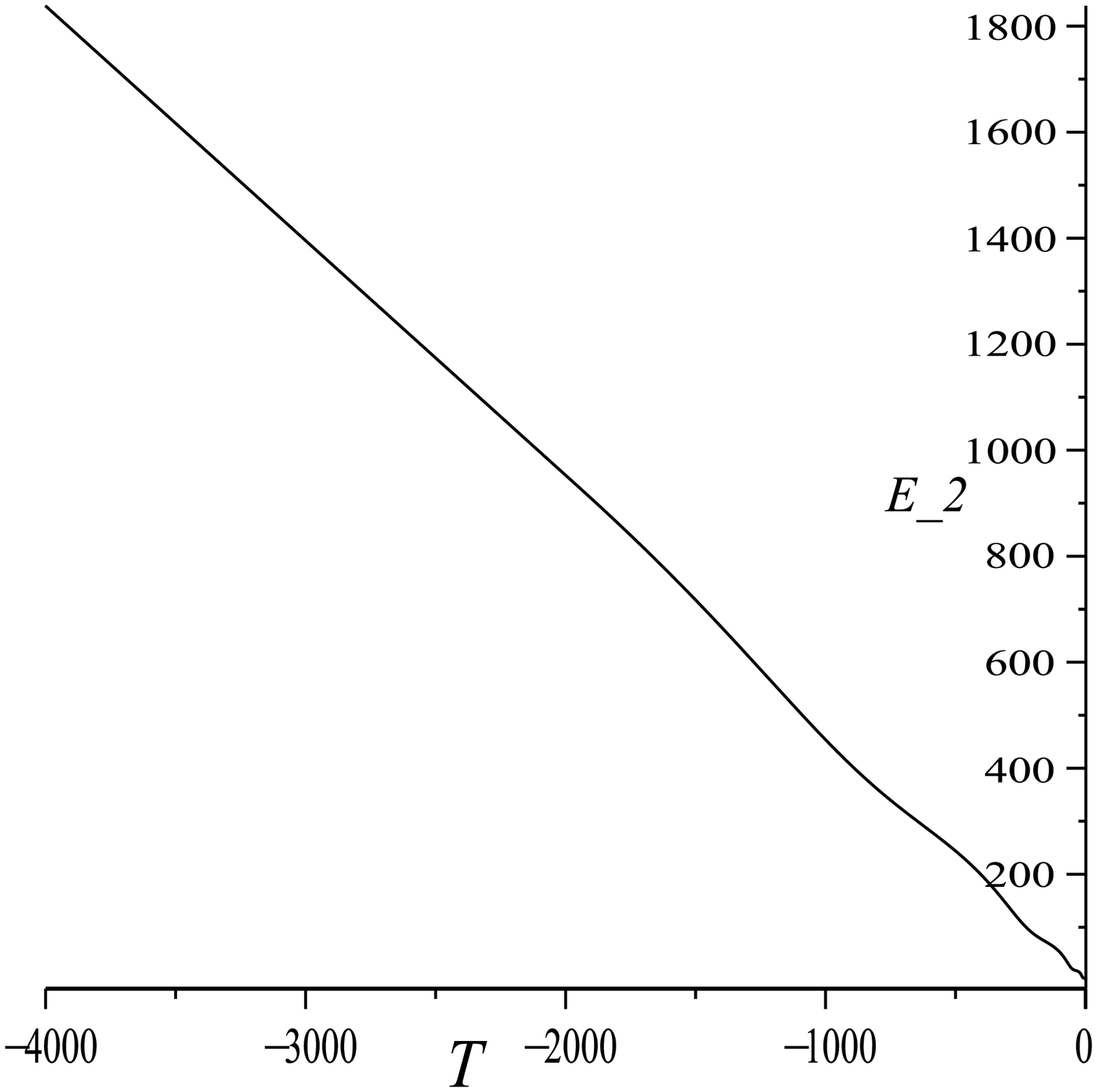}
\caption{{\small The quantum evolution for a Schwarzschild-anti deSitter black hole interior. a) The triad component $E_{3}$ plotted in the vicinity of the classical singularity. b) The triad component $E_{3}$ plotted from $T=-350$ to $T\approx -70$. c) The triad component $\pfive$ up to $T=-1\times10^{8}$. d) The triad combination $E_{2}^{2}/E_{3}$ which is equal to the classical metric component $g_{RR}$. e) A close-up of the previous plot, near the classical singularity. f) A comparison of the classical (black) vs. quantum (red) evolution of $E_{3}$ for positive $T$. g) A comparison of the classical (black) vs. quantum (red) evolution of $E_{2}^{2}/E_{3}$ for positive $T$. h) The triad component $E_{2}$. The horizon is located at $T\approx 1.7$ and the following parameters were used: $\gamma=0.274$, $M=1$, $\Lambda\approx -0.1$. }}
\label{fig:sphads}
\end{figure}

A set of sample results for the case of negative cosmological constant is shown in figure \ref{fig:sphads}.

From the two sets of figures, it can be seen that although there are some small qualitative differences in the evolution of some of the components, the singularity is avoided in both cases. The component $\pc$ never vanishes. The late time behavior of the two solutions is similar, approaching a Nariai-type universe similar to the scenario studied with $\Lambda=0$ \cite{ref:BandV}. For reference, a Nariai-type universe in the large $T$ limit, in the coordinate system chosen here can be written as
\begin{equation}
 ds^{2}=-\frac{a_{0}}{b_{0}^{\:2}T^{2}}dT^{2} + \frac{b_{0}^{\:2}T^{2}}{a_{0}}dR^{2}+a_{0}\,d\theta^{2} +a_{0}\,\sin^{2}\theta\,d\phi^{2}\,, \label{eq:nariaitype}
\end{equation}
with $a_{0}$ and $b_{0}$ constants. All orthonormal Riemann components are constants in this space-time. Furthermore, the radius of two-spheres is $\sqrt{a_{0}}$. The true Nariai universe (with all constant orthonormal Riemann components equal) is recovered in the limit $b_{0}=1$. From the graph of $E_3$, we can see that the radius of the two-sphere is of
the order of the Planck length. In the asymptotic regime, we assume that classical general relativity holds and that it is therefore justified to use the Einstein field equations, although caution must be used given that some the dimensions are near Planck size.

\subsection{The planar black hole interior}
In this section we examine the evolution of black holes with planar symmetry. We begin with a general plane-symmetric connection and densitized triad \cite{ref:neville}: \vspace{0.1cm}
\begin{subequations}
\begin{align}
A=&\left(A^{X}_{\;x} \tau_{X} + A^{Y}_{\;x}\tau_{Y}\right)dx + \left(A^{Y}_{\;y} \tau_{Y} + A^{X}_{\;y}\tau_{X}\right)dy + A^{Z}_{\;\zeta} \tau_{Z}\,d\zeta, \label{eq:planarA} \\
E=&\left(E^{x}_{\;X} \tau_{X} + E^{x}_{\;Y} \tau_{Y}\right) \frac{\partial}{\partial x} +\left(E^{y}_{\;Y} \tau_{Y} + E^{y}_{\;X} \tau_{X}\right) \frac{\partial}{\partial y} +E^{\zeta}_{\;Z}\tau_{Z} \frac{\partial}{\partial \zeta} \,, \label{eq:planarE}
\end{align}
\end{subequations}
with $\zeta$ the space-like coordinate corresponding to the time-like coordinate in the exterior domain of the black plane. All coefficients are functions of the interior time, $T$, only.

Before continuing, there is one subtlety that needs to be addressed for the subsequent black hole scenarios which is not present for the spherical black hole. In the previous case, the fact that the scenario studied consisted of pure (Lambda) gravity with imposed spherical symmetry was sufficient to guarantee that the system under study was the Schwarzschild-(A)dS black hole, at least in the purely classical case. Such a guarantee is not present, even in the classical case, for other symmetries. That is, there is no Birkhoff's theorem on which we can rely on. Admittedly, in the quantized evolution even in the spherical case we do not have a uniqueness theorem. However, the fact that Birkhoff's theorem holds (at least approximately) up until the quantum effects become strong should help insure that we are studying the analog of the Schwarzschild-(A)dS until the late-time quantum evolution takes over. To address this issue in the non-spherical cases, we test the classical evolution equations by comparing their results in the classical case with the known exact solutions. We set a tight tolerance in the evolution equations so that the numerically evolved classical equations agree with the exact solution within the tolerance throughout the entire evolution. We then ``quantize'' the equations of motion and evolve the quantum system using this tolerance. 

The simplest planar black hole interior metric is given by \cite{ref:vanzo}
\begin{equation}
ds^{2}=  -\frac{dT^{2}}{\frac{\kappa}{T}+\frac{\Lambda}{3}T^{2}} + \left(\frac{\kappa}{T}+\frac{\Lambda}{3}T^{2}\right)d\zeta^{2} +T^{2}\,\left(dx^{2} + dy^{2}\right)\,, \label{eq:planarmet}
\end{equation}
where $\kappa$ is a constant related to the mass-per-unit-area (recall $\Lambda < 0$ here). The classical time coordinate $T$ possesses the range 
$0 < T <  {^{3}\hspace{-0.2cm}\sqrt{\frac{3 \kappa}{-\Lambda}}}\,$ where the upper limit corresponds to the event horizon. This metric will be used as a consistency check for our evolution scheme as well as to provide the initial conditions for both the classical and quantum evolutions at the horizon. However, it should be emphasized that the momentum components of $E$ and the configuration components of $A$ are treated as free parameters spanning the minisuperspace of the system and the geometry is not a priori fixed, save for the initial data surface. 

The Gauss constraint, utilizing (\ref{eq:planarA}) and (\ref{eq:planarE}) yields only the following one condition:
\begin{equation}
-A^{Y}_{\;x} E^{x}_{\;X}+A^{X}_{\;x}E^{x}_{\;Y}-A^{Y}_{\;y}E^{y}_{\;X}+A^{X}_{\;y}E^{y}_{\;Y}=0\,. \label{eq:planarGC}
\end{equation}
Satisfying this constraint will also automatically satisfy the vector constraint. We can gauge fix the system in a way to satisfy this constraint and be compatible with the metric (\ref{eq:planarmet}) (via (\ref{eq:mettriadreln})) with the prescription:
\begin{equation}
 A^{X}_{\;x}=A^{Y}_{\;y},\; E^{x}_{\;X}=E^{y}_{\;Y},\;  A^{Y}_{\;x}=0,\; A^{X}_{\;y}=0,\; E^{y}_{\;X}=0,\; E^{x}_{\;Y}=0. \label{eq:planaransatz}
\end{equation}
The equality of the two diagonal components is inspired by the planar symmetry of the system whereas setting the off-diagonal terms to zero satisfies the Gauss constraint\footnote{For the planar case the fiducial spin connection $\Gamma^{i}_{\;a}$ vanishes and therefore there is no inconsistency in choosing $A^{i}_{\;a}$ as diagonal.} (\ref{eq:planarGC}). Before continuing, we mention that although $A^{X}_{\;x}=A^{Y}_{\;y}$ and $E^{x}_{\;X}=E^{y}_{\;Y}\,$, we treat them here as independent degrees of freedom and therefore they are evolved as such. We will show below that with the appropriate initial conditions of $A^{X}_{\;x}(T_{0})=A^{Y}_{\;y}(T_{0})$ and $E^{x}_{\;X}(T_{0})=E^{y}_{\;Y}(T_{0})$, the equations of motion will preserve this equality for all subsequent time.


With the above ansatz, the resulting classical Hamiltonian reads:
\begin{align}
 H=&-\frac{2N}{\sqrt{\pone\pfour\pfive}}\left[\pone\pfour\cone\cfour + \pone\pfive\cone\cfive+ \pfour\pfive\cfour\cfive\right] \nonumber \\ 
&+\frac{N\Lambda}{2} \sqrt{\pone\pfour\pfive}\;, \label{eq:classplanarham}
\end{align}
where quantities have been re-scaled in order to absorb the length factors that arise from the truncated spatial integrations. We choose to work in a coordinate system compatible with (\ref{eq:planarmet}) and therefore take\footnote{An ambiguity exists regarding whether to take $\pone$ or $\pfour$ (or some combination) in this expression. In this classical case, it makes no difference. We discuss the issue below in the quantum case.} $N=\frac{\sqrt{\pfive}}{\pone}$. Utilizing (\ref{eq:mettriadreln}) and (\ref{eq:planarmet}), this allows us to write the line element as
\begin{equation}
 ds^{2}=-\frac{\pfive}{\pone{}^{2}} dT^{2} + \frac{\pone{}^{2}}{\pfive} d\zeta^{2} + \pfive\left(dx^{2} + dy^{2}\right).
\end{equation}
This choice leads to equations of motion:
\begin{align}
&\dot{\cone}+ \frac{2}{\pone{}^{3/2}\sqrt{\pfour}}\left[\pfour\cone\cfour+ \pfive\cone\cfive\right] \nonumber \\
&-\frac{3}{\pone{}^{3/2}\sqrt{\pfour}} \left[\pone\pfour\cone\cfour +\pone\pfive\cone\cfive + \pfour\pfive\cfour\cfive\right] \nonumber \\
&+ \frac{\Lambda \pfive\pfour}{\pone{}^{3/2} \sqrt{\pfour}}=0\,, \label{planarnonQ1} 
\end{align}
\begin{align}
&\dot{\cfour}+\frac{2}{\pone{}^{3/2}\sqrt\pfour}\left[\pone\cone\cfour+\pfive\cfour\cfive\right] \nonumber \\
&-\frac{1}{\pone{}^{3/2} \pfour{}^{3/2}}\left[\pone\pfour\cone\cfour + \pone\pfive\cone\cfive +\pfour\pfive\cfour\cfive\right] \nonumber \\
&-\frac{\Lambda \pfive}{\sqrt{\pone\pfour}}=0\, , \label{eq:planarnonQ2} 
\end{align}
\begin{align}
&\dot{\pone} -\frac{2}{\pone{}^{3/2}\sqrt{\pfour}}\left[\pone\pfour\cfour-\pone\pfive\cfive\right] =0\, , \label{eq:planarnonQ3} 
\end{align}
\begin{align}
&\dot{\pfour} -\frac{2}{\pone{}^{3/2} \sqrt{\pfour}} \left[\pone\pfour\cone+ \pfour\pfive\cfive\right] =0\,, \label{eq:planarnonQ4}  
\end{align}
\begin{align}
&\dot{\cfive} +\frac{2}{\pone{}^{3/2} \sqrt{\pfour}} \left[\pone\cone\cfive + \pfour\cfour\cfive\right] -\frac{\Lambda \sqrt{\pfour}}{\sqrt{\pone}}=0\,, \label{eq:planarnonQ5} 
\end{align}
\begin{align}
&\dot{\pfive} -\frac{2}{\pone{}^{3/2} \sqrt{\pfour}} \left[\pone\pfive\cone + \pfour\pfive\cfour\right]=0\,. \label{eq:planarnonQ6}
\end{align}

\begin{figure}[h!t]
\centering
\vspace{0.3cm} a) \includegraphics[height=1.25in, width=2.0in]{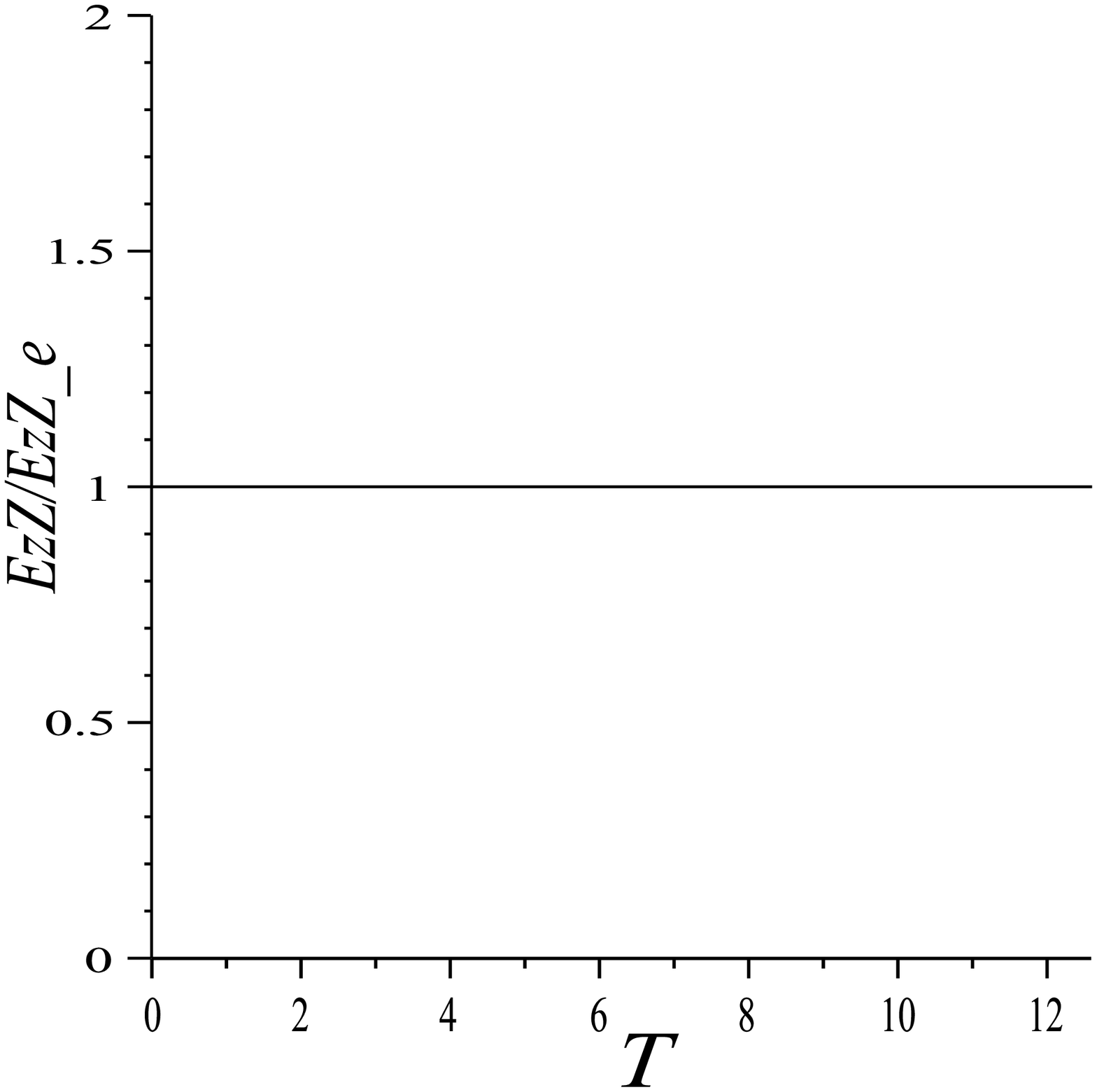}\vspace{0.3cm}
\hspace{0.5cm} b) \includegraphics[height=1.25in, width=2.0in]{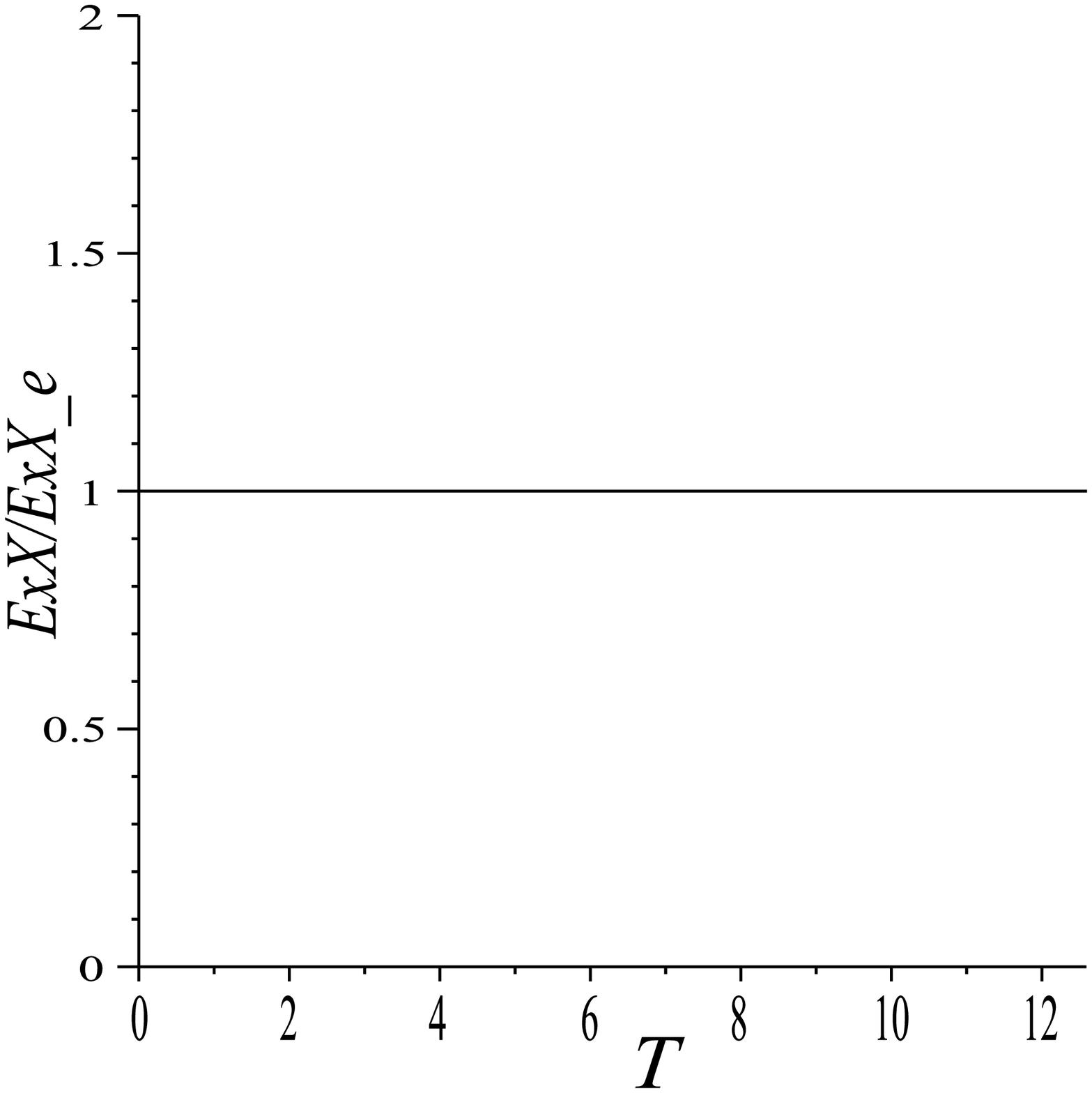} \vspace{0.3cm} \\
c) \includegraphics[height=1.25in, width=2.0in]{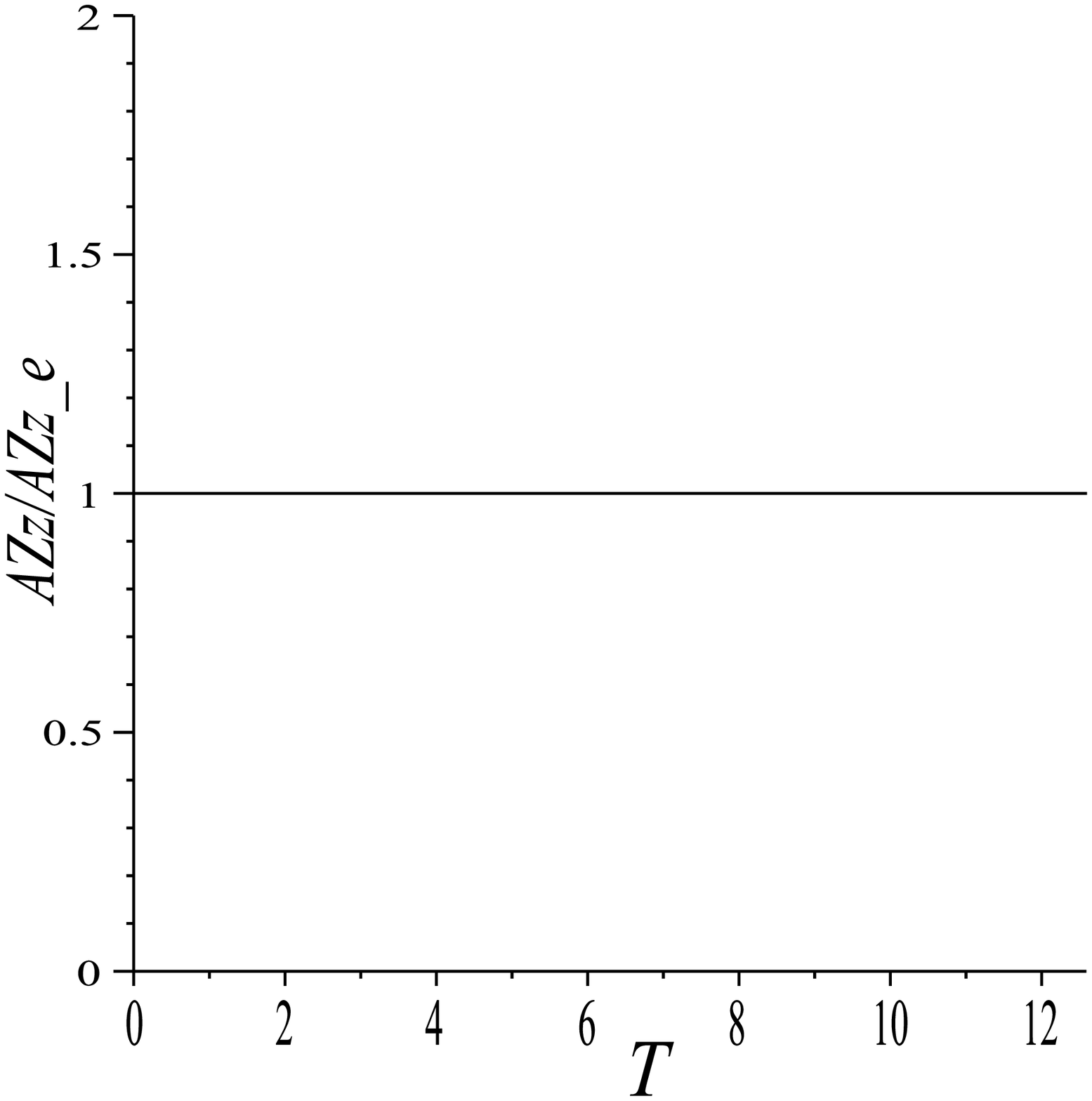} \vspace{0.3cm}
\hspace{0.5cm} d) \includegraphics[height=1.25in, width=2.0in]{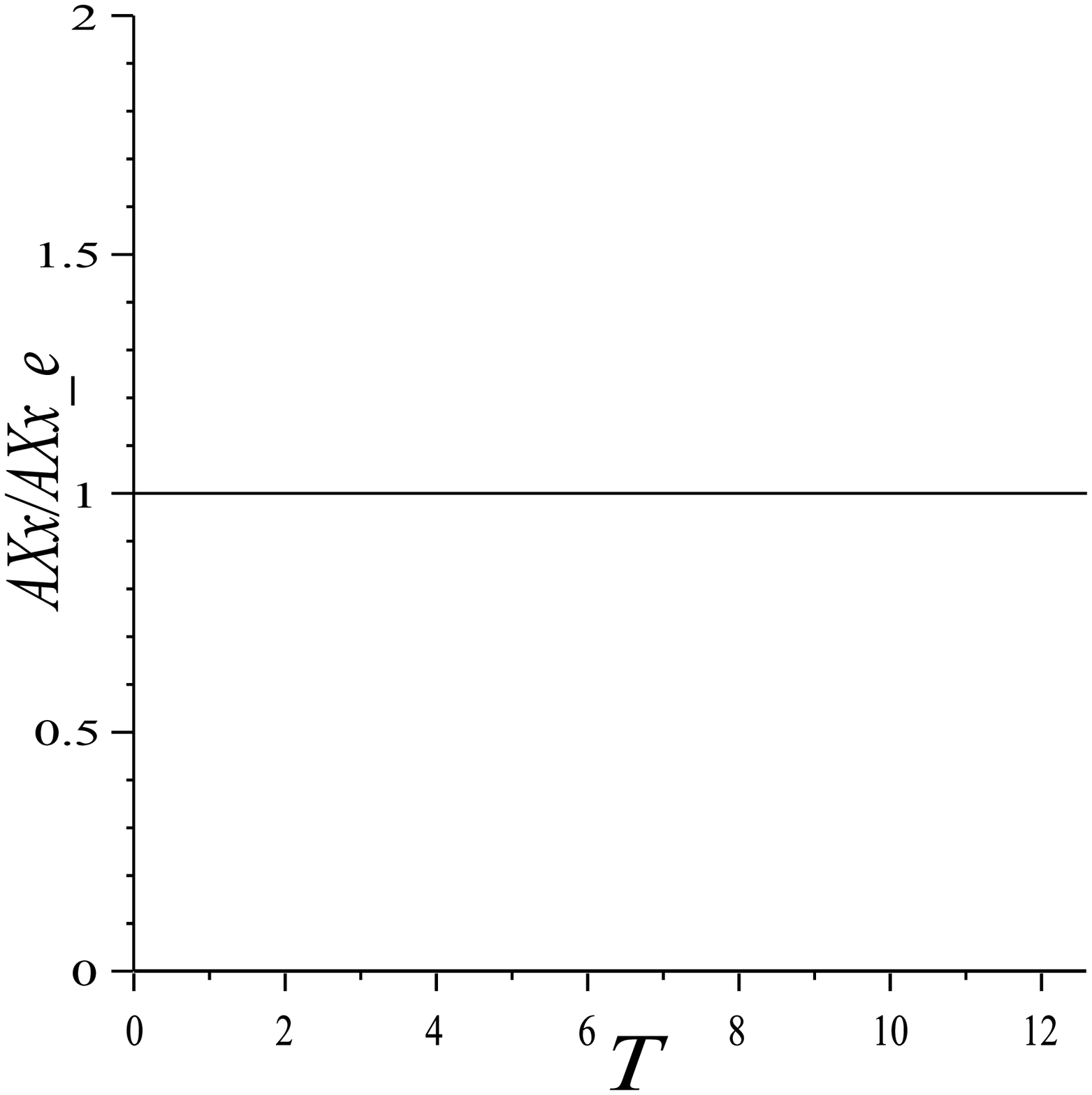}\vspace{0.3cm} \\
\hspace{-0.0cm} e) \includegraphics[height=1.25in, width=2.0in]{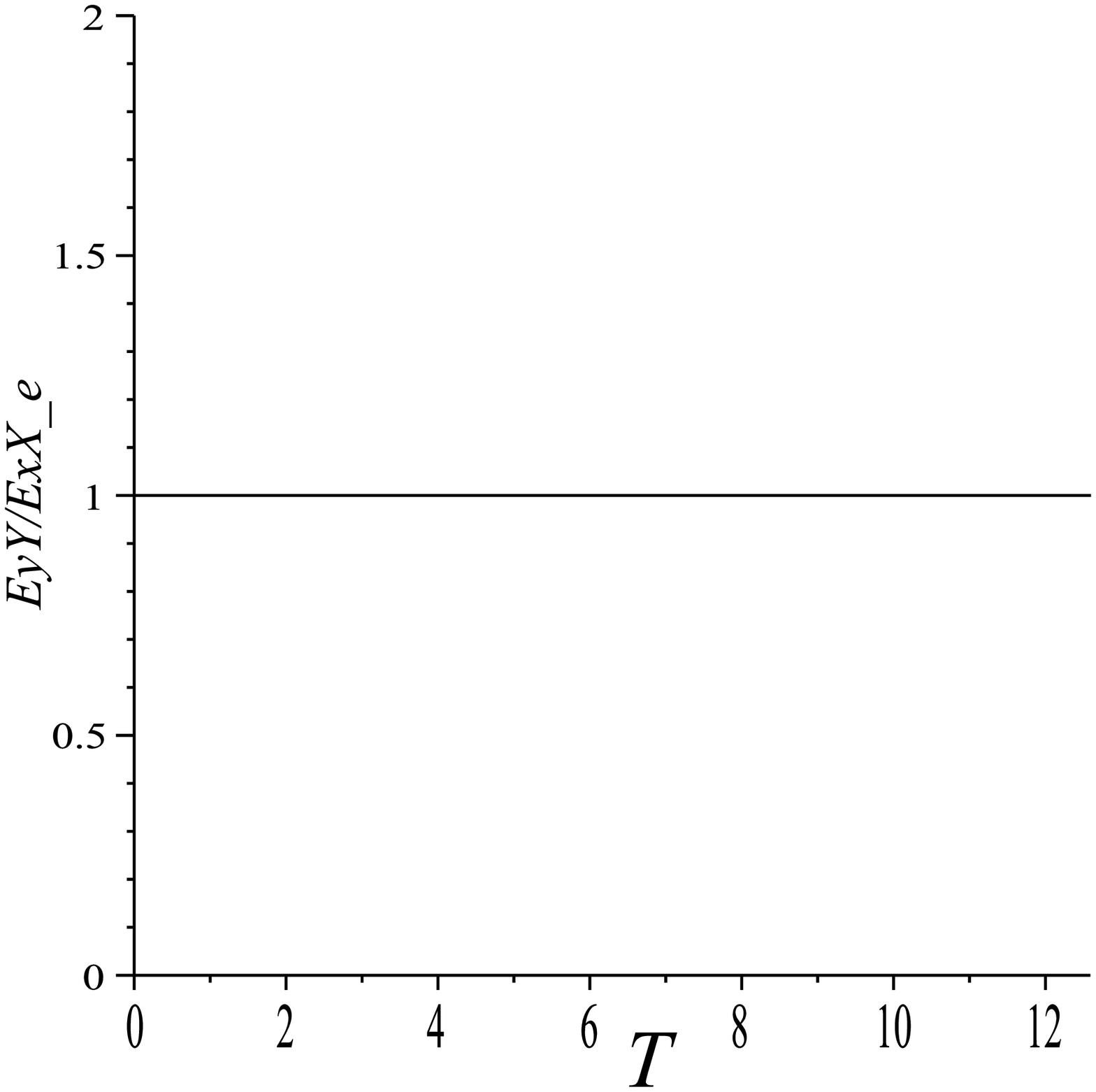} \hspace{0.55cm} \vspace{0.0cm}
f)  \includegraphics[height=1.5in, width=2.0in]{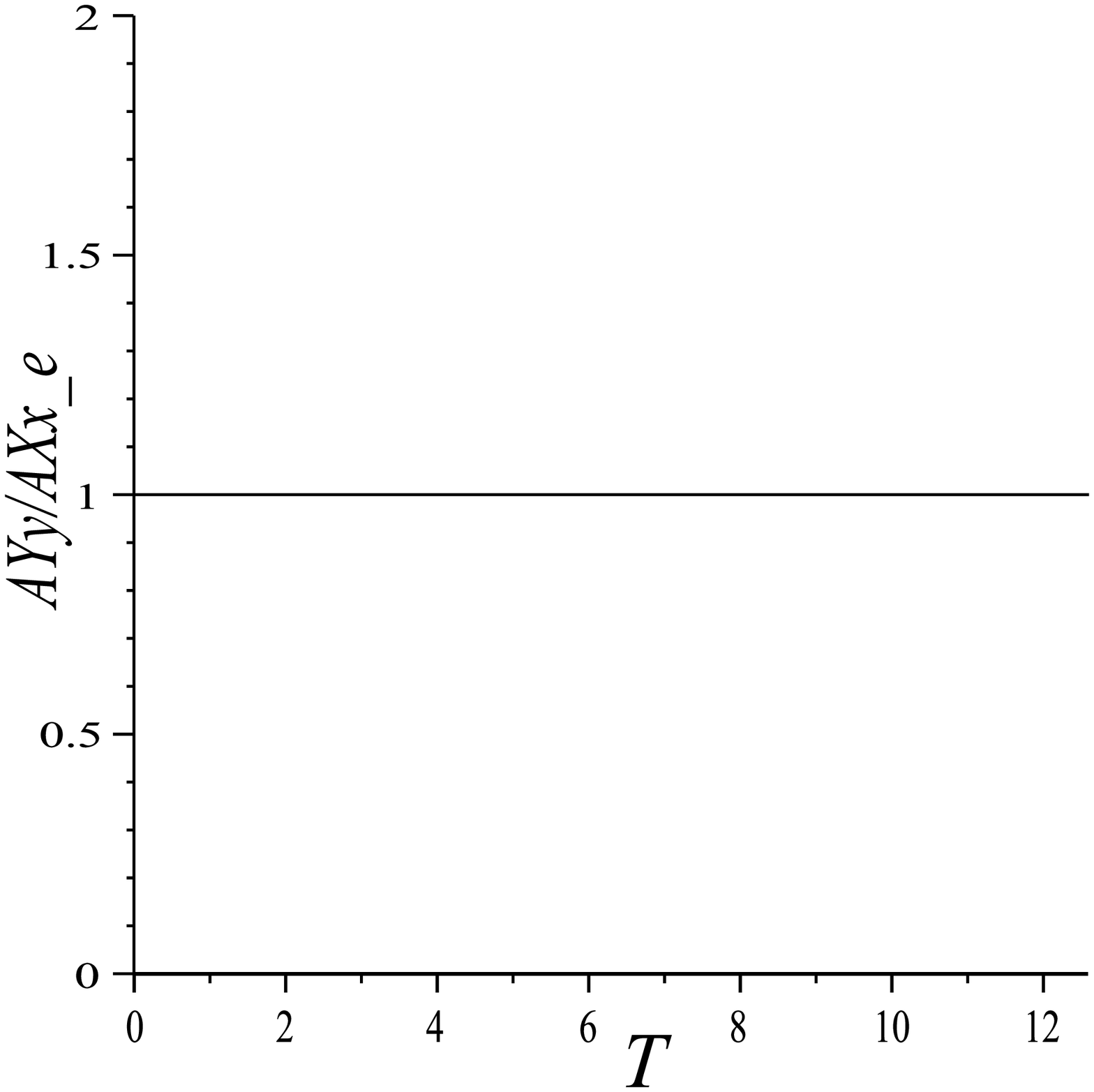}\vspace{0.0cm}
\caption{{\small A comparison of the numerically evolved classical solution to the exact classical solution (denoted with subscripts ``e''). We show here the ratios of a) $\pfive/\pfive{}_{e}\,$, b) $\pone/\pone{}_{e}\,$, c)$\cfive/\cfive{}_{e}\,$, d) $\cone/\cone{}_{e}\,$. The ratio is very close to unity, indicating that the numerical solution is equal to the exact solution throughout the entire evolution. In e) and f) we show a close-up of the ratio of $\pfour/\pone{}_{e}\,$ and $\cfour/\cone{}_{e}\,$ respectively. These plots indicate that the equality of these components is preserved throughout the evolution, compatible with the ansatz (\ref{eq:planaransatz}). The horizon is located at $T\approx 12.5$.}} \label{fig:evolvecheck}
\end{figure}

Although complex, these equations can actually be solved analytically. However, as mentioned above, we wish to test the computational evolution scheme and therefore we solve the above numerically, using the exact classical solution to set the values of the triad on the initial data surface just inside the horizon. The initial connection components are set from first solving the classical Hamiltonian equations for $\cfive$, as we know the classical $\pone$ and $\pfive$ from the form of the metric (\ref{eq:planarmet}). In figure \ref{fig:evolvecheck} we show the ratio of the numerical results to the exact analytic value and can see that the ratio is very close to unity throughout the evolution. It can be seen that near the singularity as well, the ratio is acceptably close to unity. We are therefore indeed evolving the planar black hole metric of (\ref{eq:planarmet}). It can also be seen from these figures that the equalities $\pone=\pfour$ and $\cone=\cfour$ are preserved throughout the evolution.


Next we study the corresponding quantum system. The quantization of the Hamiltonian proceeds as in the previous spherical case. However, there are some subtleties which must be addressed first. The classical three-space metric is given by (\ref{eq:planarmet}). However, when this metric is written in terms of the Ashtekar variable $E$, there is an ambiguity due to the equality of $\pone$ and $\pfour$. For example, all of the following combinations are equivalent:
\begin{subequations}
\begin{align}
ds^{2}=& -\frac{\pfive}{\pone{}^2}\,dT^{2} + \frac{\pone{}^2}{\pfive}\,d\zeta^{2} + \pfive\left(dx^{2} + dy^{2} \right), \label{eq:pambig1}\\
ds^{2}=& -\frac{\pfive}{\pfour{}^2}\,dT^{2} + \frac{\pfour{}^2}{\pfive}\,d\zeta^{2} + \pfive\left(dx^{2} + dy^{2} \right), \label{eq:pambig2}\\
ds^{2}=& -\frac{2\pfive}{\pone{}^{2}+\pfour{}^2}\,dT^{2} + \frac{\pone{}^{2}+\pfour{}^2}{2\pfive}\,d\zeta^{2} + \pfive\left(dx^{2} + dy^{2} \right), \label{eq:pambig3}\\
ds^{2}=& -\frac{4\pfive}{\left(\pone+\pfour\right)^{2}}\,dT^{2} + \frac{\left(\pone+\pfour\right)^{2}}{4\pfive}\,d\zeta^{2} + \pfive\left(dx^{2} + dy^{2} \right), \label{eq:pambig4} \\
ds^{2}=& -\frac{\pfive}{\left(\pone\pfour\right)}\,dT^{2} + \frac{\left(\pone\pfour\right)}{\pfive}\,d\zeta^{2} + \pfive\left(dx^{2} + dy^{2} \right), \label{eq:pambig5}
\end{align}
\end{subequations}
or any other combination which yields the same coordinate dependence for the metric. In the classical evolution, this ambiguity makes no difference to the subsequent evolution. However, in the quantum evolution there is a difference as the metric is used to define the length of the holonomy paths via the planar analog of (\ref{eq:rtarea}) and (\ref{eq:tparea}). It seems intuitive that one should pick one of the symmetric choices, similar to (\ref{eq:pambig3}), (\ref{eq:pambig4}) or (\ref{eq:pambig5}). In this case, it turns out that all of these symmetric schemes yield the same results and therefore the bulk of our results here utilize (\ref{eq:pambig5}), which yields the most simple Hamiltonian. The evolution utilizing (\ref{eq:pambig1}) or (\ref{eq:pambig2}) is, however, qualitatively different from the symmetric schemes. Below we show a sample evolution of this scenario as well. The important point is that \emph{all} scenarios tested removed the classical singularity but there are sometimes significant differences in the evolution.

The quantization procedure is similar to the spherical case.
Briefly, we define $\delta$s by:
\begin{equation}
 {\sf{A}_{xy}}=\pfive\, \delta_{x}^{\;2}=\Delta,\;\;\; {\sf{A}_{\zeta x}}={\sf{A}_{\zeta y}}=\sqrt{\pone \pfour}\,\delta_{\zeta}\delta_{x} = \Delta\,,
\end{equation}
where, as before, $\Delta$ is the minimum area predicted by loop quantum gravity. This yields:
\begin{equation}
 \delta_{x}=\sqrt{\frac{\Delta}{\pfive}}\,,\;\;\; \delta_{\zeta}=\frac{\sqrt{\Delta}\sqrt{\pfive}}{\sqrt{\pone\pfour}} \,.
\end{equation}
The quantum Hamiltonian reads:
\begin{align}
  H=&-\frac{2N}{\gamma^{2} \sqrt{\pone\pfour\pfive}} \left[\pone\pfour\ccone\ccfour \right. \nonumber \\
 & +  \left. \pone\pfive\ccone\ccfive +\pfour\pfive\ccone\ccfive \right] \nonumber \\
&+\frac{N\Lambda}{2\gamma^{2}} \sqrt{\pone\pfour\pfive}\;\,.
\end{align}
We choose the gauge
\begin{equation}
 N=\frac{\gamma^{2} \sqrt{\pfive}}{\sqrt{\pprod}}\,,
\end{equation}
which corresponds to using the coordinate system in (\ref{eq:planarmet}) in the particular ``momentum gauge'' of (\ref{eq:pambig5}). This leads to the following ``quantum'' equations of motion:

 \begin{align}
&\dot{\cone}-\frac{\pfive}{\Delta^{3/2}\pfour{}^{3/2}\pone{}^{5/2}}  \left[\sqrt{\pfive\pfour}\cfive\Delta\pone{}^{3/2}\sin\left(\frac{\cone\sqrt{\Delta}}{\sqrt{\pfive}}\right) \cos\left(\frac{\cfive\sqrt{\Delta\pfive}}{\sqrt{\pprod}}\right) \right. \nonumber \\
&-\pone{}^{2}\pfour\sqrt{\Delta}\sin\left(\frac{\cone\sqrt{\Delta}}{\sqrt{\pfive}}\right)\sin\left(\frac{\cfive\sqrt{\Delta\pfive}}{\sqrt{\pprod}}\right) \nonumber \\
& +\pfour{}^{3/2}\sqrt{\pfive\pone}\cfive\Delta \sin\left(\frac{\cfour\sqrt{\Delta}}{\sqrt{\pfive}}\right) \cos\left(\frac{\cfive\sqrt{\Delta\pfive}}{\sqrt{\pprod}}\right) \nonumber  \\
&+\left.\pfour{}^{2}\pone\sqrt{\Delta} \sin\left(\frac{\cfour\sqrt{\Delta}}{\sqrt{\pfive}}\right)\sin\left(\frac{\cfive\sqrt{\Delta\pfive}}{\sqrt{\pprod}}\right)\right] =0\,,
 \end{align}
 \begin{align}
&  \dot{\pone}- \frac{2\sqrt{\pfive}\cos\conefactor}{\pfour\sqrt{\Delta}}\left[\pfour\sin\left(\frac{\cfour\sqrt{\Delta}}{\sqrt{\pfive}}\right) + \sqrt{\pprod}\sin\left(\frac{\cfive\sqrt{\Delta\pfive}}{\sqrt{\pprod}}\right)\right]=0\,,
\end{align}
\begin{align}
&\dot{\cfive}-\frac{1}{2\sqrt{\pfive}\pprod} \left[ 2\Delta\pone\pfour\cone \cos\left(\frac{\cone\sqrt{\Delta}}{\sqrt{\pfive}}\right) \sin\left(\frac{\cfour\sqrt{\Delta}}{\sqrt{\pfive}}\right)\right. \nonumber \\
&-4\pprod\sqrt{\Delta}\sqrt{\pfive} \sin\left(\frac{\cone\sqrt{\Delta}}{\sqrt{\pfive}}\right) \sin\left(\frac{\cfour\sqrt{\Delta}}{\sqrt{\pfive}}\right)  \nonumber \\
&+2\pone\pfour\Delta\sin\left(\frac{\cone\sqrt{\Delta}}{\sqrt{\pfive}}\right) \cos\left(\frac{\cfour\sqrt{\Delta}}{\sqrt{\pfive}}\right) +\Lambda\Delta^{3/2}\sqrt{\pfive}\pone\pfour\nonumber \\
&-4\pone{}^{3/2}\sqrt{\pfour\pfive}\sqrt{\Delta}\sin\left(\frac{\cone\sqrt{\Delta}}{\sqrt{\pfive}}\right) \sin\left(\frac{\cfive\sqrt{\Delta\pfive}}{\sqrt{\pprod}}\right) \nonumber \\
&+2\pone{}^{3/2}\sqrt{\pfour}\Delta\cone \cos\conefactor\sin\cfivefactor \nonumber \\
&-2\pone\pfive\cfive\Delta\sin\conefactor\cos\cfivefactor \nonumber \\
&-4\pfour{}^{3/2}\sqrt{\pone\pfive}\sqrt{\Delta}\sin\cfourfactor\sin\cfivefactor \nonumber \\
&+2\pfour{}^{3/2}\sqrt{\pone}\cfour\Delta\cos\cfourfactor\sin\cfivefactor \nonumber \\
& - \left.2\pfour\pfive\cfive\Delta\sin\cfourfactor\cos\cfivefactor  \right]=0\,,
 \end{align}
\begin{align}
 \dot{\pfive} - \frac{2\pfive{}^{3/2}\cos\cfivefactor}{\pprod\sqrt{\Delta}} \left[\pone\sin\conefactor+ \pfour\sin\cfourfactor\right]=0\,,
\end{align}
where the over-dot denotes differentiation with respect to the parameter $T$. The $y-Y$ equations are related to the $x-X$ equations via the interchange of $x\rightarrow y$ and $X\rightarrow Y$ and vice-versa.

\begin{figure}[h!t]
\centering
\vspace{-1.6cm} a) \includegraphics[height=1.05in, width=2.0in]{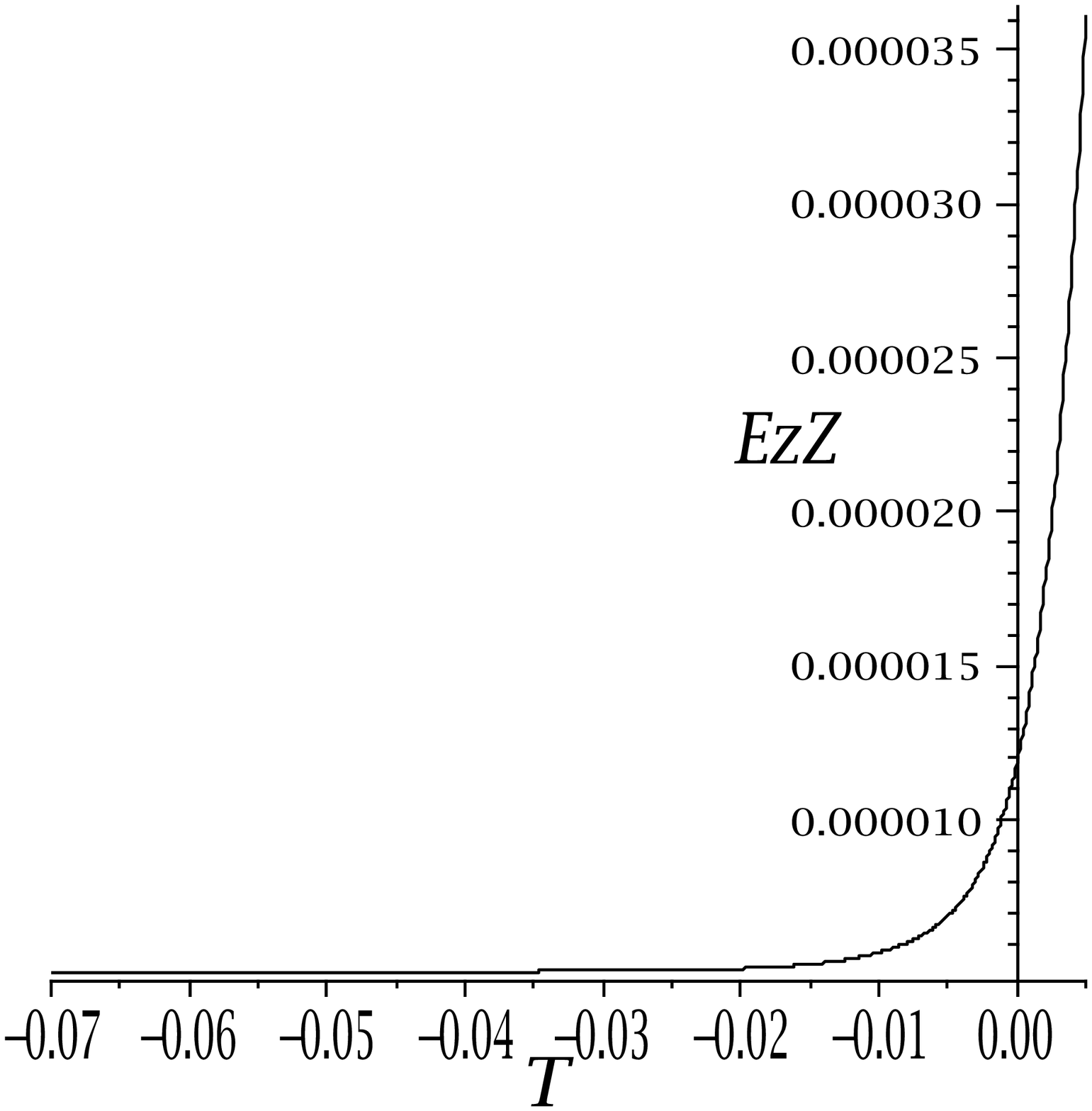}\vspace{0.3cm}
\hspace{0.3cm} b) \includegraphics[height=1.05in, width=2.0in]{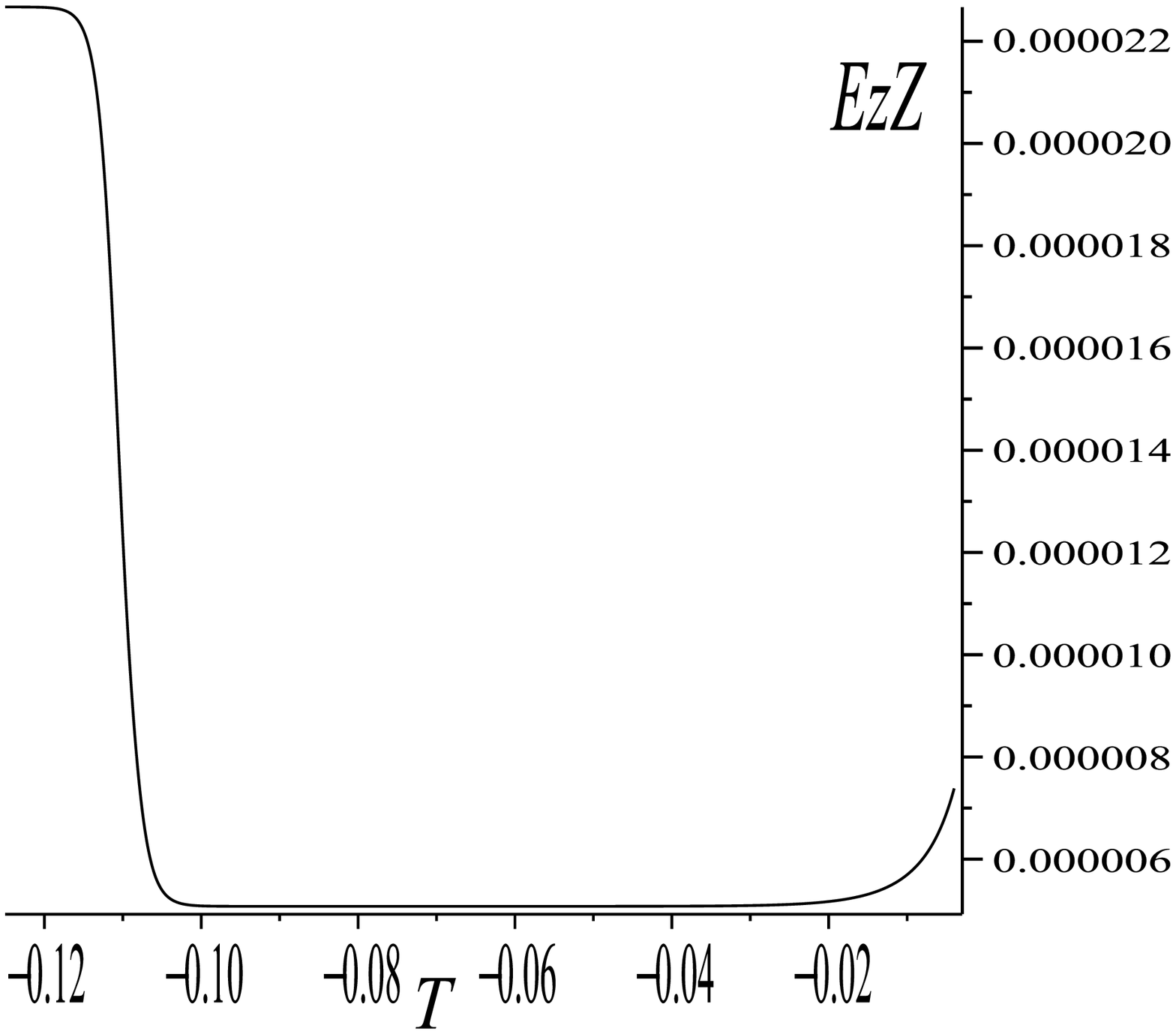} \vspace{0.3cm} \\
c) \includegraphics[height=1.05in, width=2.0in]{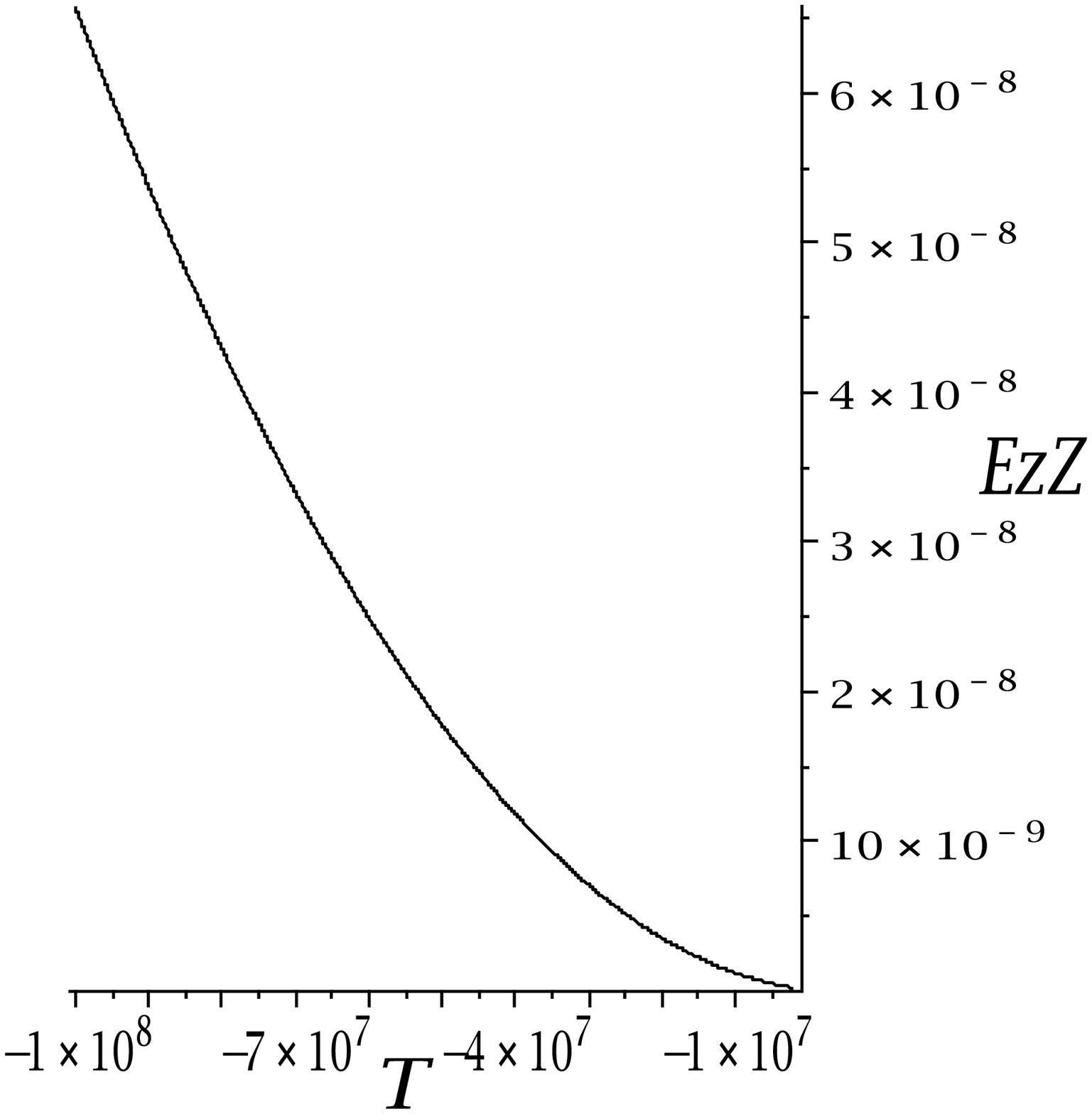} \vspace{0.3cm}
\hspace{0.3cm} d) \includegraphics[height=1.05in, width=2.0in]{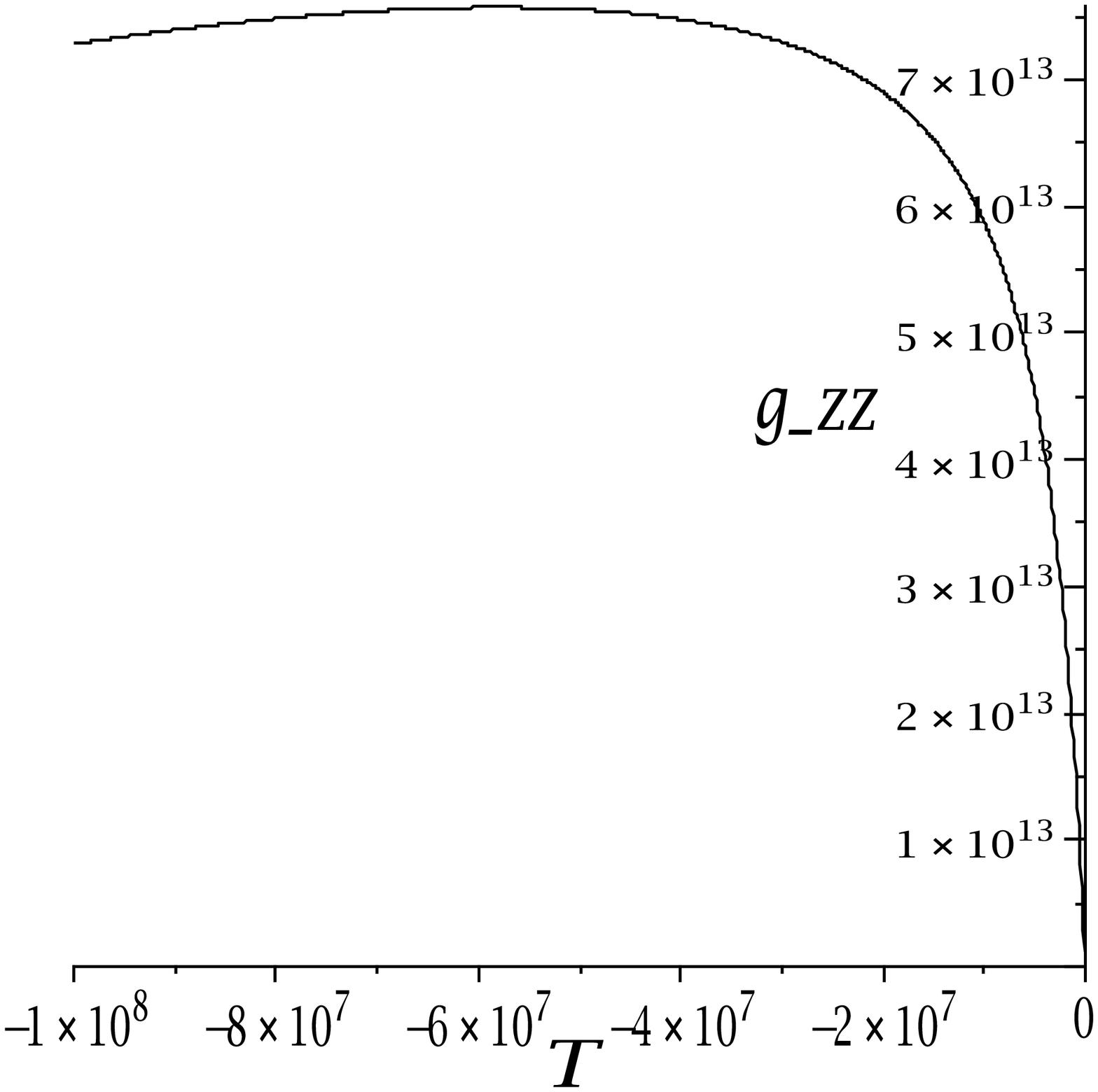}\vspace{0.3cm} \\
\hspace{0.3cm} e) \includegraphics[height=1.05in, width=2.0in]{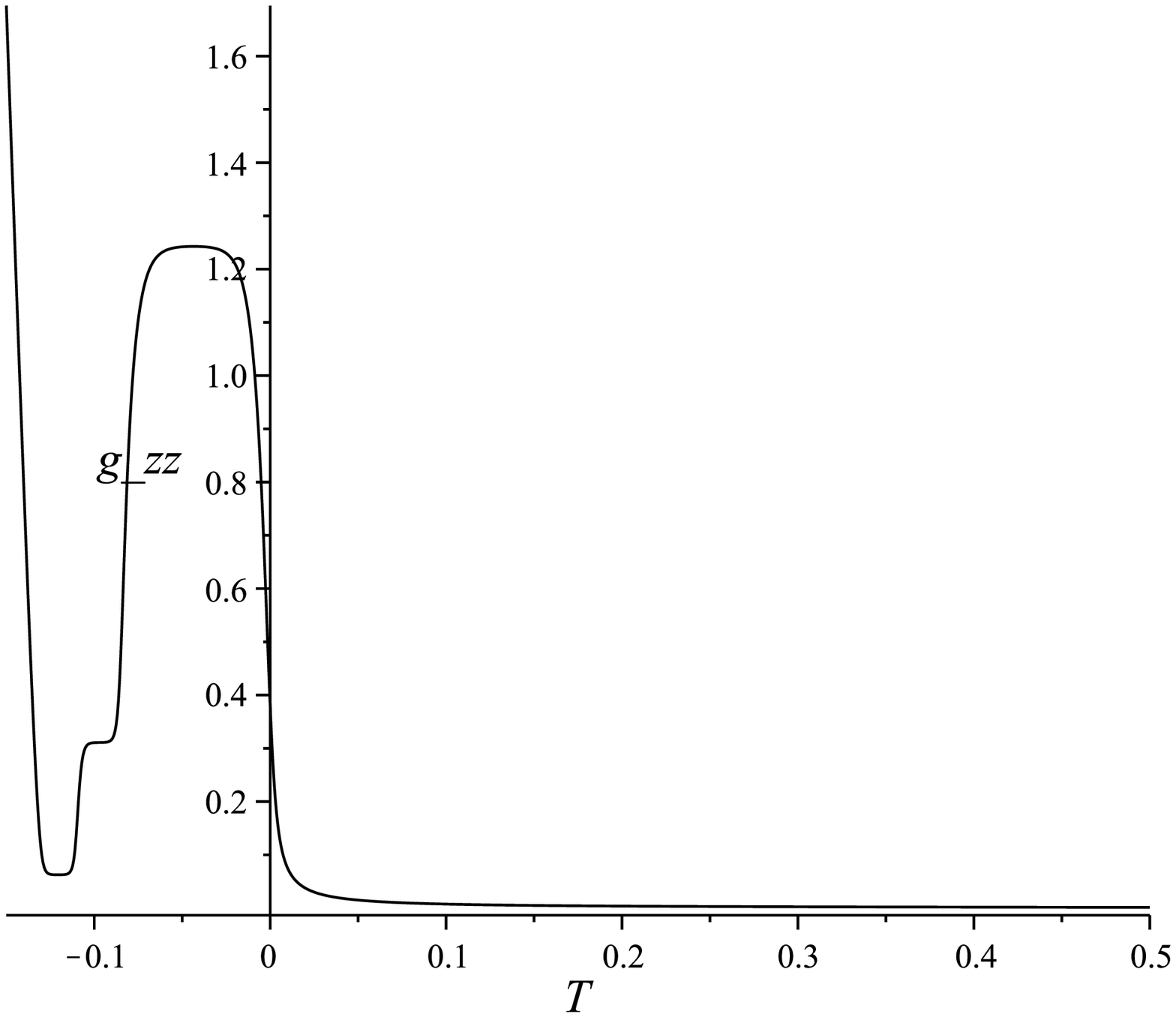}
f) \vspace{0.3cm} \includegraphics[height=1.3in, width=2.0in]{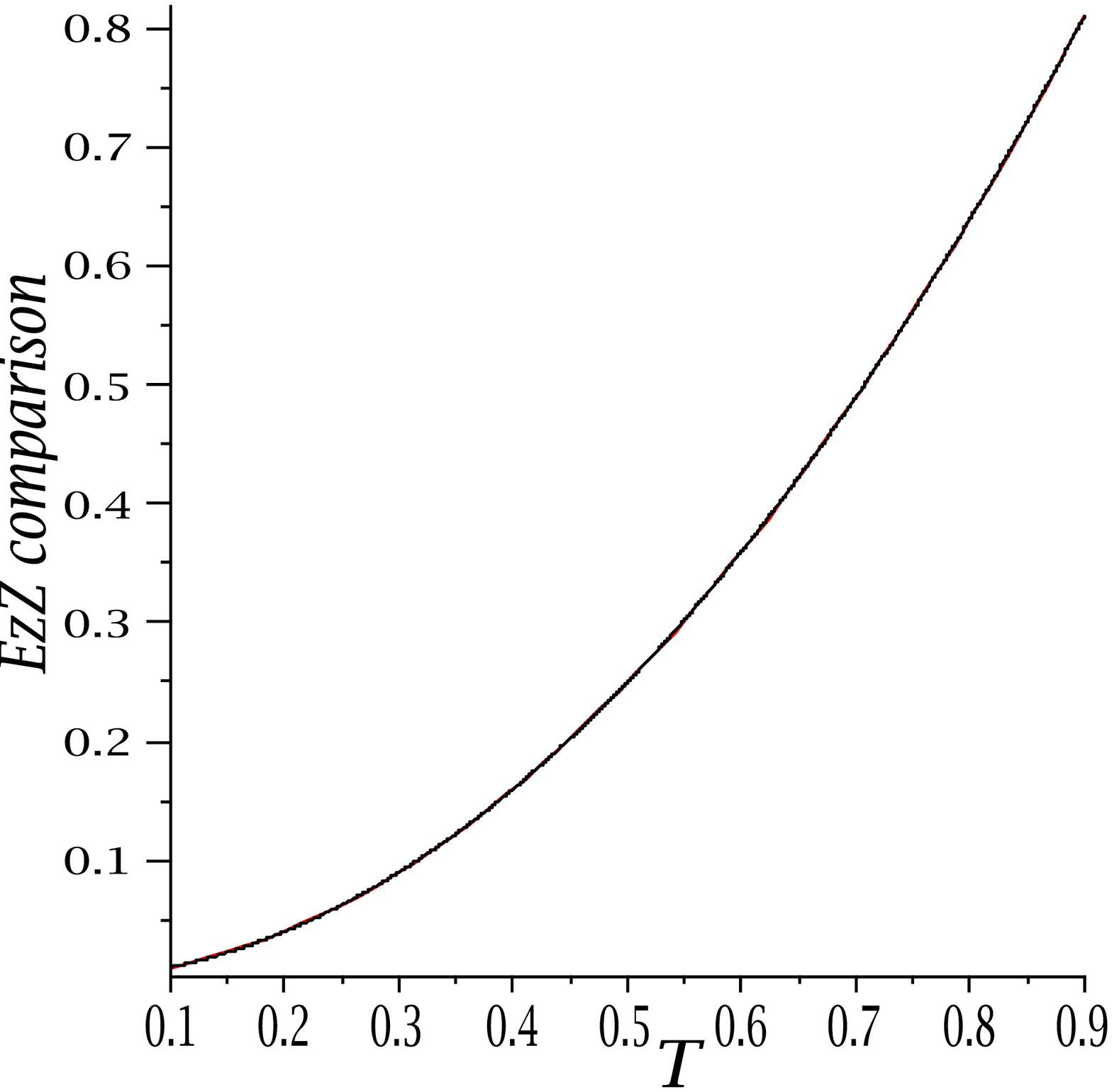}\vspace{0.3cm}\\
\hspace{0.3cm} g) \includegraphics[height=1.05in, width=2.0in]{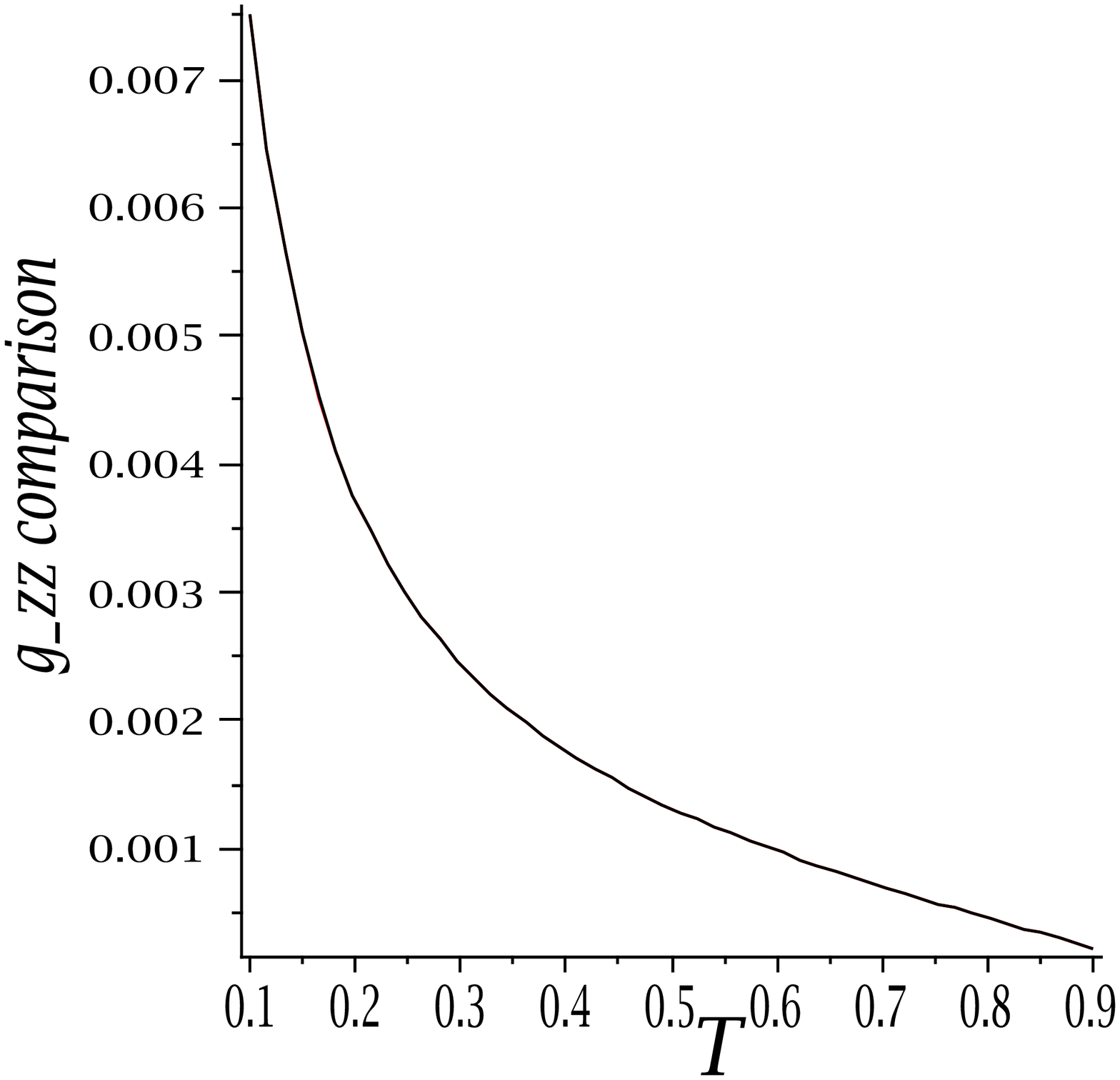} \vspace{0.3cm} 
\hspace{0.3cm} h) \includegraphics[height=1.05in, width=2.0in]{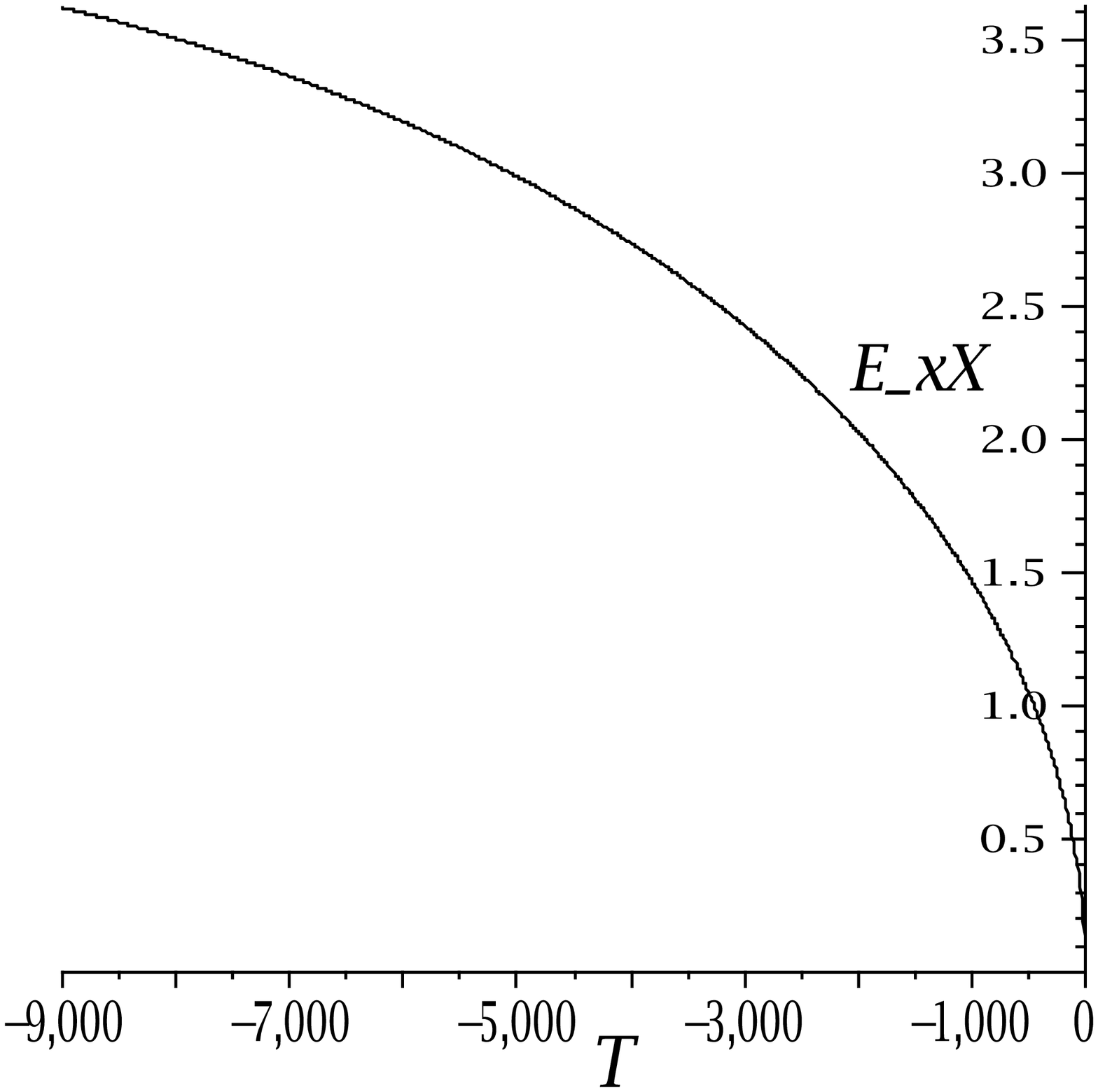} \\
\hspace{0.3cm} i) \includegraphics[height=1.05in, width=2.0in]{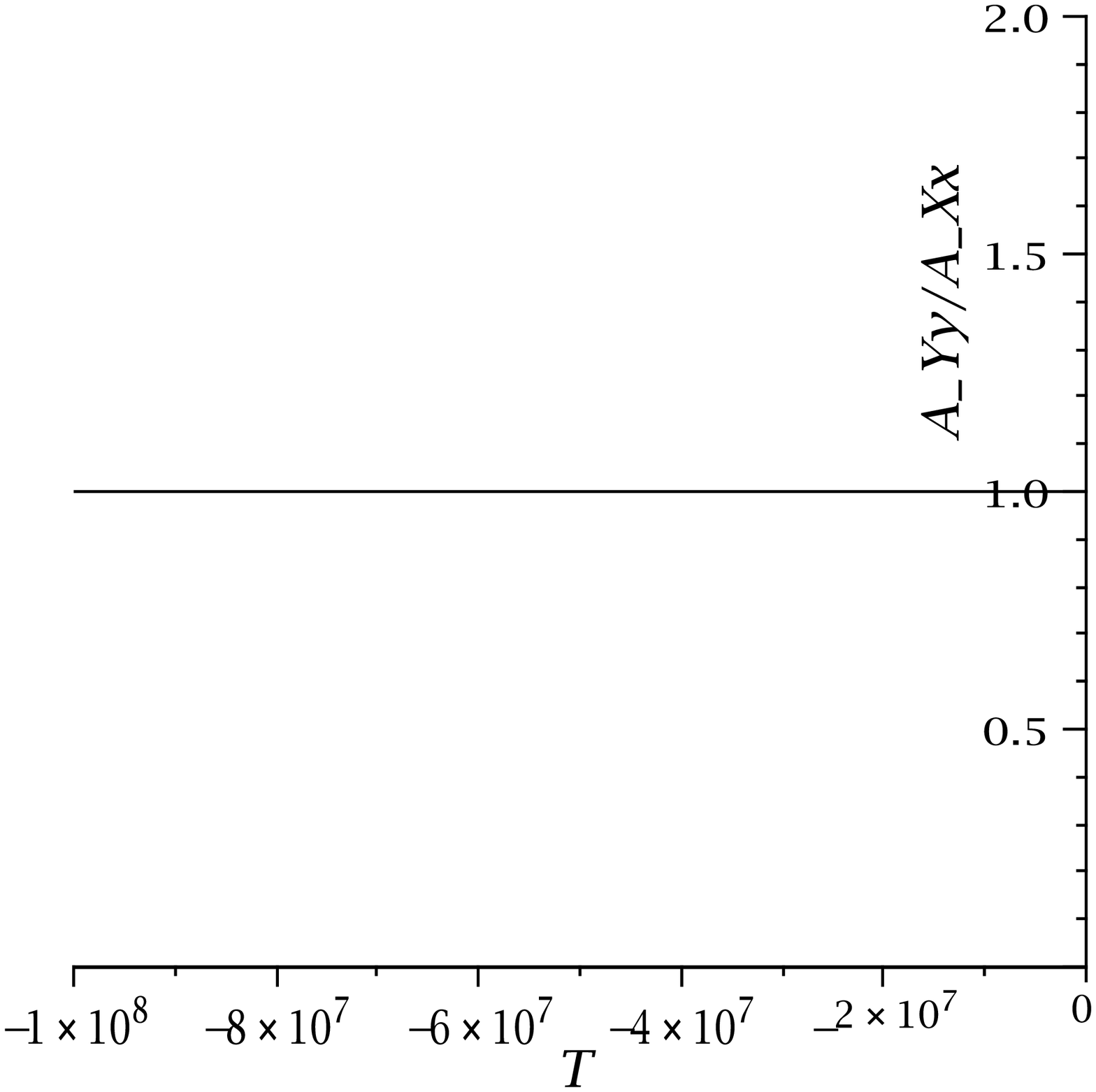}
\hspace{0.3cm} j) \includegraphics[height=1.05in, width=2.0in]{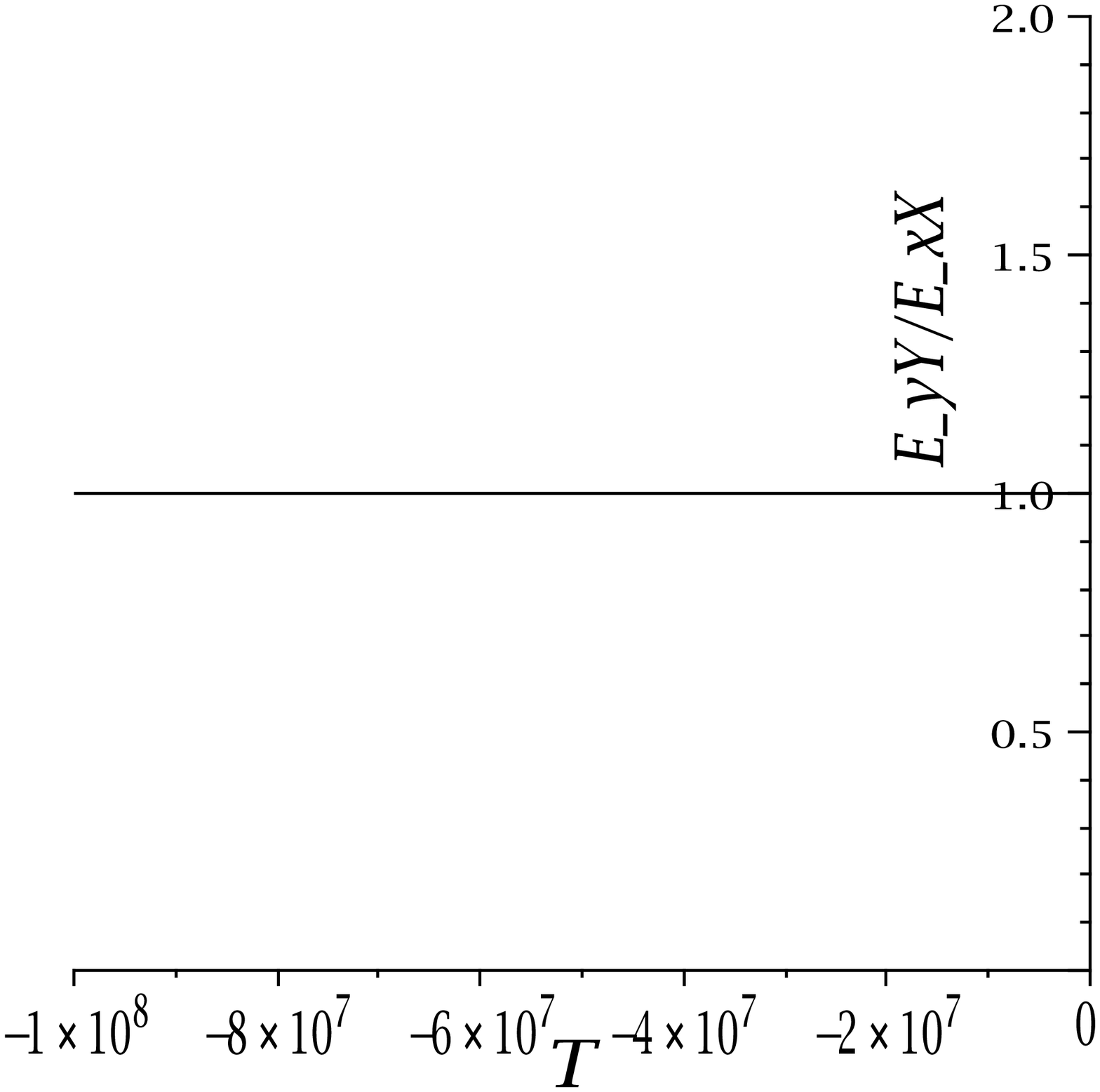}
\caption{{\small The quantum evolution for a planar black hole interior. a) The triad component $\pfive$ plotted in the vicinity of the classical singularity. b) The triad component $\pfive$ plotted in the vicinity of the bounce. c) The triad component $\pfive$ up to $T=-1\times10^{8}$. d) The triad combination $(\pone\pfour)/\pfive$ which is equal to the classical metric component $g_{\zeta\zeta}$. e) A close-up of the previous plot, near the classical singularity. f) A comparison of the classical (black) vs. quantum (red) evolution of $\pfive$ for positive $T$. g) A comparison of the classical (black) vs. quantum (red) evolution of $(\pone\pfour)/\pfive$ for positive $T$. h) The triad component $\pone$.  i) The connection component ratio $A^Y_y/A^X_x$. j) The triad component ratio $E^y_Y/E^x_X$. The horizon is located at $T=1$ and the following parameters were used: $\gamma=0.274$, $\kappa=0.00075$, $\Lambda\approx -0.0022$.}}
\label{fig:planarquant1}
\end{figure}

As before, we utilize classical values at a point near the horizon as the initial conditions. Results for one set of parameters is shown in figure \ref{fig:planarquant1}.

Note from (\ref{eq:pambig3}) that the classical singularity at $T=0$ corresponds to $\pfive = 0$. However, from figures \ref{fig:planarquant1}a, b, and c it can be seen that the evolution now proceeds through $T=0$ without $\pfive$ vanishing. In fact, $\pfive$ does not vanish for any value of $T$ but instead oscillates and then smoothly starts growing again as we go to large negative times. Therefore the singularity present in the classical theory is avoided in the corresponding quantum evolution. In figure \ref{fig:planarquant1}d we plot the densitized triad combination $\pone\pfour/\pfive$, which corresponds to the classical metric component $g_{\zeta\zeta}$. From figure \ref{fig:planarquant1}e, which is a close-up near $T=0$, it can be seen that this is also finite near the classical singularity. Figures \ref{fig:planarquant1}f and g compare the quantum evolution of $\pfive$ and $\pone\pfour/\pfive=g_{\zeta\zeta}$ to the exact (classical) solution from $T\approx 0$ to the near-horizon. It can be seen that the quantum quantities are almost coincident with the classical quantities in this regime as the two plots on graphs overlap significantly. The quantum effects therefore start to become important only very close to $T=0$.  Finally, figure \ref{fig:planarquant1}h shows the evolution of the triad component $\pone$ ($=\pfour$) and figures i and j show the consistency of   $E^y_Y=E^x_X$ and $A^Y_y = A^X_x $ throughout the evolution.

From Fig. 4c), $\pfive$ appears to grow without bound as T becomes more negative. If this is
the case, then the space cannot be Nariai-type like the other cases.

\begin{figure}[h!t]
\centering
\vspace{-0.1cm} a) \includegraphics[height=1.25in, width=2.0in]{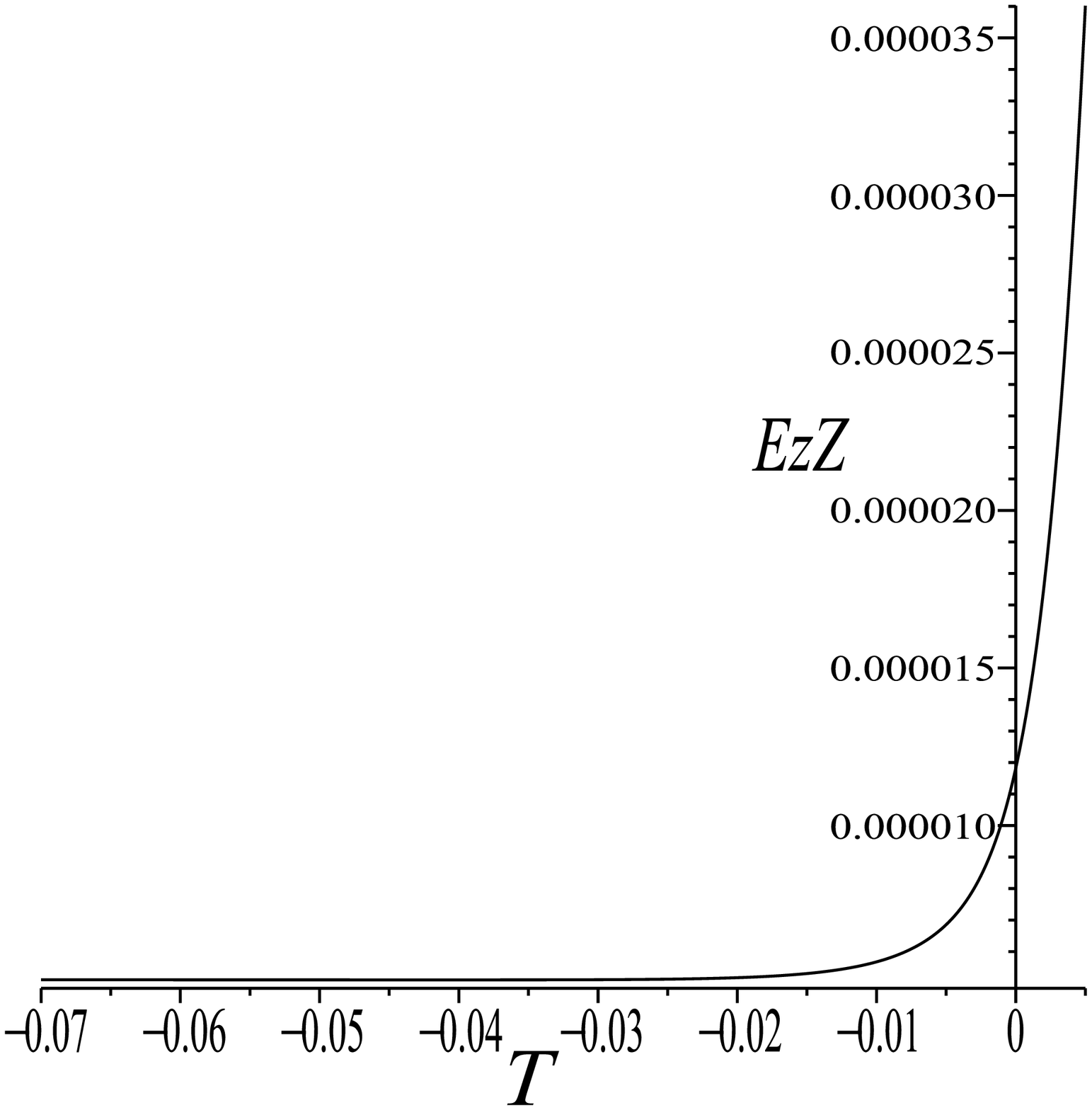}\vspace{0.3cm}
\hspace{0.3cm} b) \includegraphics[height=1.25in, width=2.0in]{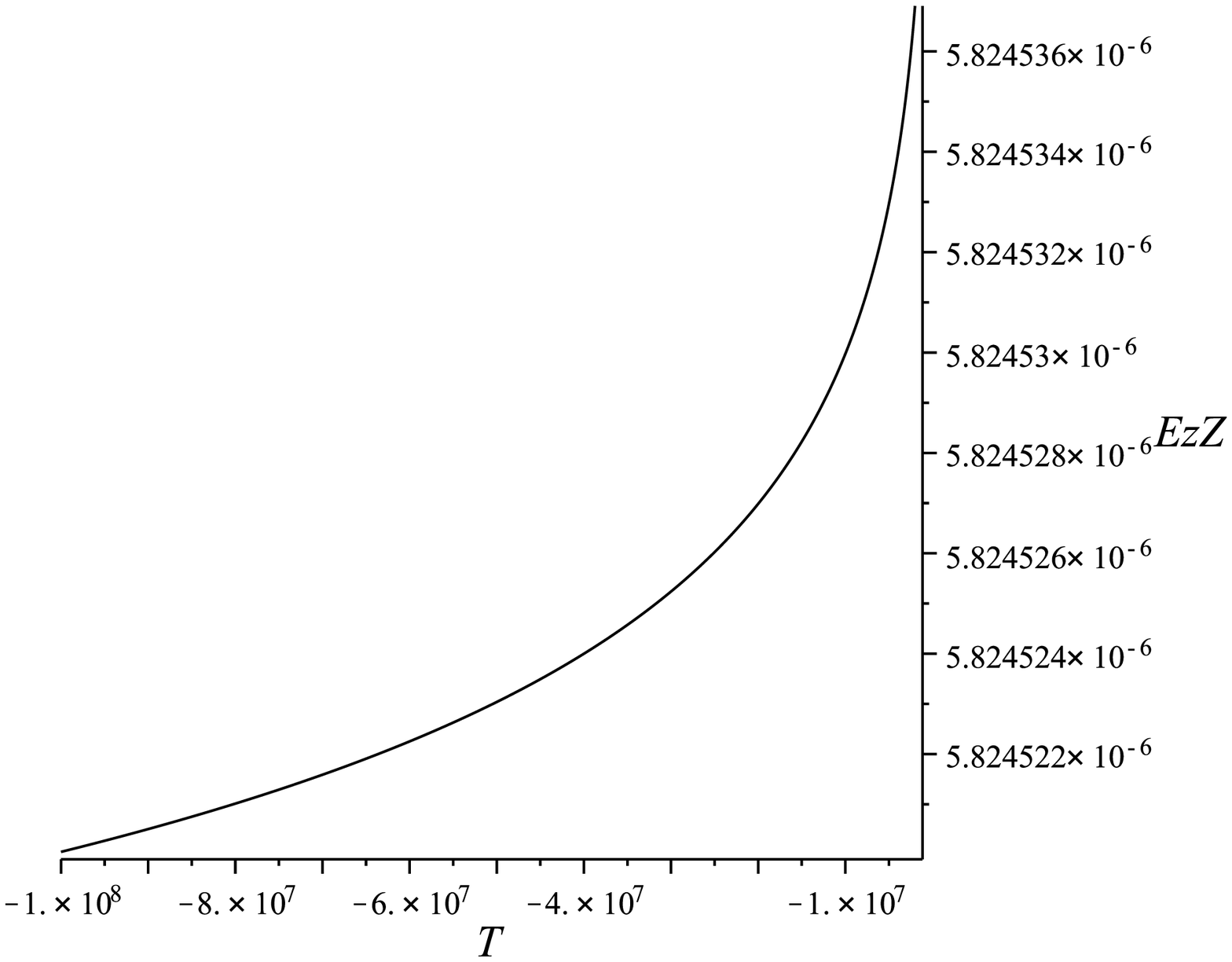} \vspace{0.3cm} \\
c) \includegraphics[height=1.25in, width=2.0in]{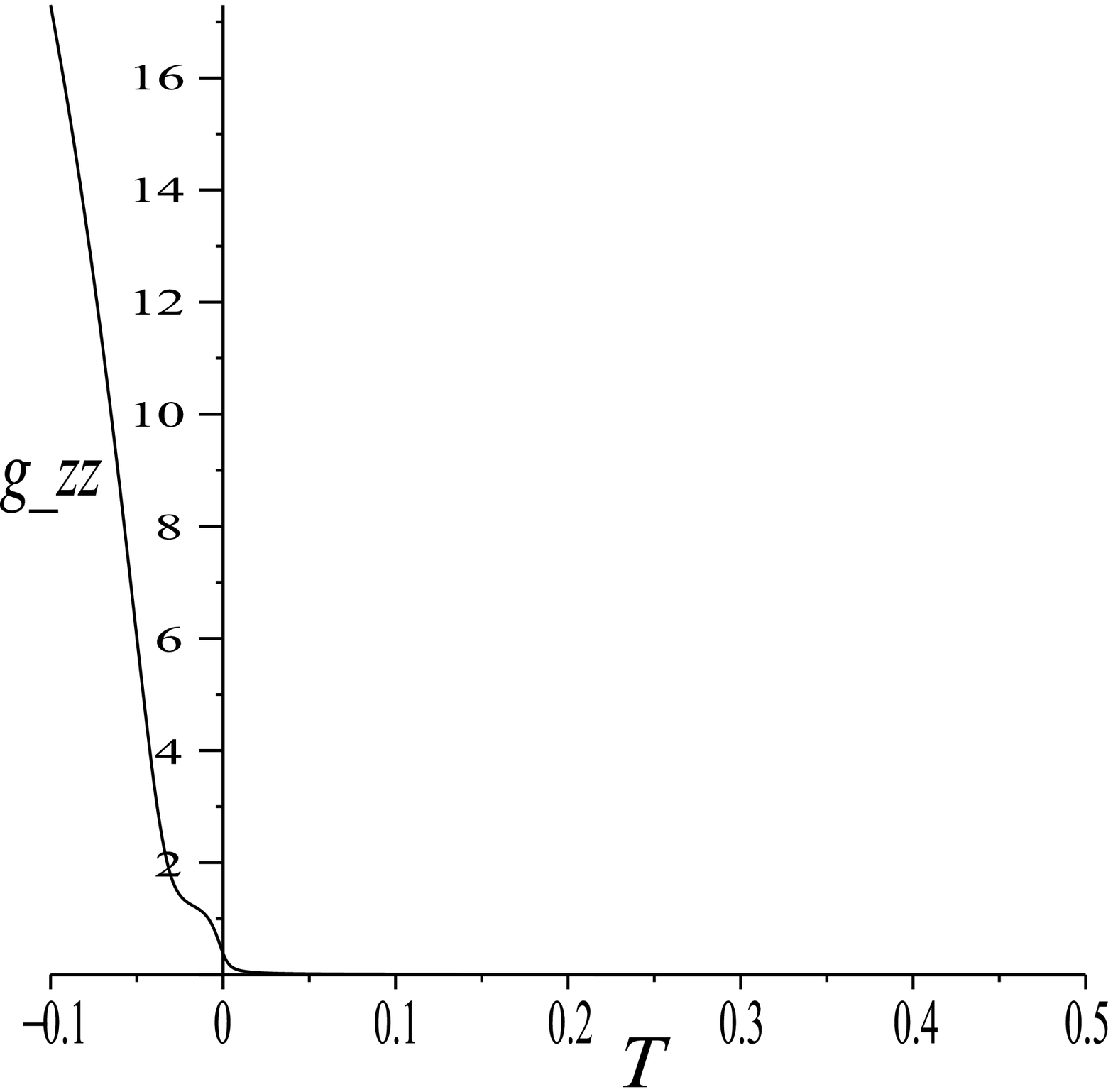} \vspace{0.3cm}
\hspace{0.3cm} d) \includegraphics[height=1.25in, width=2.0in]{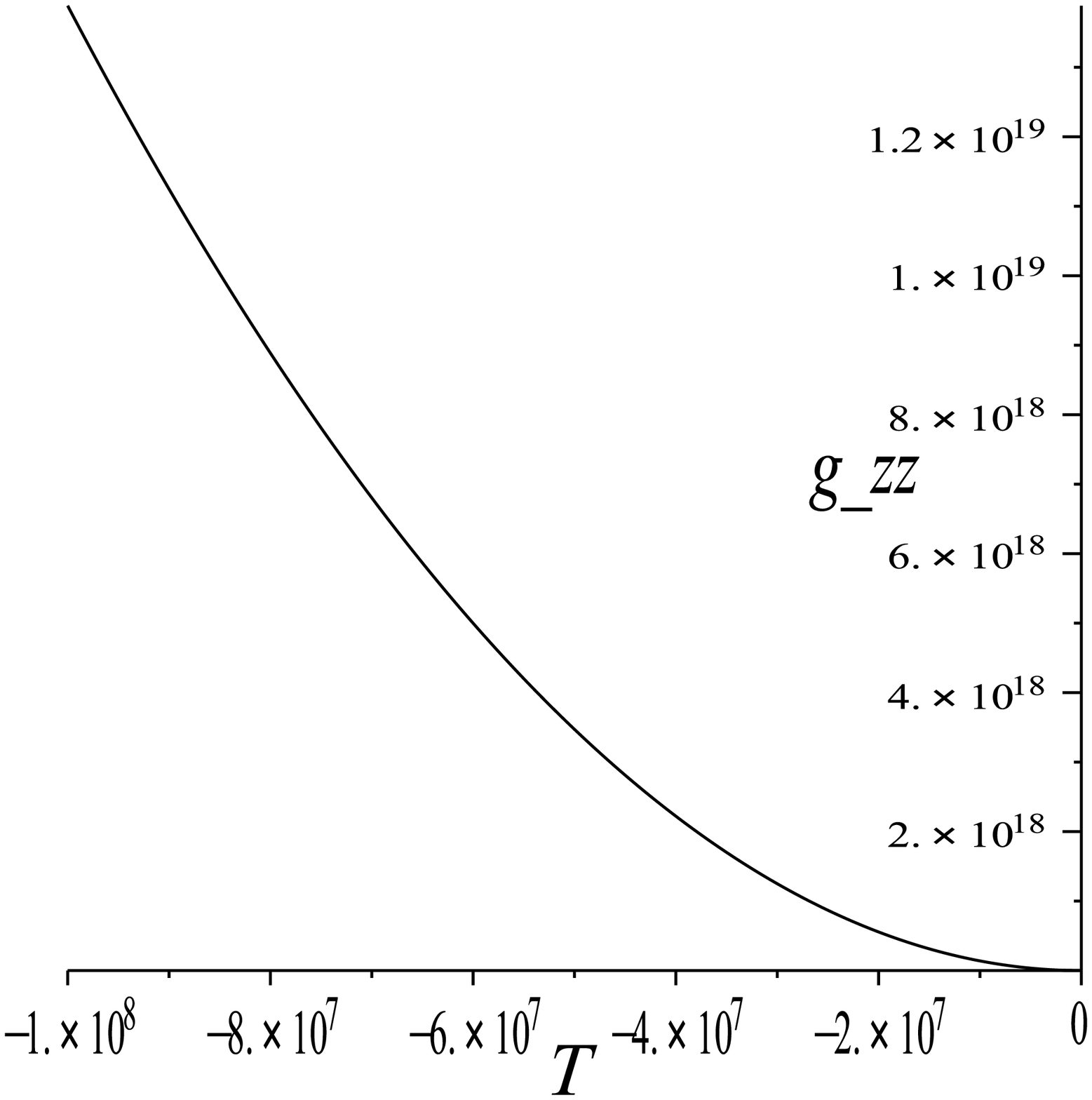}\vspace{0.3cm} 
\caption{{\small The quantum evolution for a planar black hole interior using a non-symmetric evolution scheme. a) The triad component $\pfive$ plotted in the vicinity of the classical singularity. b) The triad component $\pfive$ plotted for large negative $T$. c) The triad combination $\pone{}^2/\pfive$ which is equal to the classical metric component $g_{\zeta\zeta}$ in the vicinity of the classical singularity. d) The triad combination $\pone{}^2/\pfive$ for large negative values of $T$. The horizon is located at $T=1$. }}
\label{fig:nonsymplane}
\end{figure}

In figure \ref{fig:nonsymplane} we also show a sample evolution based on (\ref{eq:pambig1}) for comparison. It may be seen that again the singularity is avoided. However, with the same parameters as the previous evolution scheme, the rate at which the asymptotic solution is achieved is different than with the symmetric scenario. 

In closing this section we note that the results for the flat cylindrical black hole should be similar to the planar case. The flat cylinder arises from an identification of one of the coordinates, $x$ or $y$ in the metric (\ref{eq:planarmet}).

\subsection{The higher genus black hole interiors}
In this section we discuss the interior of higher genus black holes. A reasonably general line element describing such interiors is provided by \cite{ref:tor2}
\begin{equation}
 ds^{2}= -\frac{dT}{\frac{2M}{T}-\beta+\frac{\Lambda}{3}T^{2}} + \left(\frac{2M}{T}-\beta+\frac{\Lambda}{3}T^{2}\right)dR^{2} + T^{2}\,d\theta^{2} + T^{2}c\sinh^{2}(\sqrt{d}\theta)\,d\phi^{2}\,, \label{eq:highergmetric}
\end{equation}
with $\Lambda < 0$.
The two cases of interest here are the following: \\
i) $\beta=0$, $d=0,\,$ $\underset{d\rightarrow 0}\lim\,c=\frac{1}{d}$: In this case constant $(T,\,R)$ surfaces are tori. \\
ii) $\beta=-1$, $c=d=1$: In this case constant $(T,\,R)$ surfaces are surfaces of constant negative curvature of genus $g > 1$, depending on the identifications chosen. \\
An event horizon exists when $\left(\frac{2M}{T}-\beta+\frac{\Lambda}{3}T^{2}\right)=0$. Black holes with higher genus topology have also been studied in \cite{ref:vanzo} and \cite{ref:tor1}.

The connection and densitized triad, for the $SO(2,1)$ symmetry of the constant $(T,\,R)$ subspaces, are found to be
\begin{subequations}
 \begin{align}
A=&a_{3} \tau_{R}\,dR + \left(a_{1} \tau_{\theta} + a_{2} \tau_{\phi}\right)d\theta +\left( a_{2}\tau_{\theta} - a_{1}\tau_{\phi}\right)\sqrt{c}\sinh(\sqrt{d}\theta)\,d\phi \nonumber \\
& +\tau_{R}\sqrt{c}\sqrt{d}\cosh(\sqrt{d}\theta)\,d\phi\,, \label{eq:hyperA} \\
E=& -E_{3} \tau_{R} \sqrt{c}\sinh(\sqrt{d}\theta) \frac{\partial}{\partial R} - \left(E_{1}\tau_{\theta} +E_{2}\tau_{\phi}\right)\sqrt{c}\sinh(\sqrt{d}\theta) \frac{\partial}{\partial \theta} \nonumber \\
& +\left(E_{1}\tau_{\phi}-E_{2}\tau_{\theta}\right) \frac{\partial}{\partial\phi}\,. \label{eq:hyperE} 
 \end{align}
\end{subequations}
In the limit $d=0,\,$ and $\underset{d\rightarrow 0}\lim\,c=\frac{1}{d}$ this also yields the symmetry required for the toroidal case. That such a pair is indeed invariant under the appropriate symmetry group can be confirmed via the existence of a function $W$ such that
\begin{equation}
 \mathcal{L}_{\xi}A = D\,W\,,\;\;\; \mathcal{L}_\xi E =
[E,\,W]\,, \label{eq:liechecks}
\end{equation}
with the operator $D$ being the gauge covariant derivative, $\partial_{a} W + [W,A_{a}]$ ( $[\;\;\; ,\;\;\; ]$ is
the su(2) Lie bracket). $\mathcal{L}_{\xi}$ is the Lie derivative in the direction of the Killing vectors generating the symmetry, which for $SO(2,1)$ are:
\begin{equation}
 \xi^{1}:=\sin(\phi) \partial_{\theta} + \cos(\phi)\coth(\theta) \partial_{\phi}\,,\;
\xi^{2}:=-\cos(\phi) \partial_{\theta} + \sin(\phi)\coth(\theta) \partial_{\phi}\,,\;
\xi^{3}:=\partial_{\phi}\,. 
\end{equation}
It can be verified that ${_{1}W}=\cos(\phi)/\sinh(\theta) \tau_{R}$ and ${_{2}W}= \sin(\phi)/\sinh(\theta) \tau_{R}$ satisfy (\ref{eq:liechecks}). In the planar case, the situation was trivial, as setting $W=0$ satisfied (\ref{eq:liechecks}) which is then simply a statement that $A$ and $E$ were independent of $x$ and $y$.
 
The Gauss constraint again yields a single condition:
\begin{equation}
 2\sqrt{c}\sinh\left(\sqrt{d}\theta\right)\left[\bee E_{1}-a_{1}\pb\right]=0\,, \label{eq:hypergc}
\end{equation}
which will be satisfied by setting $a_{1}=E_{1}=0$. The vector constraint will also be satisfied by this choice. The resulting classical Hamiltonian is given by
\begin{equation}
 H=-\frac{N}{2}\frac{\sqrt{c} \left[\cosh(\sqrt{d}\theta_{1})-1\right]}{\sqrt{d}\sqrt{\pc}\gamma^{2}} \left[\left(\bee^{2} -\gamma^{2}d\right)\pb+2\pc\cee\bee -\Lambda\pc\pb\right]\,, \label{eq:hyperham}
\end{equation}
where $\theta_{1}$ represents the upper limit of the $\theta$ integration. In order to deal directly in the coordinate system of (\ref{eq:highergmetric}), we choose the following:
\begin{equation}
 N=\frac{\gamma^{2}\sqrt{d}\sqrt{\pc}}{\sqrt{c}\left[\cosh(\sqrt{d}\theta_{1})-1 \right]\pb}\, . \label{eq:hyperN}
\end{equation}
At this stage, the quantum prescription is similar to the scenarios in the previous sections. That is, the length of the holonomy paths are taken as
\begin{equation}
 \delta_{2}=\sqrt{\frac{\Delta}{E_{3}}}\,, \;\;\;\; \delta_{3}=\frac{\sqrt{E_{3}\, \Delta}}{E_{2}}\,,
\end{equation}
which is appropriate as (\ref{eq:highergmetric}) admits proper areas of
\begin{equation}
 {\sf{A}}_{R\theta}=E_{2}\delta_{2}\delta_{3} = \Delta\,, \;\;\;\;{\sf{A}}_{\theta\phi}=\delta_{2}^{2}  E_{3} = \Delta\,. \label{eq:hyperareas}
\end{equation}
The above Hamiltonian is then quantized via
\begin{equation}
 \bee \rightarrow \frac{\sin(\bee\delta_{2})}{\delta_{2}}\,, \;\;\;\; \cee \rightarrow \frac{\sin(\cee\delta_{3})}{\delta_{3}}\,.
\end{equation}
At this stage we should note that the particular choice of $N$ given in  (\ref{eq:hyperN}) possesses an interesting advantage over other choices of the lapse function. Note that with this choice of $N$, the term in the Hamiltonian which depends on the parameter $d$ \emph{decouples} from the degrees of freedom. That is, with this choice, the evolution equations for the S-adS, toroidal and higher genus scenarios are similar. The difference between these cases then lies only in the initial conditions, as, for example, both the initial conditions for $\pb$ and $\bee$ involve the parameter $\beta$, which is different for the three cases. We therefore do not present the equations of motion again here but simply show the results of the evolutions for both the toroidal and higher genus cases in figures \ref{fig:tor} and \ref{fig:higherg} below.

\begin{figure}[h!t]
\centering
\vspace{-1.6cm} a) \includegraphics[height=1.25in, width=2.0in]{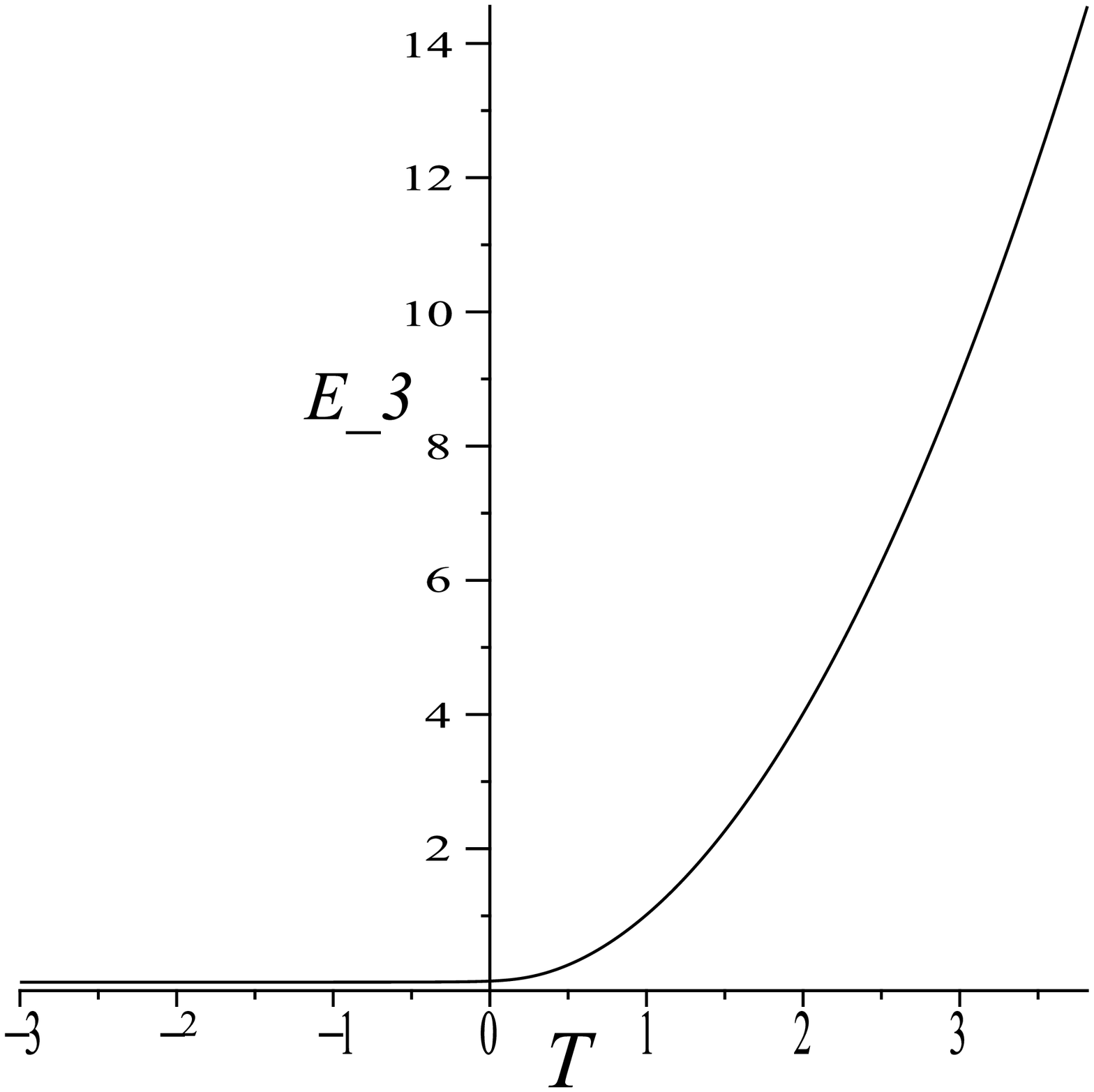}\vspace{0.3cm}
\hspace{0.3cm} b) \includegraphics[height=1.25in, width=2.0in]{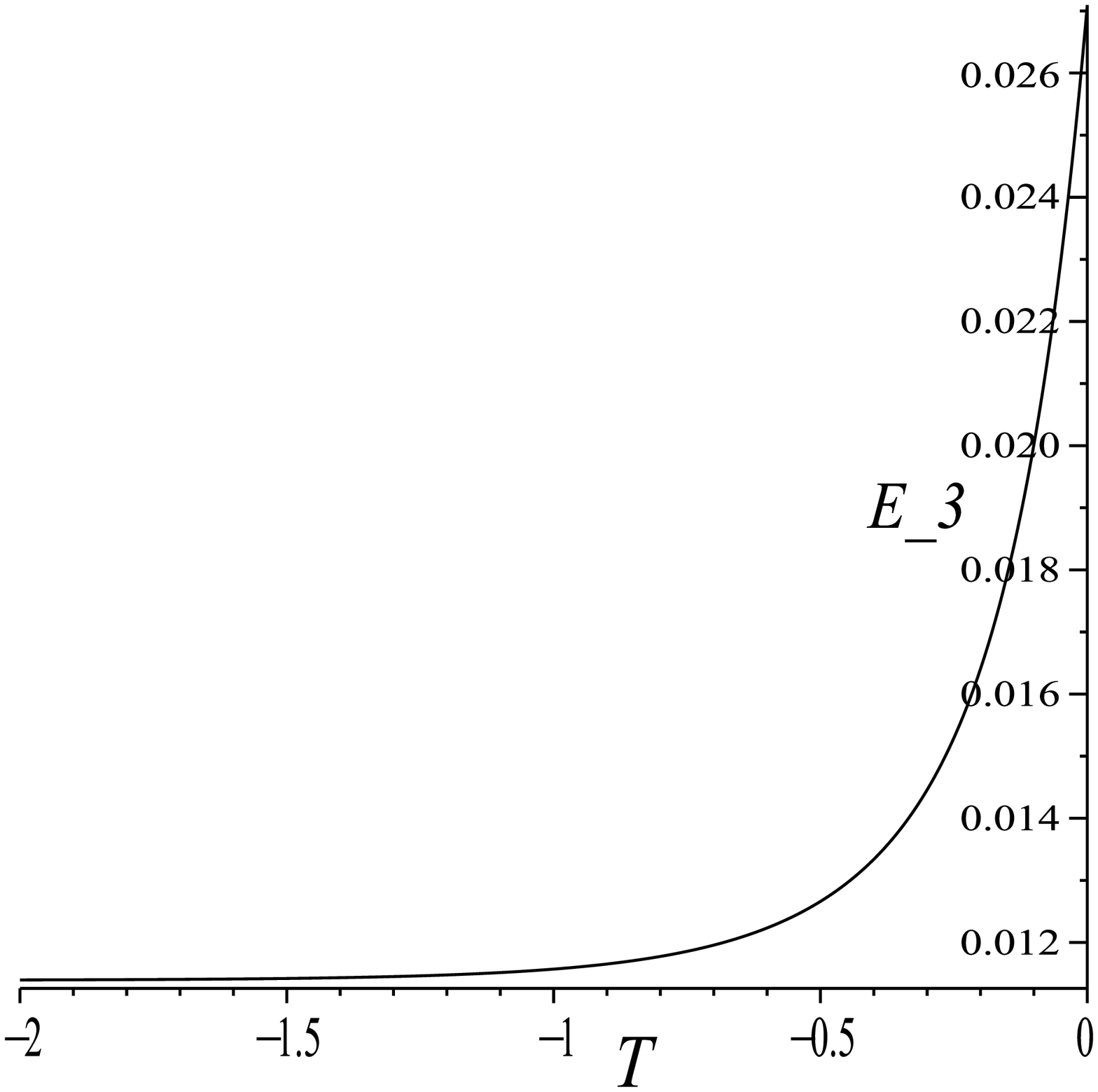} \vspace{0.3cm} \\
c) \includegraphics[height=1.25in, width=2.0in]{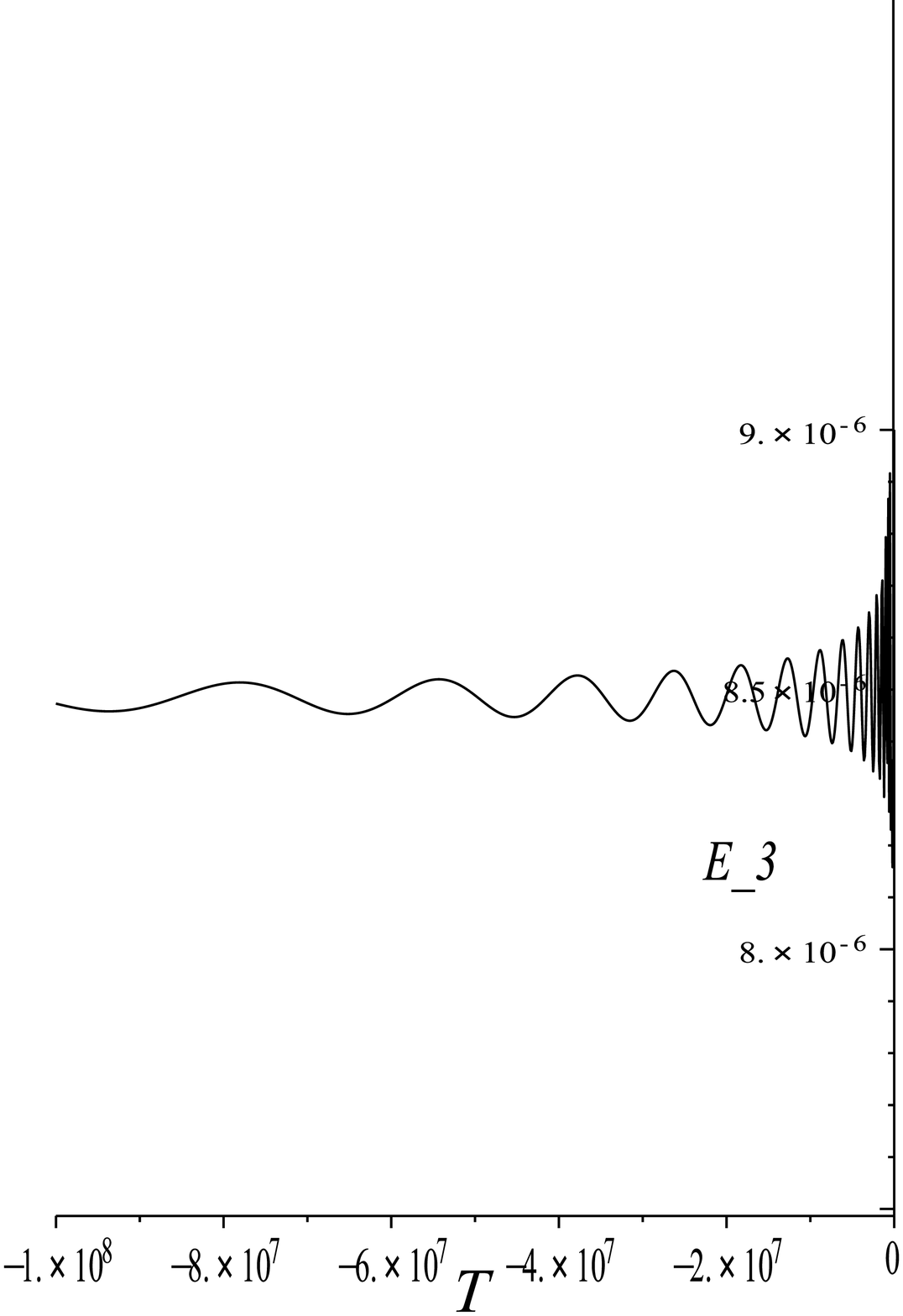} \vspace{0.3cm}
\hspace{0.3cm} d) \includegraphics[height=1.25in, width=2.0in]{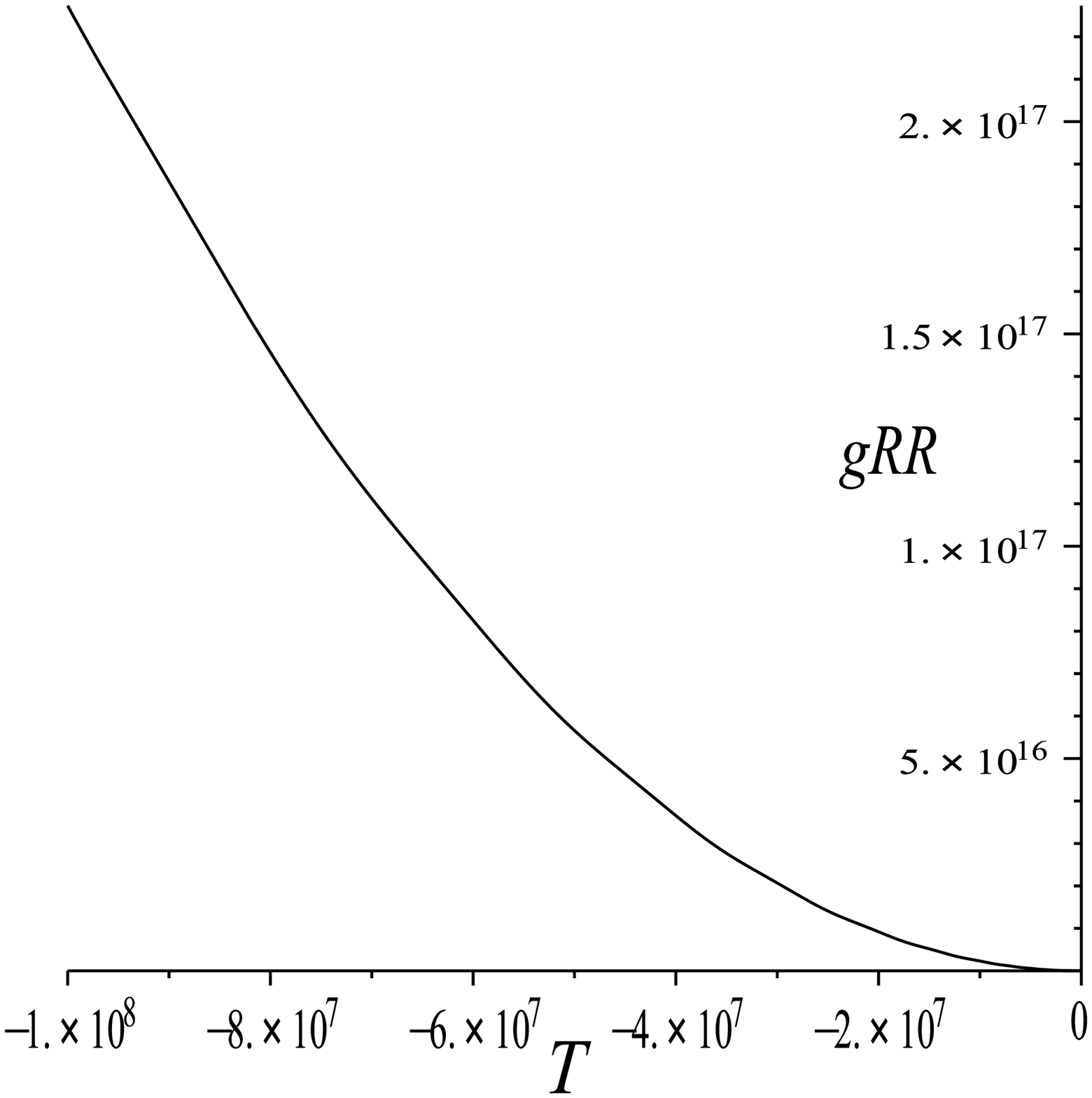}\vspace{0.3cm} \\
\hspace{0.0cm} e) \includegraphics[height=1.25in, width=2.0in]{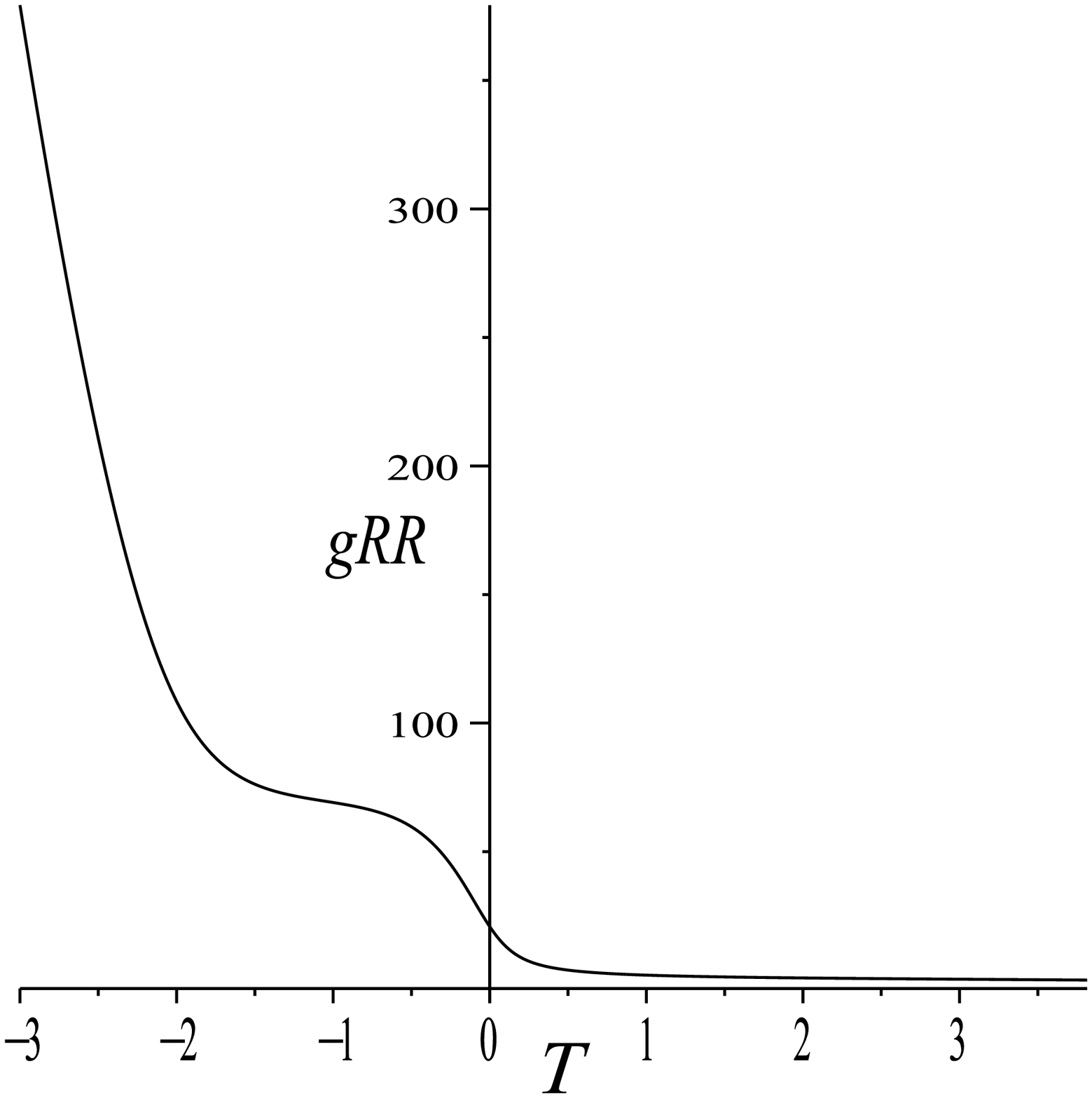}
\hspace{0.3cm} f) \vspace{0.3cm} \includegraphics[height=1.5in, width=2.0in]{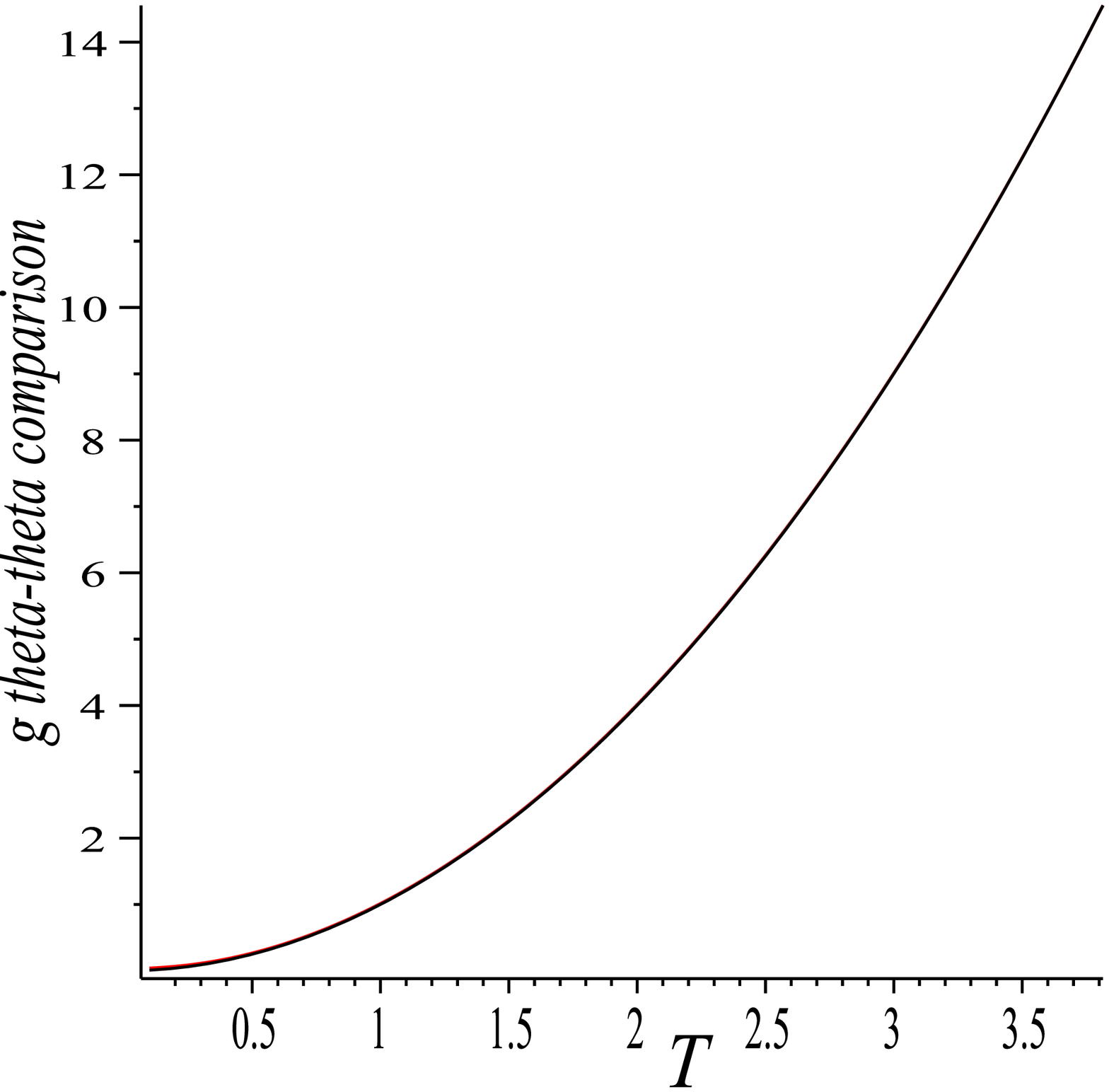}\vspace{0.3cm}\\
\hspace{0.0cm} g) \includegraphics[height=1.25in, width=2.0in]{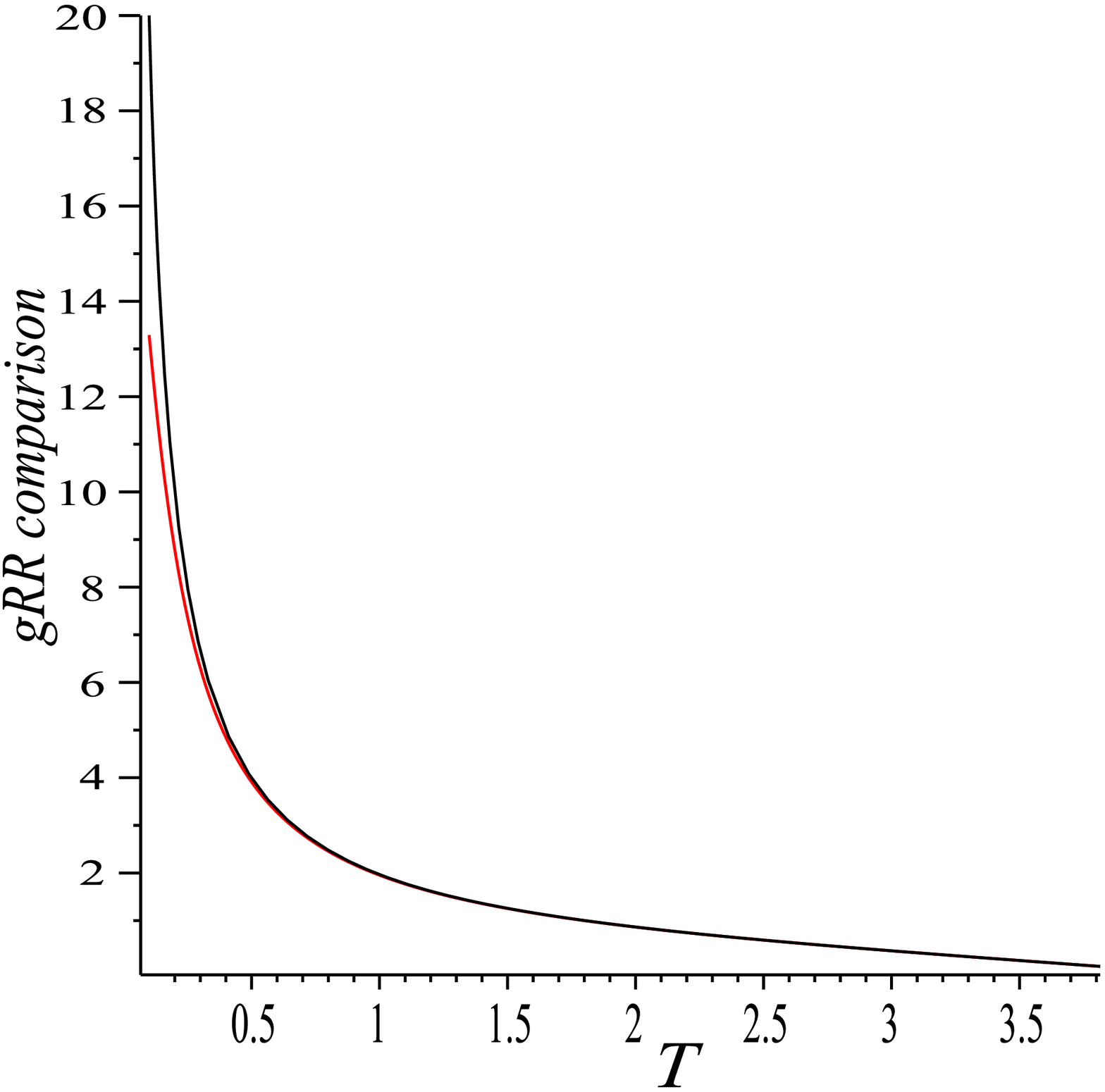} \vspace{0.3cm} 
\hspace{0.3cm} h) \includegraphics[height=1.25in, width=2.0in]{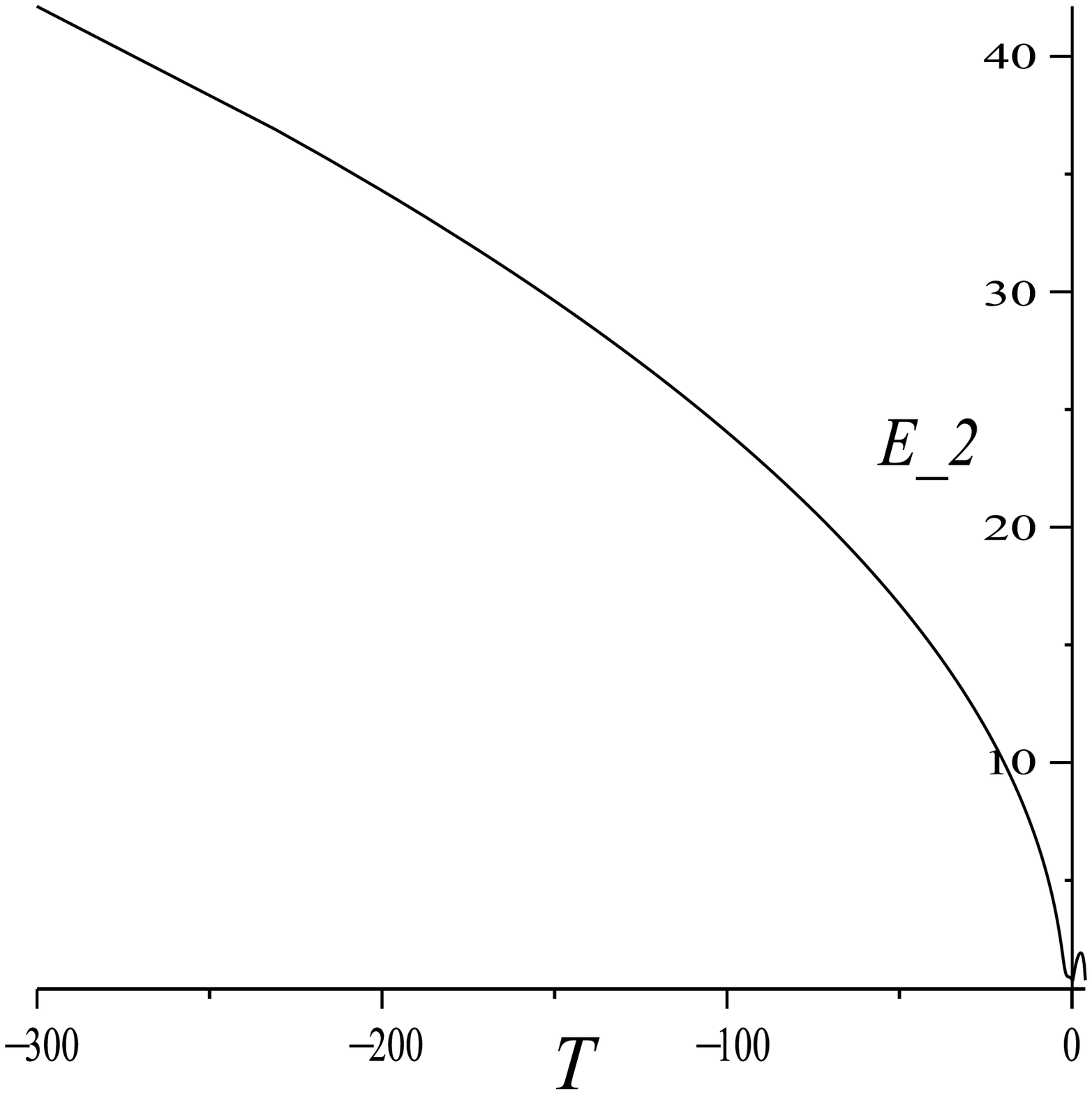}
\caption{{\small The quantum evolution for a toroidal black hole interior. a) The triad component $E_{3}$ plotted in the vicinity of the classical singularity. b) The triad component $E_{3}$ plotted from $T=-2$ to $T\approx 0$. c) The triad component $E_{3}$ up to $T=-1\times 10^{8}$. d) The triad combination $E_{2}^{2}/E_{3}$ which is equal to the classical metric component $g_{RR}$. e) A close-up of the previous plot, near the classical singularity. f) A comparison of the classical (black) vs. quantum (red) evolution of $E_{3}$ for positive $T$. g) A comparison of the classical (black) vs. quantum (red) evolution of $E_{2}^{2}/E_{3}$ for positive $T$. h) The triad component $E_{2}$. The inner horizon is located at $T\approx 3.8$ and the following parameters were used: $\gamma=0.274$, $M=1$, $\Lambda \approx -0.1$. }}
\label{fig:tor}
\end{figure}

\begin{figure}[h!t]
\centering
\vspace{-1.6cm} a) \includegraphics[height=1.25in, width=2.0in]{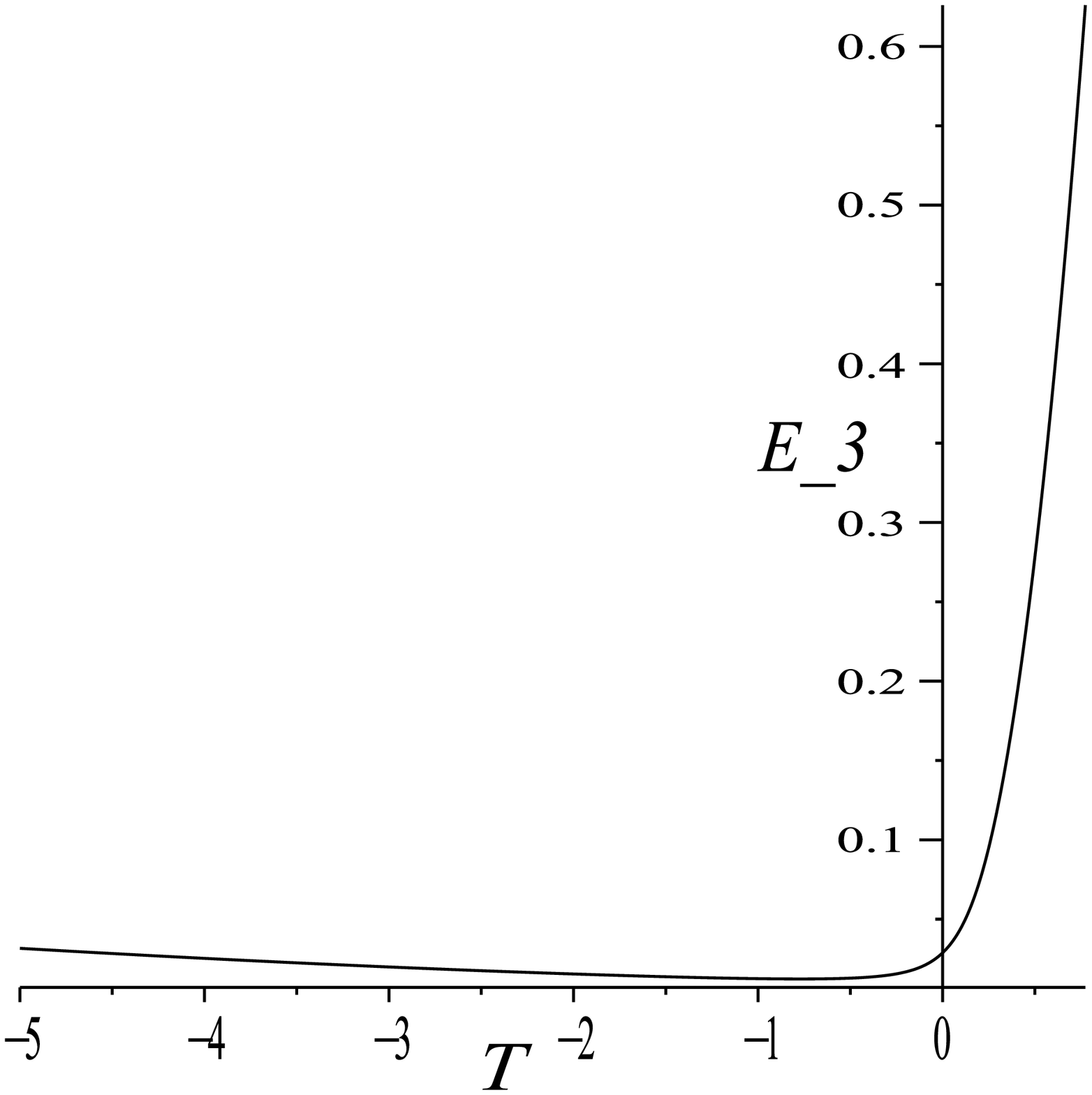}\vspace{0.3cm}
\hspace{0.3cm} b) \includegraphics[height=1.25in, width=2.0in]{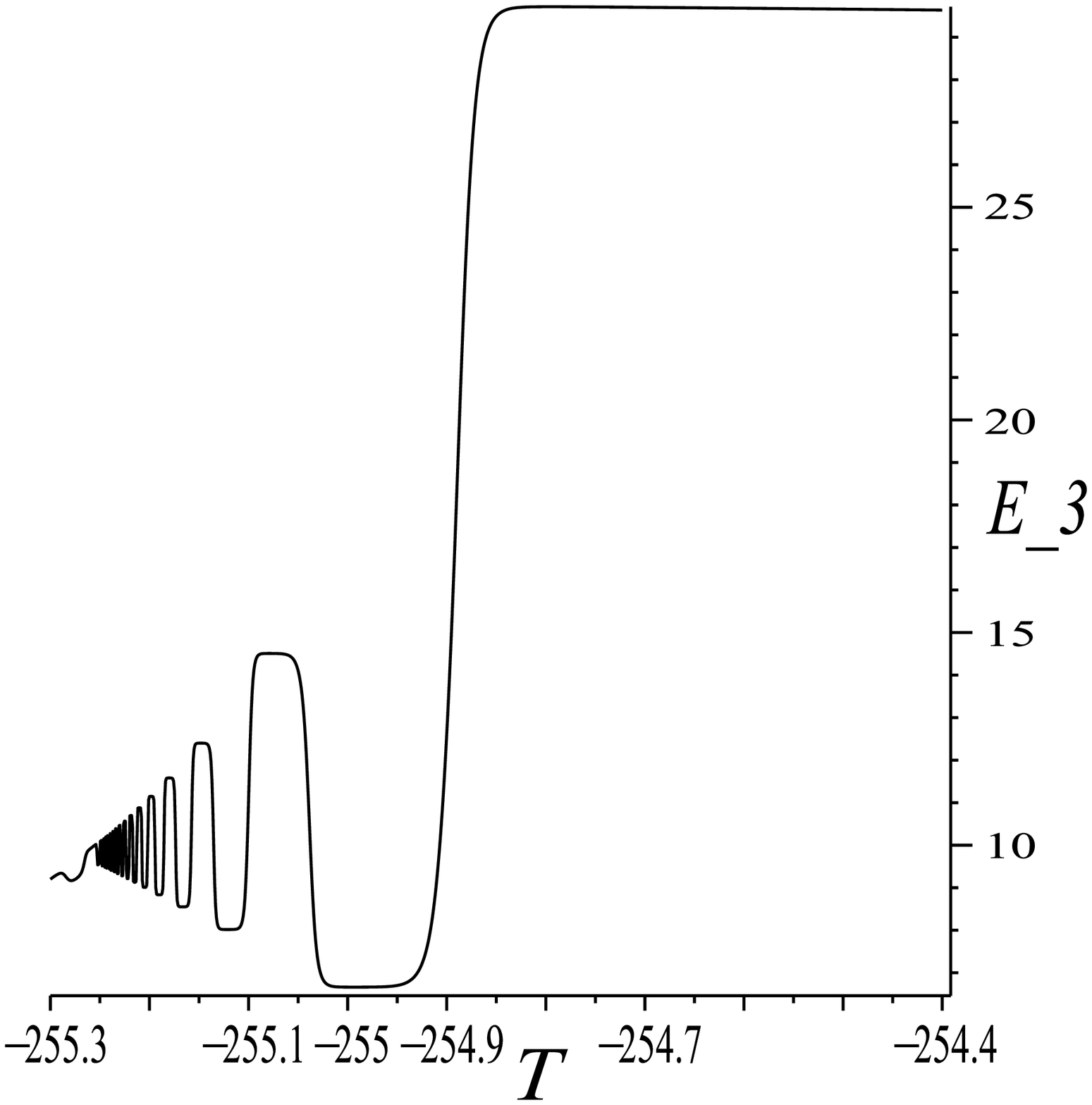} \vspace{0.3cm} \\
c) \includegraphics[height=1.25in, width=2.0in]{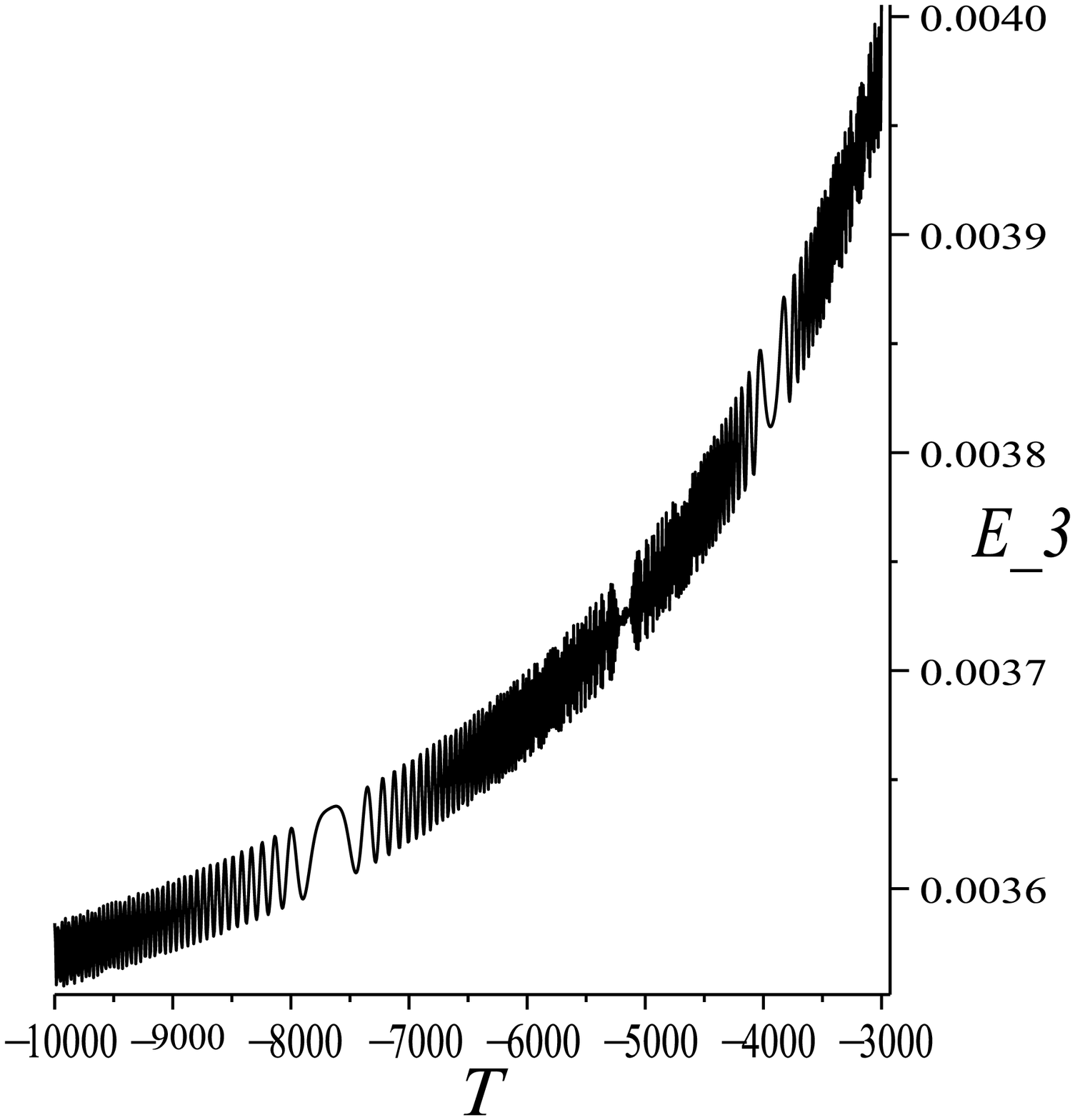} \vspace{0.3cm}
\hspace{0.3cm} d) \includegraphics[height=1.25in, width=2.0in]{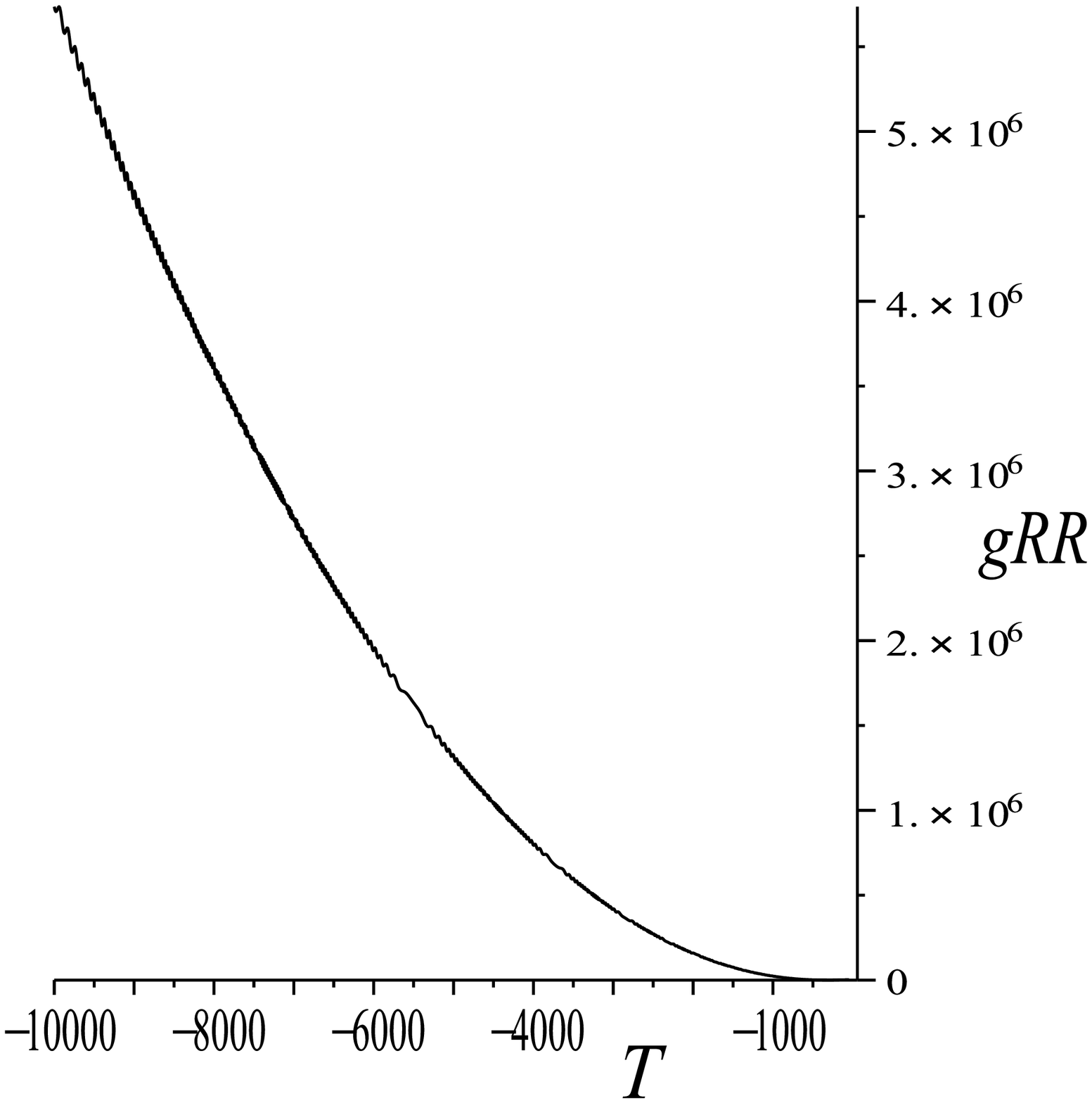}\vspace{0.3cm} \\
\hspace{0.0cm} e) \includegraphics[height=1.25in, width=2.0in]{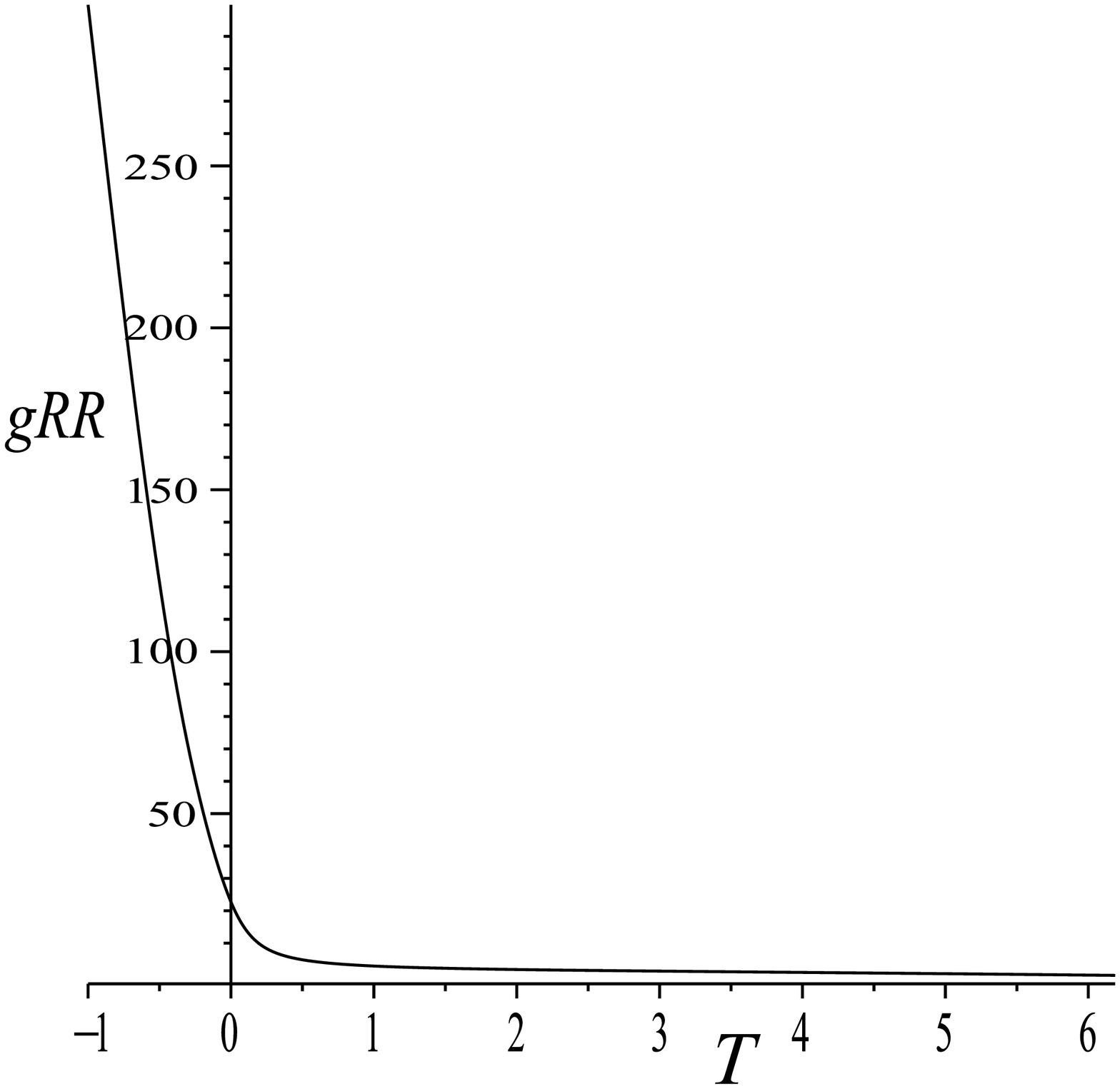}
\hspace{0.3cm} f) \vspace{0.3cm} \includegraphics[height=1.5in, width=2.0in]{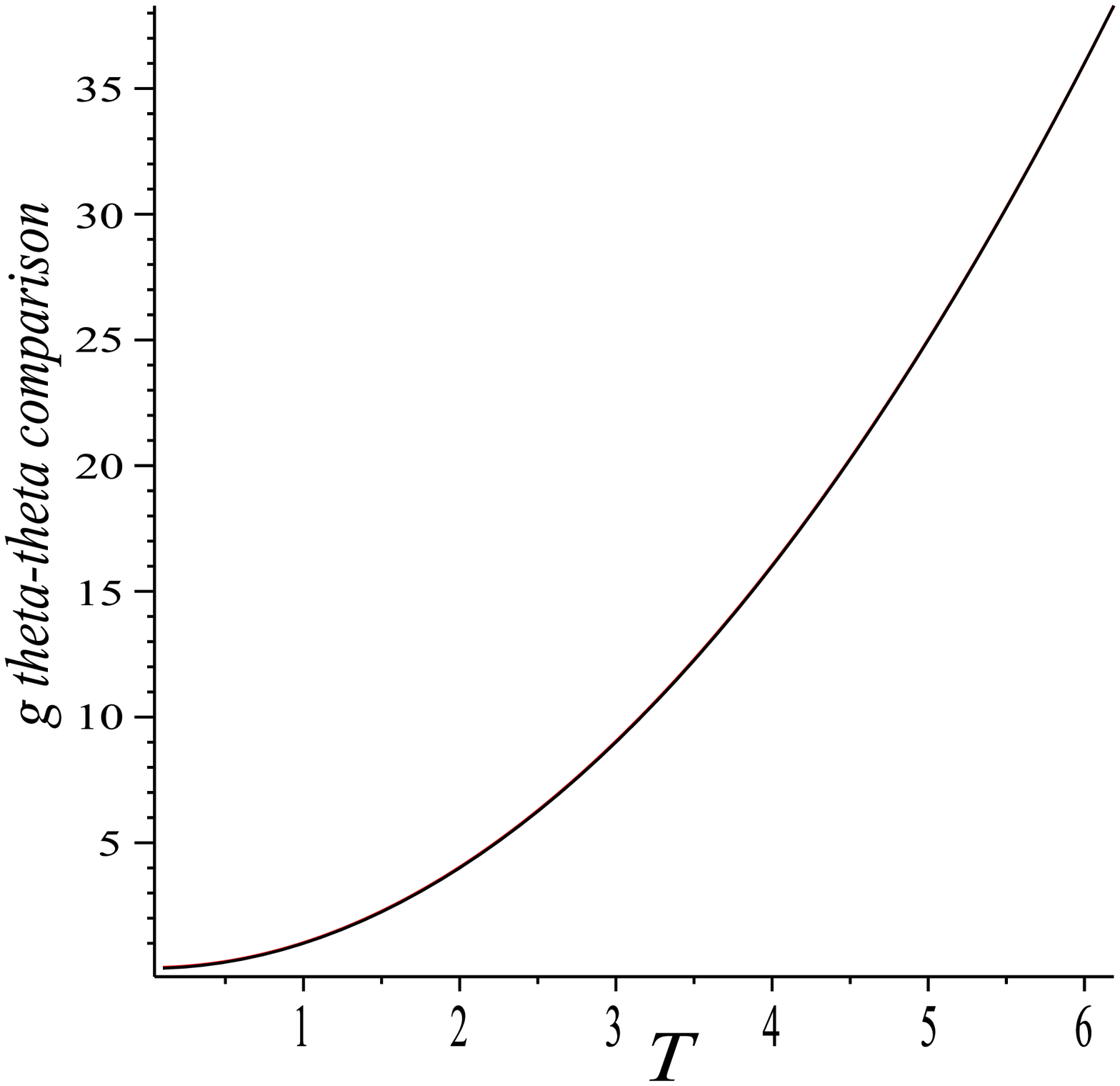}\vspace{0.3cm}\\
\hspace{0.0cm} g) \includegraphics[height=1.25in, width=2.0in]{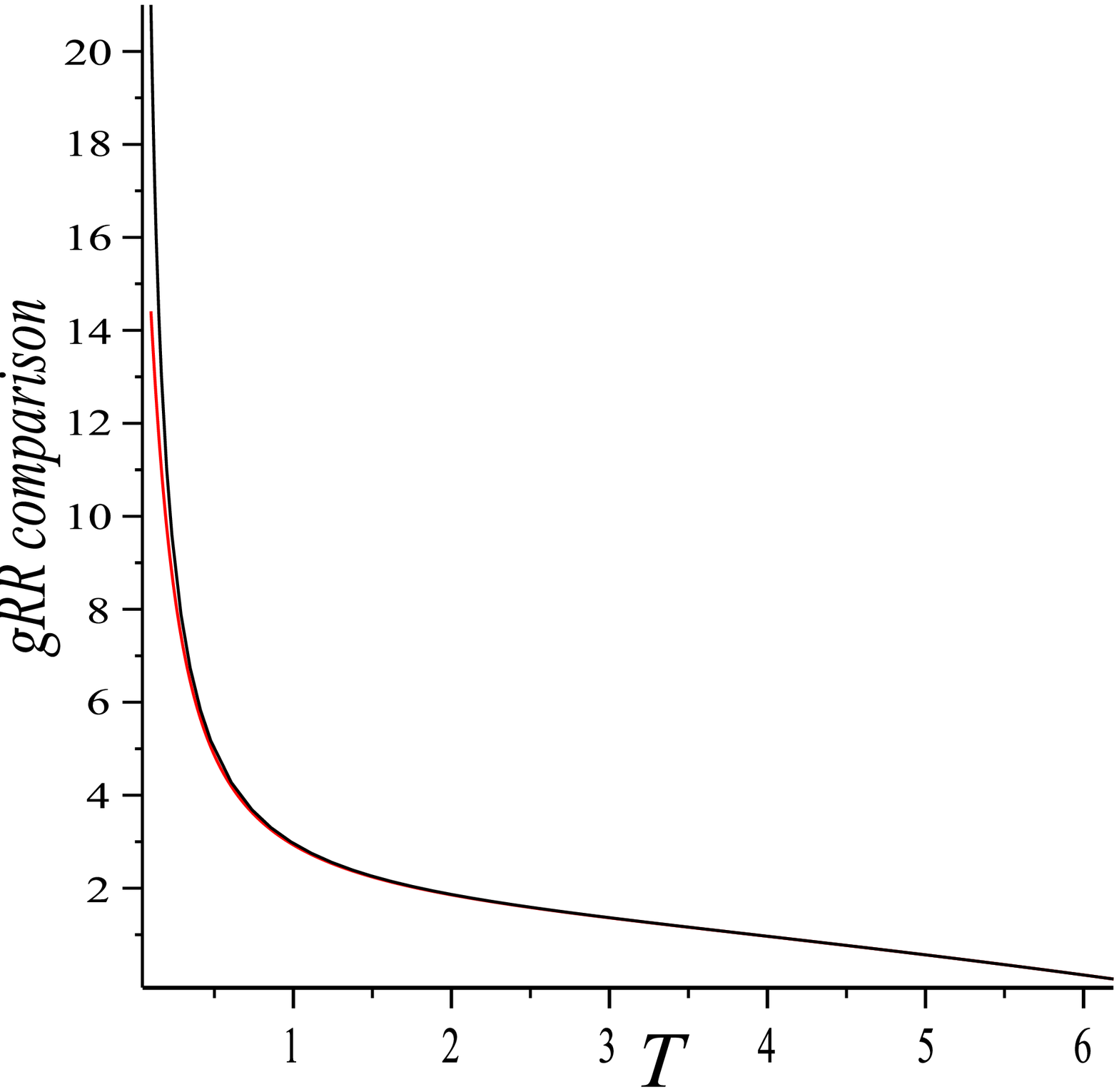} \vspace{0.3cm} 
\hspace{0.3cm} h) \includegraphics[height=1.25in, width=2.0in]{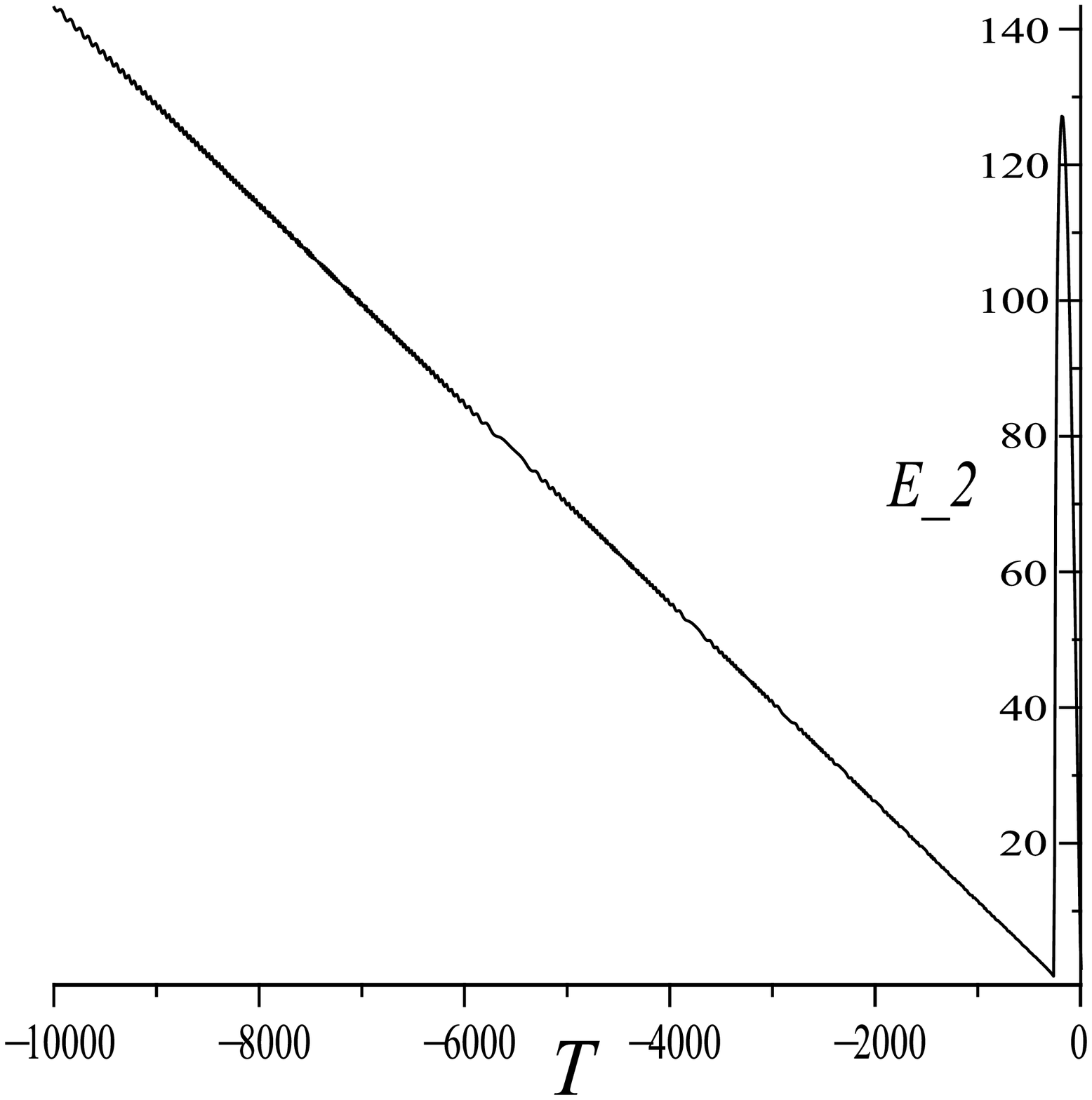}
\caption{{\small The quantum evolution for a higher-genus hyperbolic black hole interior. a) The triad component $E_{3}$ plotted in the vicinity of the classical singularity. b) The triad component $E_{3}$ plotted from $T=-255.5$ to $T\approx -254$. c) The triad component $E_{3}$ up to $T=-10000$. d) The triad combination $E_{2}^{2}/E_{3}$ which is equal to the classical metric component $g_{RR}$. e) A close-up of the previous plot, near the classical singularity. f) A comparison of the classical (black) vs. quantum (red) evolution of $E_{3}$ for positive $T$. g) A comparison of the classical (black) vs. quantum (red) evolution of $E_{2}^{2}/E_{3}$ for positive $T$. h) The triad component $E_{2}$. The inner horizon is located at $T\approx 6.2$ and the following parameters were used: $\gamma=0.274$, $M=1$, $\Lambda\approx -0.1$. }}
\label{fig:higherg}
\end{figure}

The higher genus hyperbolic evolution was the most computationally intense and therefore the evolution was halted at $T=-10000$ for the $g>1$ case. This evolution is also the ``noisiest'' and produced the highest frequency of the resulting oscillation, as can be seen from figure \ref{fig:higherg}. If these oscillations are a genuine product of the evolution, then it is perhaps more appropriate to state that the higher genus case approaches an asymptotic state which oscillates about a Nariai space-time, as the average of $E_{3}$ is seemingly approaching a constant.

The late-time evolution of the torus was also a Nariai-type universe, similar to (\ref{eq:nariaitype}), with the appropriate toroidally compatible 2-space symmetry. We performed several toroidal evolutions and noticed that in all of them, the frequency of the late-time oscilations was also greater than in their spherical and planar counter-parts. 

In \cite{ref:bvstable} B\"{o}hmer and Vandersloot show that the Schwartzschild interior
solution is a stable solution of the LQG equations. We suspect that all of the interior solutions we have
discussed will likewise be stable.

\section{Concluding remarks}
In this manuscript we have studied the evolution of the Schwarzschild-deSitter, Schwarz\-schild-anti deSitter, planar, toroidal, and higher genus hyperbolic black holes in the mini-superspace approximation to loop quantum gravity. We have utilized the improved quantization scheme where the length of the holonomy paths are not constant, but instead are related to the proper area dicated by the corresponding classical densitized triad. We have found that in all cases, the classical singularity is avoided and, at some point during the evolution a ``quantum bounce'' occurs, usually followed by a series of less severe bounces. The exact details of the bounces and evolutions depend on the symmetry but the indication from all of these evolutions is that Nariai-type solutions seem to be generic attractors for the asymptotic space-time on the other side of the bounces for most cases studied. This is also an indication that loop quantum gravity is consistent in replacing classically singular space-times with non-singular ones, regardless of symmetry, at least at the level of approximation employed here. It would be interesting to improve on this scheme even further to see which of these features remain intact as one lessens the degree of approximation. Progress along this front is already being made. For example, see discussions in \cite{ref:newbojo} and \cite{ref:newmodesto}.

\section*{Acknowledgments}
We would like to thank K. Vandersloot for helpful comments and the anonymous referees for comments and suggestions that have improved the manuscript.

\linespread{0.6}
\bibliographystyle{unsrt}

\begin{thebibliography}{10}
{\small

\bibitem{ref:lrrLQG}
C. Rovelli,
\newblock {\em Living Rev. Relativity} {\bf 1}  (1998),  1.
http://www.livingreviews.org/lrr-1998-1. 

\bibitem{ref:thiem}
T. Thiemann,
\newblock {\em gr-qc/0110034}.

\bibitem{ref:rovbook}
C. Rovelli,
\newblock {\em Quantum Gravity} (Cambridge University Press, Cambridge, 2004).

\bibitem{ref:rev1}
A. Perez,
\newblock {\em gr-qc/0409061}.

\bibitem{ref:rev2}
A. Ashtekar and J. Lewandowski,
\newblock {\em Class. Quant. Grav.} {\bf 21} (2004) R53.

\bibitem{ref:entrev1}
C. Rovelli,
\newblock {\em Phys. Rev. Let.} {\bf 77} (1996) 3288.

\bibitem{ref:ashbaez}
A. Ashtekar, J.~C. Baez and K. Krasnov,
\newblock {\em Adv. Theor. Math. Phys.} {\bf 4} (2000) 1.

\bibitem{ref:daskaul}
S. Das, R.~K. Kaul and P. Majumdar,
\newblock {\em Phys. Rev.} {\bf D63} (2001) 044019.

\bibitem{ref:chatmaj}
A. Chatterjee and P. Majumdar,
\newblock {\em gr-qc/0303030}.

\bibitem{ref:domlew}
M. Domagala and J. Lewandowski,
\newblock {\em Class. Quant. Grav.} {\bf 21} (2004) 5233.

\bibitem{ref:entrev2}
K.~A. Meissner,
\newblock {\em Class. Quant. Grav.} {\bf 21} (2004) 5245.

\bibitem{ref:entrev3}
A. Ghosh and P. Mitra,
\newblock {\em Phys. Let.} {\bf B616} (2005) 114.

\bibitem{ref:entrev3_5}
T. Tamaki and H. Nomura,
\newblock {\em Phys. Rev.} {\bf D72} (2005) 107501.

\bibitem{ref:dms}
O. Dreyer, F. Markopoulou and L. Smolin,
\newblock {\em Nucl.Phys.} {\bf B744} (2006) 1.

\bibitem{ref:entrev4}
A. Ghosh and P. Mitra,
\newblock {\em Indian J. Phys.} {\bf 80} (2006) 867.

\bibitem{ref:entrev5}
M.~H. Ansari,
\newblock {\em gr-qc/0603121}.

\bibitem{ref:entrev6}
A. Corichi, J. D\'{i}az-Polo and E. Fern\'{a}ndez-Borja,
\newblock {\em gr-qc/0605014}.

\bibitem{ref:KBD}
S. Kloster, J. Brannl\"{u}nd and A. DeBenedictis,
\newblock {\em Class. Quant. Grav.} {\bf 25} (2008) 065008.

\bibitem{ref:manuel1}
J.~M. Garcia-Islas,
\newblock {\em Class. Quant. Grav.} {\bf 25} (2008) 245001. 

\bibitem{ref:manuel2}
J.~M. Garcia-Islas,
\newblock {\em Class. Quant. Grav.} {\bf 25} (2008) 238001

\bibitem{ref:vanzo}
L. Vanzo,
\newblock {\em Phys. Rev.} {\bf D56} (1997) 6475.

\bibitem{ref:mans}
R.~B. Mann and S.~N. Solodukhin,
\newblock {\em Nucl.Phys.} {\bf B523} (1998) 293.

\bibitem{ref:liko}
T. Liko,
\newblock {\em Phys. Rev.} {\bf D77} (2008) 064004.

\bibitem{ref:LQClrr}
M. Bojowald,
\newblock {\em Living Rev. Relativity} {\bf 11}  (2008) 4.
http://www.livingreviews.org/lrr-2008-4.

\bibitem{ref:modestobhsing}
L. Modesto,
\newblock {\em Phys. Rev.} {\bf D70} (2004) 124009.

\bibitem{ref:modestolqbh}
L. Modesto,
\newblock {\em Class. Quant. Grav.} {\bf 23} (2006) 5587.

\bibitem{ref:modestoint}
L. Modesto,
\newblock {\em gr-qc/0611043}.

\bibitem{ref:modestosing2}
L. Modesto,
\newblock {\em Proc. XVII SIGRAV} (2006).

\bibitem{ref:AandB}
A. Ashtekar and M. Bojowald,
\newblock {\em Class. Quant. Grav.} {\bf 26} (2006) 391.

\bibitem{ref:BandV}
C.~G. B\"{o}hmer and K. Vandersloot,
\newblock {\em Phys. Rev.} {\bf D76} (2007) 104030.

\bibitem{ref:chiou2}
D-W. Chiou,
\newblock {\em Phys. Rev.} {\bf D78} (2008) 064040.

\bibitem{ref:CGPint}
M.~ Campiglia, R. Gambini, and J. Pullin,
\newblock {\em AIP Conf. Proc.}: Third Mexican Meeting on Mathematical and Experimental Physics, (2008) 52.

\bibitem{ref:rg1}
R.~G. Cai and Y.~Z. Zhang,
\newblock  {\em Phys. Rev.} {\bf D54} (1996) 4891.

\bibitem{ref:tor1}
J.~P.~S. Lemos and V.~T. Zanchin,
\newblock {\em Phys. Rev.} {\bf D54} (1996) 3840.

\bibitem{ref:tor2}
W.~L. Smith and R.~B. Mann,
\newblock {\em Phys. Rev.} {\bf D56} (1997) 4942.

\bibitem{ref:tor3}
J.~P.~S. Lemos,
\newblock {\em Phys. Rev.} {\bf D57} (1998) 4600.

\bibitem{ref:rg2}
R.~G. Cai, J.~Y. Ji and K.~S. Soh,
\newblock {\em Phys. Rev.} {\bf D57} (1998) 6547.

\bibitem{ref:rg3}
R.~G. Cai and K.~S. Soh,
\newblock {\em Phys. Rev.} {\bf D59} (1999) 044013.

\bibitem{ref:tor4}
S. Surya, K. Schleich and D. Witt,
\newblock {\em Phys. Rev.Let.} {\bf 86} (2001) 5231.

\bibitem{ref:mena}
F.~C. Mena, J. Nat\'{a}rio and P. Tod,
\newblock {\em 	arXiv:0707.2519} [gr-qc].

\bibitem{ref:BL}
I. Booth and T. Liko,
\newblock {\em Phys. Let.} {\bf B670} (2008) 61.

\bibitem{ref:APS}
A. Ashtekar, T. Pawlowski and P. Singh,
\newblock {\em Phys. Rev.} {\bf D74} (2006) 084003.

\bibitem{ref:APS2}
A. Ashtekar, T. Pawlowski, P. Singh and K. Vandersloot,
\newblock {\em Phys. Rev.} {\bf D75} (2007) 024035.

\bibitem{ref:vander2}
K. Vandersloot,
\newblock {\em Phys. Rev.} {\bf D75} (2007) 023523.

\bibitem{ref:LQC_difference}
M. Bojowald,
\newblock {\em Phys. Rev. Lett.} {\bf 86} (2001) 5227.

\bibitem{ref:chiou}
D-W. Chiou,
\newblock {\em Phys. Rev.} {\bf D76} (2007) 124037.

\bibitem{ref:newbojo}
M. Bojowald,
\newblock {\em arXiv:0811.4129} [gr-qc].

\bibitem{ref:zhou}
Z. Zhou,
\newblock {\em Helv. Phys. Acta} {\bf 65} (1992) 767.

\bibitem{ref:neville}
D.~E. Neville,
\newblock {\em Phys. Rev.} {\bf D73} (2006) 124004.

\bibitem{ref:bvstable}
C.~G. B\"{o}hmer and K. Vandersloot,
\newblock {\em Phys. Rev.} {\bf D78} (2008) 067501.

\bibitem{ref:newmodesto}
L. Modesto,
\newblock {\em arXiv:0811.2196} [gr-qc].

}

\end{thebibliography}

\end{document}